%% file: 6csoptical_final.tex
\documentclass{article}
\usepackage{kmn,latexsym,subfigure,epsfig,graphics}
\newcommand{\mc}{\multicolumn}

\input macro.tex
\def \cosone{$\Omega_{\rm M}=1$ and $\Omega_{\Lambda}=0$}

\def\HI{H{\footnotesize I}\,}
\begin{document}

\title[The 6C* sample II: Spectrophotometric data]{A sample of 6C radio sources
designed to find objects at redshift $> 4$: II --- spectrophotometry
and emission line properties}

\author[Jarvis et al.]
{Matt J.\,Jarvis$^{1,2}$\thanks{Email: jarvis@strw.leidenuniv.nl}, Steve
Rawlings$^{1}$, Mark Lacy$^{1,3,4}$, 
Katherine M.\,Blundell$^{1}$,
\and
Andrew J.\,Bunker$^{5,6}$, Steve Eales$^{7}$, Richard Saunders$^{8}$,
Hyron Spinrad$^{6}$, 
\and
Daniel Stern$^{6,9}$ and Chris J.\,Willott$^{1}$ \\
$^{1}$Astrophysics, Department of Physics, Keble Road, Oxford, OX1
3RH, UK \\
$^{2}$Sterrewacht Leiden, Postbus 9513, 2300 RA Leiden, the
Netherlands \\
$^{3}$Institute of Geophysics and Planetary Physics, L-413 Lawrence
Livermore National Laboratory, Livermore, CA 94550, USA \\
$^{4}$Department of Physics, University of California, 1 Shields
Avenue, Davis CA 95616, USA\\
$^{5}$Institute of Astronomy, University of Cambridge, Madingley Road,
Cambridge CB3 0HA, UK \\
$^{6}$Astronomy Department, University of California at Berkeley, CA 94720\\
$^{7}$Department of Physics and Astronomy, University of Wales 
College of Cardiff, P.O. Box 913, Cardiff, CF2 3YB, UK\\
$^{8}$Astrophysics Group, Cavendish Laboratory, Madingley Road, Cambridge
CB3 0HE, UK \\
$^{9}$Jet Propulsion Laboratory, California Institute of Technology, Mail
Stop 169-327, Pasadena, CA 91109 \\
}
\maketitle
\begin{abstract}
This is the second in a series of three papers which present and
interpret basic observational data on the 6C* 151-MHz radio sample: a
low-frequency selected sample which exploits filtering criteria based
on radio properties (steep spectral index and small angular size) to
find radio sources at redshift $z > 4$ within a 0.133\,sr patch of
sky. We present results of a programme of optical spectroscopy which
has yielded redshifts in the range $0.5 \ltsim z \ltsim 4.4$ for the
29 sources in the sample, all but six of which are secure.  We find that the filtering criteria used
for 6C* are very effective in excluding the low-redshift,
low-luminosity radio sources: the median redshift of 6C* is $z \approx
1.9$ compared to $z \approx 1.1$ for a complete sample matched in
151-MHz flux density. By combining the emission-line dataset
for the 6C* radio sources with those for the 3CRR, 6CE and 7CRS
samples we establish that $z \geq 1.75$ radio galaxies follow a rough
proportionality between Ly$\alpha$- and 151 MHz-luminosity which, like
similar correlations seen in samples of lower-redshift radio sources,
are indicative of a primary link between the power in the source of
the photoionising photons (most likely a hidden quasar nucleus) and
the power carried by the radio jets. 
We argue that radio sources modify their environments and that the
range of emission-line properties seen is determined more by the range
of source age than by the range in ambient environment.
The smallest $z > 1.75$ radio galaxies
have all the properties expected if the size distribution of luminous
high-redshift steep-spectrum radio sources reflects a broad range
($\sim 2$ dex) of source ages with a narrower range ($\ltsim\, 1.5$ dex)
of environmental densities, namely: (1) high-ionisation lines,
e.g. Ly$\alpha$, of relatively low luminosity; (2) boosted
low-ionisation lines, e.g. CII]; (3) spatially compact emission-line
regions; and (4) \HI-absorbed Ly$\alpha$ profiles. This is in accord
with the idea that all high-redshift, high-luminosity radio sources
are triggered in similar environments, presumably recently collapsed
massive structures.
\end{abstract}

\section{Introduction}\label{sec:intro}
Observations of active galaxies provide a relatively
straightforward way of probing conditions within massive structures in
the young Universe, and may play an important role in placing
constraints on cosmological models for structure formation.  Radio
galaxies are particularly useful in this respect as their optical and
infra-red emission tend not to be dominated by an active nucleus as is
the case for quasars, hence the structural properties of their
host galaxies are less difficult to determine. The general belief that
powerful radio galaxies harbour hidden quasar nuclei leads to another
advantage of using radio galaxies as cosmological probes: the hidden
nuclei and/or shocks associated with jets can excite strong narrow
emission lines, making redshift determination much easier than is
afforded by the relatively weak absorption lines of `normal' massive
elliptical galaxies.

Radio galaxies at redshifts $z > 4$ provide information on the
Universe at an epoch $\ltsim 10^{9}$ years (\cosone, $H_{\circ}=50~$km
s$^{-1}$ Mpc$^{-1}$) after the Big Bang. The
principal difficulties involved in finding high-redshift radio sources
from flux-density-limited samples have been highlighted previously
(e.g. Blundell et al. 1998). Such objects are absent
from the brightest radio samples (e.g. the 3CRR sample: Laing, Riley
\& Longair 1983) because of what is effectively an upper limit to the
low-frequency radio luminosity of radio galaxies. However, such
luminous sources could be detected up to very high redshift in fainter
(151-MHz flux-density, $S_{151} \approx 1$\,Jy) flux-density limited
samples, for example the most radio luminous 3CRR source (3C9) could
in principle be detected out to $z \sim 10$ in a 1~Jy sample. If
luminous radio sources were present at this epoch, they have not been
detected so far. One reason for this is the rapid decrease in the
fraction of high-luminosity, high-redshift radio sources as the
flux-density limit of the sample is reduced. This means that compiling
redshift complete flux-density-limited samples with a significant
number of high-redshift sources over moderate sky areas
becomes extremely difficult. Indeed, at $S_{151} \sim 1$\,Jy, the
source count is $\sim 10^{4}$\, per steradian, and even in the absence
of a decline in the co-moving space density of such sources at
high redshift (c.f. Dunlop \& Peacock 1990; Willott et al. 2001a;
Jarvis et al. 2001a), we would not expect to find more than $\approx
1$\% of these sources at $z > 4$.

To ameliorate the problem of the increasing source counts at faint
flux-density, various groups have employed additional filtering
criteria to find high-redshift radio sources. The most widely used is
the steep spectral index criterion [e.g. Chambers, Miley \& van
Breugel (1990); R\"ottgering et al. (1994); Lacy et al.\ 1994;
Chambers et al. (1996)], which utilises the steep and often curved
radio spectra of many radio-luminous sources. The correlations between
radio luminosity, redshift and spectral shape are rather complicated
(see e.g.\ Blundell, Rawlings \& Willott 1999) but the simplest
effects to appreciate are: (i) that the spectral index
measured for a source at high redshift will be steeper than the same
source observed at a lower redshift if, as is typically the case, that source has a concave radio
spectrum; (ii) more luminous sources tend to have steeper spectra
above frequencies at which absorption effects become important.
Filtering based on steep radio spectra is one of the
selection criteria used to filter the 6C* sample which was defined in
Blundell et al. (1998). 

A second filtering criterion used is an angular size limit. This
criterion has been used because sources at high redshift are intrinsically
smaller than those at low-redshift in the same flux-density-limited
sample. (e.g. Neeser et al. 1995; Blundell et al. 1999). The physical
cause of this `linear size evolution' has been the cause of recent
debate: Neeser et al. (1995) suggested that the environments of radio
galaxies change systematically with epoch, while Blundell et
al. (1999) argued that the sources are shorter because they are
observed when they are younger [c.f. Kaiser, Dennett-Thorpe \&
Alexander (1997) and Blundell \& Rawlings (1999)].

The 6C* sample was devised to provide a filtered sample biased towards
detecting radio galaxies at $z > 4$ within a sample of spectroscopic
targets that was small enough that complete redshift information could
conceivably be obtained.  The 6C* sample led to the discovery of one
of the highest redshift radio sources currently known (6C*0140+326 at
$z = 4.41$; Rawlings et al. 1996) and more recently it has been used
to place constraints on the radio luminosity function of the
most-luminous radio-loud objects at high redshift, thus placing
constraints on the evolution in the co-moving space density of these
sources (Jarvis et al. 2001a).

In this paper we present the optical spectroscopy of the 6C*
sample. In Sec.~\ref{sec:6C*} we outline the properties of the 6C*
sample, and an appendix gives brief notes on sources in the original
6C* list (Blundell et al. 1998) but which are now excluded. In
Sec.~\ref{sec:spectra} we outline the observational techniques
employed and present the optical spectra for 28 of the 29 sources
remaining in the sample, giving notes on each source in
Sec.~\ref{sec:notes}. In Sec~\ref{sec:6C*missing} we investigate the
redshift space probed by 6C* and compare it to the distribution of
sources one might find in a sample without the filtering criteria but
at the same limiting flux-density as 6C*. In Sec.~\ref{sec:linking} we
combine the 6C* spectrophotometric dataset with similar datasets from
the 3CRR sample (see summary in Willott et al. 1999), 6CE (Rawlings,
Eales \& Lacy 2001) and 7CRS (Willott et al. in prep; Blundell et
al. in prep) samples to investigate how various optical properties of
high-redshift radio sources are linked to their radio properties. In
Sec.~\ref{sec:age} we seek to explain the results of
Sec~\ref{sec:linking} in terms of a simple physical model for radio
source evolution.

Our spectroscopic programme on the 6C* sample has been supplemented by
imaging data (principally in the $K-$band) taken throughout the past decade. In many cases,
the spectroscopic observations which have led to the spectra shown in
this paper have used identifications based on these data. We defer
presentation of the imaging data (principally in $R$- and $K$-bands)
on the 6C* sample to the final paper in this series (Jarvis et al. 2001b; hereafter Jea01).

We use $H_{\circ} = 50~{\rm km\,s}^{-1}\,{\rm Mpc}^{-1}$, $\Omega_{\rm
M}=1$ and $\Omega_{\Lambda}=0$ throughout; none of the principal
results are sensitive to this particular choice of cosmological
parameters. Radio flux-density is measured in Jy, all radio
luminosities quoted are measured in units of W\,Hz$^{-1}$\,sr$^{-1}$,
and the convention for radio spectral index ($\alpha$) is that
$S_{\nu} \propto \nu^{-\alpha}$, where $S_{\nu}$ is the flux-density
at frequency $\nu$. Narrow-line luminosities are measured in units of
W.

\section{The 6C* sample}\label{sec:6C*}

\begin{figure*}[!ht] 
{\hbox to \textwidth{\epsfxsize=0.95\textwidth
\epsfbox{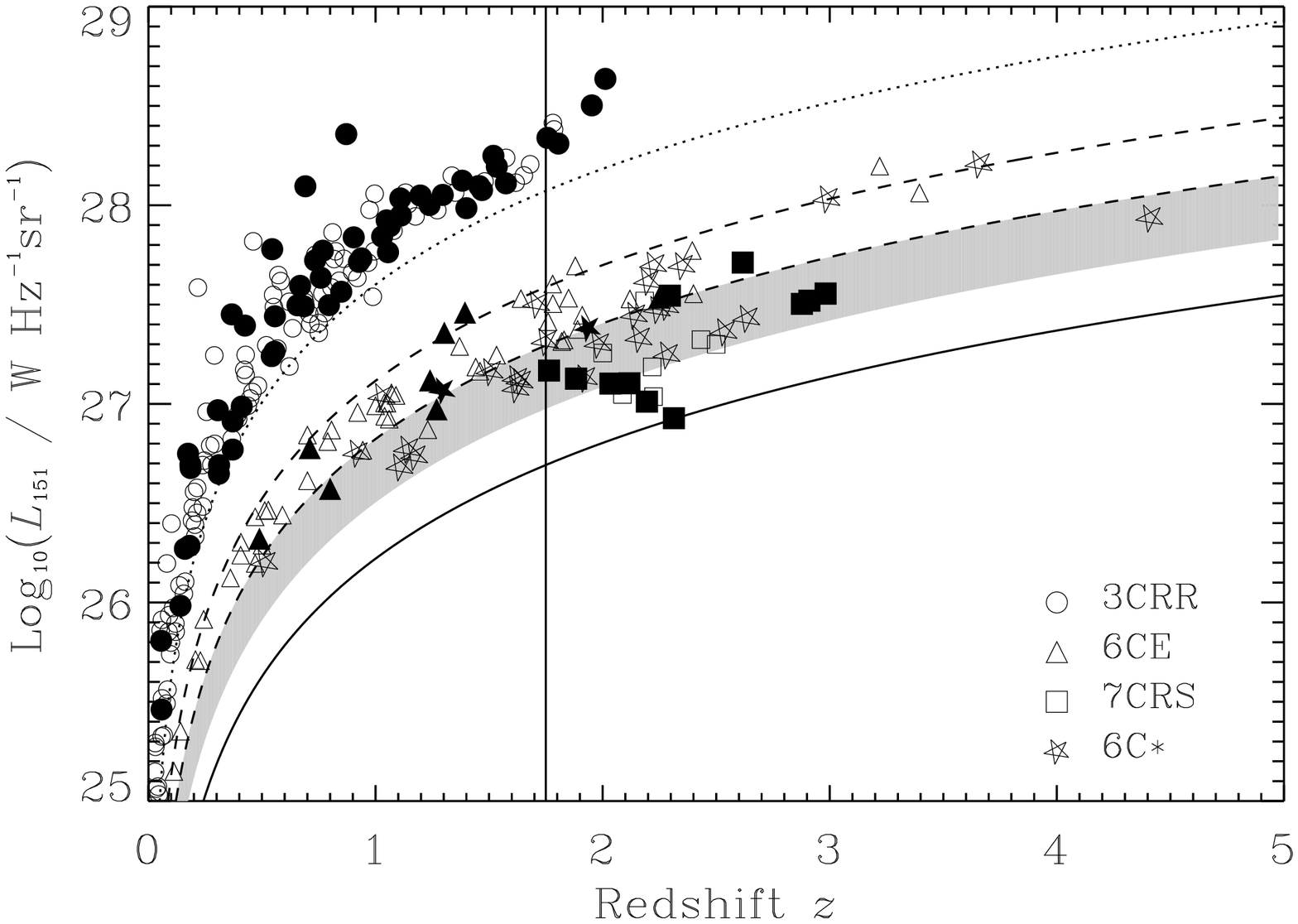}}}
{\caption{\label{fig:pzplane} Rest-frame 151-MHz luminosity
($L_{151}$) versus redshift $z$ plane for the 3CRR (circles), 6CE
(triangles) and 6C* (stars) samples; the vertical line at $z = 1.75$
is the lower limit of the high-redshift sample considered in
Sec.~\ref{sec:linking} and objects from regions I and II of the 7CRS sample
with $z > 1.75$ are also plotted (squares). The open symbols
are radio galaxies and the filled symbols are quasars.  The rest-frame
151-MHz luminosity $L_{151}$ has been calculated according to a
polynomial fit to the radio spectrum [the relevant radio data can be
found in Blundell et al. (1998) and Blundell et al. in prep].  
The curved lines
show the lower flux-density limit for the 3CRR sample (dotted line; Laing et
al. 1983) and the 7CRS (solid line; Blundell et al. in prep;
Willott et al. in prep). The dashed lines correspond to the limits
for the 6CE sample (Rawlings et al. 2001) and the shaded region shows the 6C* flux-density
limits (all assuming a low-frequency radio spectral index of
0.5). Note that the area
between the 3CRR sources and 6CE sources contains no sources, this is
the area which corresponds to the absence of a flux-density-limited
sample between the 6CE ($S_{151} \leq 3.93\:$Jy) and 3CRR ($S_{178}
\geq 10.9\:$Jy) samples. The reason why some of the sources lie very
close to or below the flux-density limit of the samples represented by
the curved lines is because the rest-frame 151-MHz spectral indices lie very close to or
below the assumed spectral index of the curves of $\alpha = 0.5$.}}
\end{figure*}

The selection criteria used to filter the 6C* sample, discussed
fully in Blundell et al. (1998), are summarised as follows:

\begin{itemize}
\item $00^{\rm h}20^{\rm m} \leq$ R.A. (B1950) $\leq 02^{\rm
h}10^{\rm m}$ 
 
\item $30^\circ \leq$ Dec. (B1950) $\leq 51^\circ$
 
\item 0.96\,Jy $\leq S_{151} \leq$ 2.00\,Jy
 
\item $\rm \alpha^{4850}_{151} \geq 0.981$, where $\alpha^{\rm
4850}_{\rm 151}$ is the radio spectral index measured between
151\,MHz and 4850\,MHz.  

\item $\theta_{\rm Texas} < 15^{\prime\prime}$, where $\theta_{\rm
Texas}$ is the angular size of the radio source given in the Texas
catalogue of radio sources at 365\,MHz (Douglas et al. 1996).
 
\end{itemize}

These selection criteria resulted in the original 6C* sample
comprising 34 objects. Further radio information obtained either from
the NVSS catalogue (Condon et al. 1998) or from our own subsequent
radio observations have led to five of these objects being identified
as having an angular size larger than 15\,arcsec, and these are now
excluded from the sample.  Notes and data on these five excluded
sources can be found in Appendix A. Therefore, the 6C* sample now
comprises 29 sources. In 23 cases (80\%) the spectroscopic redshifts
are unequivocal, in 6 (20\%) further spectroscopic confirmation is
ideally required, although tentative redshifts are given; these are
normally based on a single definite emission line identified with a
specific feature on the basis of a redshift estimate from the $K-$band
magnitude.

\section{The Observations}\label{sec:spectra}
The optical spectra presented in this paper were obtained from many
observing runs on three different telescopes. The vast majority of the
redshifts for this sample were obtained using spectra taken on four
runs at the WHT, from January 1991 using the FOS-II spectrometer, and
on three occasions up until February 2000 using the red- and blue-arms
of the ISIS spectrometer. 
Spectra for four of the fainter sources were obtained using the low
resolution imaging spectrometer on the Keck-II telescope between
August 1998 and October 1998. Further spectroscopy on the Lick-3m
telescope in October 1998 allowed us to confirm redshifts for those
sources lacking a secure redshift from the WHT observations. A summary
of all of these observations is presented in Table~\ref{tab:journal}.
The observations were carried out by offsetting from bright stars to
either the radio position, or to where there was an identification
based on our $R$- or $K$-band imaging (Jea01). The position angle (PA)
of the slit was aligned with the radio structure in the cases where
there was only one or no infrared counterpart. Otherwise, the PA was
oriented to pass through the two most probable counterparts.  For
presentation purposes most of the spectra have been boxcar smoothed
over three pixels. The measurements presented in
Table~\ref{tab:spectra} were all made prior to this smoothing
process. Reduction of the spectra was carried out using standard IRAF
routines. The one-dimensional spectra were extracted from the
two-dimensional data with two apertures: a full-width at
zero-intensity for flux measurement purposes, and a full-width at
half-maximum for good signal-to-noise. Cosmic rays were edited out of
the 1-D spectra by inspection of the 2-D spectra and in all cases,
apart from the single-arm observations on the Keck telescope, the spectra from
the red and blue arms were joined together by averaging over $\approx
50$\,\AA\, in the overlapping regions. 

For the pre-1998 WHT+ISIS observations the beam was split using the 540\,nm
dichroic; the blue-arm detector was the TEK1 chip on all runs up until
and including 1995 and the EEV12 chip on all runs later than 1995 with
pixel scales of 0.358 and 0.2 arcsec pixel$^{-1}$ respectively; the
red-arm detector was the EEV3 chip in 1994 and the TEK2 chip from 1995
with pixel scales of 0.336 and 0.358 arcsec pixel$^{-1}$
respectively. For the post-1998 runs the dichroic used split the beam at
610\,nm. The spectra taken on the Keck-II telescope with LRIS (Oke
et al. 1995)
used a TEK $2048 \times 2048$ CCD detector with a pixel scale of
0.212\,arcsec pixel$^{-1}$; we used a 300 lines/mm
grating which has a spectral scale of 2.44\,\AA\,pixel$^{-1}$ at a
central wavelength of $\approx 6500$\,\AA.  For the Lick-3m observations
the observations were made using the Kast double spectrograph (Miller
\& Stone 1994) with a
Reticon $400 \times 1200$ CCD detector for both blue- and
red-arms. For the blue-arm the 452/3306 grating was used which gives a
spectral scale of 2.54\,\AA\,pixel$^{-1}$, and the 300/7500 grating was
used for the red-arm observations giving a spectral scale of
4.60\,\AA\,pixel$^{-1}$.

In nearly all of the cases, definite emission lines have been detected
and in all of the others plausible emission lines are seen. The
reduced 1-D spectra are presented in Fig.~\ref{fig:spectra}.

\begin{table*}
\begin{center}
\scriptsize
\begin{tabular}{l|r|r|c|r|r|r|r|}
\hline\hline
\mc{1}{c|}{Source} & \mc{2}{c|}{Pointing Position} & \mc{1}{c|}{Telescope +} & \mc{1}{c|}{ Date} & \mc{1}{c|}{Exposure} & \mc{1}{c|}{Slit} & \mc{1}{c|}{PA} \\
\mc{1}{c|}{name} & \mc{2}{c|}{B1950} &\mc{1}{c|}{Spectrometer} & \mc{1}{c|}{} & \mc{1}{c|}{time (s)} & \mc{1}{c|}{width (arcsec)} & \mc{1}{c|}{($^{\circ}$)} \\
\hline\hline
6C*0020+440  & \bf 00 20 01.70 & \bf +44 02 32.1 & \bf WHT+ISIS  & \bf 94Jan08
& \bf 1800 & \bf 3.0 & 80 \\
6C*0024+356  & 00 24 12.68 & +35 39 47.7 &  Lick-3m+KAST   & 98Oct20 & 1200
& 2.0 & 106 \\
	     & \bf 00 24 12.97 & \bf +35 39 48.1 & \bf WHT+ISIS  & \bf 98Dec20 & \bf 1800 & \bf 2.5 & \bf 85 \\
6C*0031+403  & 00 31 46.85 & +40 19 25.0 & WHT+FOS-2 & 91Jan17 & 1200
& 4.0 & 90  \\
 	     & 00 31 46.82 & +40 19 25.9 & WHT+ISIS  & 95Jan29 & 2700
& 3.1 & 110  \\
	     & 00 31 46.82 & +40 19 25.9 & Lick-3m+KAST   & 98Oct20 & 2700
& 2.0 & 128 \\
	     & \bf 00 31 46.82 & \bf +40 19 25.9 & \bf WHT+ISIS  & \bf 98Dec20 & \bf 1800 & \bf 2.5 & \bf 128 \\
6C*0032+412  & \bf 00 32 10.73 & \bf +41 15 00.2 & \bf WHT+FOS-2 & \bf 91Jan17 & \bf 1200 & \bf 4.0 & \bf 128 \\
6C*0041+469  & \bf 00 41 16.04 & \bf +46 58 27.3 & \bf WHT+ISIS  & \bf 95Jan29 & \bf 1800 & \bf 2.7 & \bf 10 \\
	     & 00 41 16.00 & +46 58 25.3 & Keck-II+LRIS & 98Oct16 & 2400 &
1.0 & 171 \\
\hline
6C*0050+419  & 00 50 46.52 & +41 58 45.3 & WHT+FOS-2 & 91Jan17 &  900
& 4.0 & 31 \\
	     & \bf 00 50 46.90 & \bf +41 58 46.7 & \bf Keck-II+LRIS   & \bf 98Aug26 & \bf 3600 & \bf 1.0 & \bf 41 \\
	     &  00 50 46.50 & +41 58 45.3 & WHT+ISIS  & 00Feb09 & 1800
& 2.0 & 160 \\
6C*0052+471  & 00 52 09.59 & +47 09 46.2 & WHT+FOS-2 & 91Jan17 &  900
& 4.0 & 12 \\
             & \bf 00 52 09.60 & \bf +47 09 46.5 & \bf WHT+ISIS  & \bf 95Jan28 &  \bf 900 & \bf 3.1 & 10 \\
6C*0058+495  & \bf 00 58 24.75 & \bf +49 34 04.3 & \bf WHT+ISIS  & \bf 92Jan28 & \bf  900 & \bf 4.0 & \bf 35 \\
6C*0106+397  & \bf 01 06 35.31 & \bf +39 44 02.8 & \bf Keck-II+LRIS   & \bf 98Aug25 & \bf 3000 & \bf 1.0 & 128 \\
6C*0112+372  & 01 12 00.33 & +37 16 41.4 & WHT+FOS-2 & 91Jan16 & 1200
& 4.0 & 108 \\
	     & \bf 01 12 00.33 & \bf +37 16 41.4 & \bf Lick-3m+KAST & \bf 98Oct20 & \bf 2400 & \bf 2.0 & \bf 112 \\
\hline
6C*0115+394  & 01 15 03.47 & +39 28 45.9 & Lick-3m+KAST   & 98Oct20 & 5400
& 2.0 & 167 \\
	     & \bf 01 15 03.47 & \bf +39 28 45.9 & \bf WHT+ISIS  & \bf 98Dec18 & \bf 1800 & \bf 2.5 & \bf 166 \\
6C*0118+486  & 01 18 16.49 & +48 41 58.0 & WHT+FOS-2 & 91Jan17 &  600
& 4.0 & 150 \\
	     & \bf 01 18 16.50 & \bf +48 41 58.1 & \bf WHT+ISIS  & \bf 95Jan28 & \bf 900 & \bf 0.9 & \bf 155 \\
6C*0122+426  & 01 22 56.80 & +42 36 16.0 & Lick-3m+KAST   & 98Oct21 & 3600
& 2.0 & 85 \\
	     & \bf 01 22 56.80 & \bf +42 36 16.0 & \bf WHT+ISIS  & \bf 98Dec17 & \bf 1800 & \bf 2.5 & \bf 98 \\
6C*0128+394  & 01 28 34.70 & +39 27 32.0 & WHT+ISIS  & 95Jan28 &  900
& 0.9 & 20 \\
	     & \bf 01 28 34.67 & \bf +39 27 32.0 & \bf WHT+ISIS  & \bf
98Dec20 & \bf 1800 & \bf 2.5 & \bf 17 \\
6C*0132+330  & \bf 01 32 39.09 & \bf +33 01 40.4 & \bf WHT-ISIS  & \bf 98Dec18 & \bf 1800 & \bf 2.5 & \bf 174  \\
\hline
6C*0133+486  & 01 33 36.29 & +48 37 07.9 & WHT+ISIS  & 98Dec17 & 1800 &
2.5 & 24 \\
	     &  \bf 01 33 36.29 & \bf +48 37 07.9 & \bf WHT+ISIS  & \bf 98Dec19 & \bf 2700 & \bf 2.5 & \bf 24 \\
6C*0135+313  & \bf 01 35 16.13 & \bf +31 17 27.3 & \bf WHT+ISIS  & \bf 98Dec19 & \bf 1800 & \bf 2.5 & \bf 67 \\
6C*0136+388  & 01 36 59.09 & +38 48 09.2 & WHT+FOS-2 & 91Jan16 & 2700
& 4.0 & 124  \\
	     & 01 36 59.31 & +38 48 07.3  & WHT+ISIS  & 95Jan29 & 1800
& 0.5 & 131 \\
	     & \bf 01 36 59.09 & \bf +38 48 09.5 & \bf WHT+ISIS  & \bf
98Dec20 & \bf 1800 & \bf 2.5 & 132 \\
6C*0139+344  & 01 39 25.87 & +34 27 02.5 & WHT+ISIS  & 92Jan29 &  900
& 4.0 & 159 \\
	     & 01 39 25.85 & +34 27 03.2 & WHT+ISIS	 & 95Jan28 &
900 & 0.9 & 155 \\
	     & \bf 01 39 25.91 & \bf +34 26 32.5 & \bf Keck-II+LRIS & \bf 98Sep04 & \bf 3600 & \bf 1.0 & \bf 26\\
6C*0140+326  & \bf 01 40 51.53 & \bf +32 38 45.8 & \bf WHT+ISIS & \bf 94Jan08 & \bf 1200 & \bf 2.0 & \bf 128  \\
\hline
6C*0142+427  & 01 42 28.13 & +42 42 41.6 & WHT+FOS-2 & 91Jan17 & 1200
& 4.0  & 37 \\
	     & \bf 01 42 28.14 & \bf +42 42 41.9 & \bf WHT+ISIS  & \bf 94Jan08 & \bf 1800 & \bf 3.0 & \bf 37 \\
6C*0152+463  & 01 52 38.22 & +46 22 30.4 & WHT+ISIS  & 94Jan08 & 1800
& 3.0 & 0\\
	     & \bf 01 52 38.10 & \bf +46 22 28.2 & \bf WHT+ISIS  & \bf 95Jan28 & \bf 900 & \bf 3.1 & \bf 30 \\
6C*0154+450  & \bf 01 54 24.85 & \bf +45 03 46.4 & \bf WHT+FOS-2 & \bf 91Jan17 & \bf 300 & \bf 4.0 & \bf 180 \\
6C*0155+424  & 01 55 29.74 & +42 25 38.2 & WHT+ISIS  & 94Jan08 & 1800
& 3.0 & 0 \\
	     & \bf 01 55 29.74 & \bf +42 25 38.2 & \bf WHT+ISIS  & \bf
98Dec18 & \bf 1800 & \bf 2.5 & 102 \\
6C*0158+315  &  01 58 01.53 &  +31 31 45.7 &  WHT+ISIS & 98Dec19 & 1800 & 3.0 & 169 \\
& \bf 01 58 01.53 & \bf +31 31 45.7 & \bf WHT+ISIS & \bf
98Dec20 & \bf 3600 & \bf 3.0 & \bf 169 \\
\hline
6C*0201+499  & 02 01 08.41 & +49 54 38.1 & WHT+FOS-2 & 91Jan17 & 1500
& 4.0 & 178 \\
	     & \bf 02 01 08.39 & \bf +49 54 38.5 & \bf WHT+ISIS  & \bf 95Jan28 & \bf 1315 & \bf 0.9 & \bf 158 \\
6C*0202+478  & 02 02 07.03 & +47 51 42.2 & Lick-3m+KAST   & 98Oct21 & 1200
& 2.0 & 86  \\
	     & \bf 02 02 06.90 & \bf +47 51 41.70 & \bf WHT+ISIS  & \bf 98Dec19 & \bf 1800 & \bf 2.5 & \bf 83\\
6C*0208+344  & 02 08 55.47 & +34 29 57.3 & WHT+FOS-2 & 91Jan17 & 1200
& 4.0 & 131  \\
	     & \bf 02 08 55.50 & \bf +34 29 57.0 & \bf WHT+ISIS  & \bf 94Jan09 & \bf 1800 & \bf 3.0 & \bf 120 \\
6C*0209+276  & 02 09 20.01 & +47 39 16.6 & WHT+ISIS  & 94Jan09 & 1800
& 3.0 & 0  \\
	     & \bf 02 09 20.00 & \bf +47 39 16.5 & \bf WHT+ISIS & \bf 00Jan11 & \bf 1800 & \bf 2.0  & \bf 90 \\
\hline\hline
\end{tabular}
\end{center}
{\caption{\label{tab:journal} Log of the observations of the sources
present in the 6C* sample. The larger, bold text represents the observation
for the spectra presented in Fig.~\ref{tab:spectra}.  
The observations of 6C*0140+326 are detailed in Rawlings et al. (1996). }}
\end{table*}

\begin{table*}
\begin{center}
\begin{tabular}{l|c|c|c|c|l|c|c|}
\hline\hline
\mc{1}{c|}{Source} & \mc{1}{c|}{z} & \mc{1}{c|}{Line} &  \mc{1}{c|}{
$\lambda_{\mathrm rest}$} & \mc{1}{c|}{$\lambda_{\mathrm obs}$} &
\mc{1}{c|}{FWHM} & \mc{1}{c|}{Flux} & \mc{1}{c|}{Galaxy} \\

\mc{1}{c|}{name} & & & \mc{1}{c|}{(\AA)} &
\mc{1}{c|}{(\AA)} & \mc{1}{c|}{(km s$^{-1}$)} &
\mc{1}{c|}{(Wm$^{-2}$)} & \mc{1}{c|}{or Quasar} \\
\hline\hline
6C*0020+440  & 2.988 & Ly$\alpha$ & 1216 & 4850 $\pm$ 1 & $600 - 1300$ & 3.4E-19 $\pm$ 25\% & G \\
\hline
6C*0024+356  & 2.161 & Ly$\alpha$ & 1216 & 3844 $\pm$ 2 & $0 - 1100$ & 2.8E-19
             $\pm$ 10\% & G \\
             & & CII] & 2326 & 7356 $\pm$ 1 & $800 - 1100$ & 1.4E-19
             $\pm$ 15\% & \\
\hline
6C*0031+403 & 1.619 & CIV & 1549 & 4057 $\pm$ 2 & $0 - 1000$ & 8.7E-20
	     $\pm$ 10\% & G \\ 
             & & CIII] & 1909 & 4999 $\pm$ 1 & $0 - 700$ & 4.4E-20
	     $\pm$ 10\% & \\
\hline
6C*0032+412  & 3.658 & Ly$\alpha$ & 1216 & 5660 $\pm$ 1 & $1000 - 1400$ &
	     2.1E-19 $\pm$ 10\% & G \\
	     & & CIV & 1549 & 7194 $\pm$ 5 & $1000 - 1300$ & 3.2E-20
	     $\pm$ 20\% & \\
\hline
6C*0041+469  & 2.140? & Ly$\alpha$?! & 1216 & 3836 $\pm$ 12 & $1000 - 1600$
	     & 1.1E-19 $\pm$ 20\% & G \\
	     & & CIV?! & 1549 & 4872 $\pm$ 1 & $0 - 900$ & 7.8E-20
	     $\pm$ 10\% & \\
\hline
6C*0050+419  & 1.748? & HeII? & 1640 & 4507 $\pm$ 1 & $0 - 1000$	 &
7.2E-21 $\pm$ 15\% & G \\
	     & & CII]?! & 2326 & 6385 $\pm$ 2 & $0 - 300$ & 4.2E-21
$\pm$ 10\% & \\
	     & & [NeIV]? & 2424 & 6657 $\pm$ 3 & $0 - 900$ & 8.1E-21
$\pm$ 15\% & \\ 
\hline
6C*0052+471  & 1.935 & Ly$\alpha$ & 1216 & 3577 $\pm$ 8 & $> 2500$ &
	     5.1E-19 $\pm$ 15\% & Q \\
	     & & CIV & 1549 & 4549 $\pm$ 18 & $> 7000$ &
             1.1E-18 $\pm$ 60\% & \\
             & & CIII] & 1909 & 5599 $\pm$ 10 & $> 2500$ & 3.7E-19
             $\pm$ 35\% & \\
\hline
6C*0058+495  & 1.173 & CIII] & 1909 & 4142 $\pm$ 7 & $2000 - 2500$ &
	     4.1E-19 $\pm$ 15\% & G \\
	     & & CII] & 2326 & 5034 $\pm$ 2 & $1100 - 1600$ & 2.1E-19
	     $\pm$ 25\% & \\ 
	     & & MgII & 2799 & 6087 $\pm$ 2 & $800 - 1200$ &
	     2.1E-19 $\pm$ 20\% & \\
	     & & [OII] & 3727 & 8097 $\pm$ 1 & $800 - 1100$ & 1.0E-18
	     $\pm$ 10\% &\\
	     & & [SII] & 4072 & 8863 $\pm$ 1 & $600 - 900$ & 4.0E-19
	     $\pm$ 15\% & \\
\hline
6C*0106+397 & 2.284 & CIV & 1549 & 5089 $\pm$ 2 & $900 - 1300$ &
	    $>$ 2.0E-20 & G \\
            & & HeII & 1640 & 5384 $\pm$ 1 & $300 - 900$ & $>$ 2.7E-20 & \\
	    & & CIII] & 1909 & 6262 $\pm$ 1 & $300 - 700$ & $>$ 1.8E-20 & \\
\hline
6C*0112+372 & 2.535 & Ly$\alpha$ & 1216 & 4299 $\pm$ 1 & $1400 - 1800$ &
1.6E-18 $\pm$ 15\% & G \\
	    & & CIV & 1549 & 5476 $\pm$ 1 & $350 - 900$ & 3.3E-19 $\pm$ 5\% & \\ 
\hline
6C*0115+394 & 2.241 & Ly$\alpha$ & 1216 & 3940 $\pm$ 1 & $700 - 1400$ &
	    4.0E-18 $\pm$ 10\% & G \\
	    & & CIV & 1549 & 5020 $\pm$ 1 & $0 - 1100$ & 2.7E-19 $\pm$
	    15\% & \\
	    & & HeII & 1640 & 5319 $\pm$ 2 & $1000 - 1500$ & 3.4E-19
	    $\pm$ 25\% & \\
	    & & CIII] & 1909 & 6193 $\pm$ 1 & $< 1800$ & 2.1E-19
	    $\pm$ 15\% & \\
	    & & [NeIV] & 2424 & 7861 $\pm$ 1 & $0 - 700$ & 1.8E-20
	    $\pm$ 20\% & \\
\hline
6C*0118+486a  & 2.350 & Ly$\alpha$ & 1216 & 4073 $\pm$ 1 & $600 - 1300$ &
	     3.7E-19 $\pm$ 25\% & G \\
	     & & CIV & 1549 & 5200 $\pm$ 2 & $1200 - 1600$ & 1.5E-19
	     $\pm$ 20\% & \\
& & & & & & & \\
6C*0118+486b & 0.529 & [OII] & 3727 & 5702 $\pm$ 1 & $0 - 800$ & 1.5E-19
	     $\pm$ 15\% & G \\
	     & & H$\beta$ & 4861 & 7433 $\pm$ 1 & $0 - 500$ & 9.1E-20
	     $\pm$ 30\% & \\
\hline
6C*0122+426  & 2.635 & Ly$\alpha$ & 1216 & 4420 $\pm$ 1 & $600 - 1300$ & 1.8E-19
	     $\pm$ 10\% & G \\
             & & CIII] & 1909 & 3940 $\pm$ 2 & $1100 - 2000$ &  6.1E-20
	     $\pm$ 10\% & \\
\hline
6C*0128+394  & 0.929 & MgII & 2799 & 5389 $\pm$ 4 & $400 - 1200$ &
6.3E-20 $\pm$ 20\% & G \\ 
	     & & [OII] & 3727 & 7190 $\pm$ 2 & $0 - 600$ & 6.8E-20
$\pm$ 5\% & \\
\hline
6C*0132+330  & 1.710 & CIII] & 1909 & 5175 $\pm$ 1 & $0 - 700$ &
8.0E-20 $\pm$ 10\% & G \\
	     & & CII] & 2326 & 6304 $\pm$ 1 & $0 - 700$ & 7.7E-20
$\pm$ 10\% & \\
	     & & [NeIV] & 2424 & 6578 $\pm$ 1 & $0 - 700$ & 5.0E-20
$\pm$ 10\% & \\ 
\hline
6C*0133+486  & 1.029? & [OII]? & 3727 & 7559 $\pm$ 1 & $0 - 600$ &
1.8E-19 $\pm$ 15\% & G \\
\hline\hline
\end{tabular}
\end{center}
\end{table*}

\begin{table*}
\begin{center}
\begin{tabular}{l|c|c|c|c|l|c|c|}
\hline\hline
\mc{1}{c|}{Source} & \mc{1}{c|}{z} & \mc{1}{c|}{Line} &  \mc{1}{c|}{
$\lambda_{\mathrm rest}$} & \mc{1}{c|}{$\lambda_{\mathrm obs}$} & \mc{1}{c|}{FWHM} &
\mc{1}{c|}{Flux} & \mc{1}{c|}{Galaxy} \\

\mc{1}{c|}{name} & & & \mc{1}{c|}{(\AA)} &
\mc{1}{c|}{(\AA)} & \mc{1}{c|}{(km s$^{-1}$)} &
\mc{1}{c|}{(Wm$^{-2}$)} & \mc{1}{c|}{or Quasar} \\
\hline\hline
6C*0135+313  & 2.199 & Ly$\alpha$ & 1216 & 3900 $\pm$ 1 & $1500 - 2000$ &
	     1.5E-17 $\pm$ 15\% & G \\
	     & & CIII] & 1909 & 6058 $\pm$ 2 & $300 - 1000$ & 3.1E-18
	     $\pm$ 15\% & \\
\hline
6C*0136+388  & 1.108? & [OII]? & 3727 & 7858 $\pm$ 3  & $800 - 1200$ & 2.5E-16 $\pm$
10\% & G  \\
\hline	
6C*0139+344  & 1.637 & [NeIV] & 2424 & 6392 $\pm$ 3 & $400 - 800$ & & G \\ 
	     & & MgII & 2799 & 7381 $\pm$ 2 & $700 - 1000$ & & \\
	     & & [NeV] & 3426 & 9035 $\pm$ 2 & $200 - 400$ & & \\
	     & & [OII] & 3727 & 9828 $\pm$ 1 & $500 - 800$ & &  \\
\hline
6C*0142+427  & 2.225 & Ly$\alpha$ & 1216 & 3921 $\pm$ 1 & $1100 - 1800$ &
	     1.5E-18 $\pm$ 15\% & G \\
	     & & CIV & 1549 & 5001 $\pm$ 1 & $900 - 1400$ & 1.6E-19
	     $\pm$ 25\% & \\
	     & & HeII & 1640 & 5284 $\pm$ 6 & $0 - 800$ & 7.1E-20
	     $\pm$ 20\% & \\
\hline
6C*0152+463  & 2.279 & Ly$\alpha$ & 1216 & 3987 $\pm$ 3 & $0 - 1200$ &
	     3.8E-19 $\pm$ 30\% & G \\
	     & & HeII & 1640 & 5393 $\pm$ 1 & $300 - 1100$ &
	     1.2E-19 $\pm$ 25\%  & \\
\hline
6C*0154+450  & 1.295 & CIV & 1549 & 3554 $\pm$ 2 & $\sim 7000$ &
	     1.6E-17 $\pm$ 20\%  & Q \\
	     & & HeII & 1640 & 3787 $\pm$ 2 & $\sim 4000$ & 2.8E-18
	     $\pm$ 20\% & \\
	     & & CIII] & 1909 & 4361 $\pm$ 15 & $\sim 5000$ & 3.1E-18
	     $\pm$ 40\%  & \\
	     & & MgII & 2799 & 6422 $\pm$ 6 & $\sim 6000$ & 2.3E-18
	     $\pm$ 35\% & \\
	     & & [OII]? & 3727 & 8584 $\pm$ 4 & $< 3000$ & 1.4E-18
	     $\pm$ 15\% & \\
\hline
6C*0155+424    & 0.513  & [OII] & 3727 & 5639 $\pm$ 1 & $0 - 900$ &
 	       3.3E-19 $\pm$ 15\% & G \\
	       & & H$\delta$ & 4102 & 6205 $\pm$ 1 & $1000 - 1400$ &
 	       2.3E-19 $\pm$ 10\% & \\
\hline
6C*0158+315  & 1.505 & CII] & 2326 & 5813 $\pm$ 3 & $0 - 600$ &
5.9E-19 $\pm$ 15\% & G \\
	     & & [NeIV] & 2424 & 6072 $\pm$ 3 & $0 - 600$ & 6.6E-19
$\pm$ 10\% & \\
             & & MgII & 2799 & 7014 $\pm$ 1 & $200 - 850$ & 1.4E-18
$\pm$ 10\% & \\   
\hline
6C*0201+499  & 1.981 & Ly$\alpha^{\dagger}$ & 1216 & 3625 $\pm$ 2 &
$1200 - 2000$ & 2.1E-19 $\pm$ 20\% & G \\
	     & & NV & 1240 & 3692 $\pm$ 2 & $0 - 1000$ & 2.5E-20 $\pm$
	     20\% & \\ 
	     & & SiIV + OIV] & 1402 & 4172 $\pm$ 9 & $1600 - 2100$ &
	     1.0E-19 $\pm$ 15\% & \\ 
	     & & CIV & 1549 & 4615 $\pm$ 3 & $1000 - 1600$ & 4.5E-20
	     $\pm$ 40\% & \\
	     & & CII] & 2326 & 6924 $\pm$ 8 & $1200 - 1500$ & 8.1E-20
	     $\pm$ 20\% & \\
	     & & [NeIV] & 2424 & 7242 $\pm$ 21 & $1600 - 2000$ & 1.5E-19
	     $\pm$ 20\% & \\
\hline
6C*0202+478  & 1.613? & CII]?! & 2326 & 6066 $\pm$ 2 & $0 - 700$ &
7.5E-19 $\pm$ 10\% & G \\
	     & & MgII?! & 2799 & 7314 $\pm$ 3 & $0 - 700$ & 5.8E-19
$\pm$ 20\% & \\

\hline
6C*0208+344  & 1.920 & Ly$\alpha$ & 1216 & 3551 $\pm$ 2 & $1200 - 2000$ &
	     4.1E-19 $\pm$ 10\%  & G \\
	     & & CIV & 1549 & 4524 $\pm$ 1 & $600 - 1400$ & 1.0E-19
	     $\pm$ 10\% & \\
	     & & CII] & 2326 & 6799 $\pm$ 1 & $400 - 900$ & 9.9E-20
	     $\pm$ 10\% & \\
\hline
6C*0209+476  & 1.141? & [OII]? & 3727 & 7980 $\pm$ 1 & $200 - 600$ &
1.9E-19 $\pm$ 5\% & G \\
\hline\hline
\end{tabular}
\end{center}
{\caption{\label{tab:spectra} Measurements obtained from the spectra
of sources in the 6C* radio sample. The `?' denotes an uncertain line
identification and the `!' denotes a feature which is possibly
spurious. The $\dagger$ symbol means that the line is contaminated by
a cosmic ray. For many of the uncertain lines in sources with known
redshifts the line diagnostics are not presented because of the low
signal-to-noise. For sources with redshifts based on uncertain lines,
we present the emission line data for the emission lines which are
likely to be real. Errors on the line fluxes represent the 1$\sigma$
uncertainty expressed as a percentage of the best line-flux estimate,
for the strongest lines these are dominated by roughly equal
contributions from uncertainties in fixing the local continuum level,
and from the absolute flux calibration (including plausible slit
losses).  Line widths were estimated from the FWHM of the best-fit
Gaussian to each line, the lower value assumes the line-emitting
region fills the slit, and the higher value assumes that it is
broadened only by the seeing. The emission line fluxes for 6C*0106+397
are lower limits estimated by fitting a Gaussian from the maximum
continuum level, as the spectrophotometric standard used to calibrate
this observation was taken on a different night. For 6C*0118+486 we
give emission line fluxes for the foreground galaxy (6C*0118+486b) in
addition to the radio galaxy (6C*0118+486a). There are no emission
line fluxes available for 6C*0139+344 due to the lack of a
spectrophotometric standard for the spectrum presented. Emission line
data for 6C*0140+326 can be found in Rawlings et al. (1996).}}
\end{table*}

\begin{table*}
\begin{center}
\begin{tabular}{l|c|c|l|c|c|c}
\hline\hline
\mc{1}{c|}{(1)} & \mc{1}{c|}{(2)} & \mc{1}{c|}{(3)} & \mc{1}{c|}{(4)} &
\mc{1}{c|}{(5)} & \mc{1}{c|}{(6)} & \mc{1}{c|}{(7)} \\
\hline
\mc{1}{c|}{Source} & \mc{1}{c|}{$S_{151}$} &
\mc{1}{c|}{$\alpha_{151}$} & \mc{1}{c|}{$z$} &
\mc{1}{c|}{$\log_{10}L_{151}$} & \mc{1}{c|}{Line} & 
\mc{1}{c|}{$\log_{10}L_{\rm line}$} \\
\hline\hline
6C*0020+440 & 2.00 & 0.97 & 2.988 & 28.03 & Ly$\alpha$ & 36.37 \\
6C*0024+356 & 1.09 & 0.69 & 2.161 & 27.33 & Ly$\alpha$ & 35.96  \\
6C*0031+403 & 0.96 & 0.90 & 1.618 & 27.08 & CIV & 35.18  \\
6C*0032+412 & 1.29 & 1.28 & 3.658 & 28.21 & Ly$\alpha$ & 36.35\\
6C*0041+469 & 1.53 & 0.61 & 2.145? & 27.45 & Ly$\alpha$ & 35.55  \\
\hline
6C*0050+419 & 1.00 & 1.24 & 1.748? & 27.31 & HeII? & 34.16 \\
6C*0052+471 & 1.31 & 0.88 & 1.935 & 27.38 & Ly$\alpha$ & 36.12   \\
6C*0058+495 & 0.97 & 0.74 & 1.173 & 26.74 & [OII] & 35.92  \\
6C*0106+397 & 0.96 & 0.52 & 2.284 & 27.25 & CIV & 34.87 \\
6C*0112+372 & 1.03 & 0.42 & 2.535 & 27.36 & Ly$\alpha$ & 37.05  \\
\hline
6C*0115+394 & 0.96 & 1.06 & 2.241 & 27.48 & Ly$\alpha$ & 37.15  \\
6C*0118+486 & 0.98 & 1.47 & 2.350 & 27.70 & Ly$\alpha$ & 36.17  \\
6C*0122+426 & 1.05 & 0.53 & 2.635 & 27.43 & Ly$\alpha$ & 35.97  \\
6C*0128+394 & 1.94 & 0.50 & 0.929 & 26.75 & [OII] & 34.52  \\
6C*0132+330 & 1.56 & 1.28 & 1.710 & 27.50 & CIII] & 35.19  \\
\hline
6C*0133+486 & 1.89 & 1.22 & 1.029? & 27.04 & [OII]? & 35.05  \\
6C*0135+313 & 1.24 & 1.18 & 2.199 & 27.62 & Ly$\alpha$ & 37.71 \\
6C*0136+388 & 0.99 & 0.68 & 1.108? & 26.68 & [OII]? & 35.68 \\
6C*0139+344 & 1.10 & 0.74 & 1.637 & 27.11 & [OII] & - \\
6C*0140+326 & 1.00 & 0.62 & 4.410 & 27.94 & Ly$\alpha$ & 36.87  \\  
\hline
6C*0142+427 & 1.46 & 1.20 & 2.225 & 27.70 & Ly$\alpha$ & 36.72  \\
6C*0152+463 & 1.29 & 0.79 & 2.279 & 27.49 & Ly$\alpha$ & 36.15  \\
6C*0154+450 & 1.15 & 1.28 & 1.295 & 27.07 & CIV & 37.22 \\
6C*0155+424 & 1.51 & 0.89 & 0.513 & 26.20 & [OII] & 34.66  \\
6C*0158+315 & 1.51 & 0.75 & 1.505 & 27.16 & MgII & 36.31  \\
\hline
6C*0201+499 & 1.14 & 0.74 & 1.981 & 27.30 & Ly$\alpha$ & 35.75  \\
6C*0202+478 & 1.06 & 0.86 & 1.620? & 27.13 & CII]? & 36.10  \\
6C*0208+344 & 0.97 & 0.57 & 1.920 & 27.13 & Ly$\alpha$ & 36.01  \\
6C*0209+476 & 1.14 & 0.69 & 1.141? & 26.77 & [OII]? & 35.18 \\
\hline\hline
\end{tabular}
\end{center}
{\caption{\label{tab:linediag2} Summary of key information on the 6C*
sample. {\bf Column 1:} Name of the 6C* source. {\bf Column 2:}
151\,MHz flux-density measurements in Jy from the 6C survey (Hales et al. 1993). {\bf Column 3:} radio spectral index evaluated at rest-frame
151\,MHz using the polynomial fit described in Blundell et al. (1998). {\bf Column
4:} redshift, `?' signifies that this value is not yet an unequivocal
redshift. {\bf Column 5:} $\log_{10}$ of the rest-frame 151\,MHz radio
luminosity (measured in units of W Hz$^{-1}$ sr$^{-1}$), calculated
using the polynomial fit to the radio spectra. {\bf Column 6:}
Prominent emission line in the existing spectra, `?' signifies that
the line identification is uncertain. {\bf Column 7:} $\log_{10}$ of
the line luminosity (measured in units of W), `$-$' means that the data
are inadequate to obtain a line luminosity through the absence of a spectrophotometric
standard. }}
\end{table*}

\section{Notes on Individual Sources}\label{sec:notes}

The radio maps referred to in this section can be found in Blundell et
al. (1998) and the optical and $K-$band images referred to are
presented in Jea01.

\makebox{}

{\bf 6C*0020+440}  A blind spectrum was taken pointed at the mid-point
of the two hotspots where there appears to be a faint $K$-band identification. Strong Ly$\alpha$ emission and probable CIV$\lambda
1549$\,\AA\, emission point to a radio galaxy at $z=2.988$. The
$K-$band magnitude ($K = 18.9$ in a 5 arcsec aperture) is also
consistent with the source being at this redshift.

{\bf 6C*0024+356} We find a strong line at 3844\,\AA\, which we take
to be Ly$\alpha$ at $z=2.161$ and the presence of CII]$\lambda
2326$\,\AA\, confirms this redshift.

{\bf 6C*0031+403}
Emission lines corresponding to CIV$\lambda 1549$\,\AA\, and CIII]$\lambda
1909$\,\AA\, give this a redshift of $z=1.619$.
There are also hints of CII]$\lambda 2326$\,\AA\, and [NeIV]$\lambda
2424$\,\AA\, in all of the 2-D spectra taken.

{\bf 6C*0032+412} This is the second highest confirmed redshift object in the
sample at $z=3.658$, with strong Ly$\alpha$ emission and a faint but
secure emission line corresponding to CIV$\lambda 1549$\,\AA\, at this
redshift. We also present a $K$-band spectrum of this source in
Fig.~\ref{fig:6cs0032Kspec}, which shows the presence of
[OIII]$\lambda 4959$\,\AA\, and [OIII]$\lambda 5007$\,\AA\, at $z =
3.670$. 

{\bf 6C*0041+469} $K-$ and $I-$band imaging of this source reveals emission
accurately aligned with the radio structure. Spectroscopic observations along
this aligned component show tentative lines corresponding to
Ly$\alpha$, CIV and CIII] at $z = 2.140$, although further
spectroscopy will be needed to confirm this. It is worth noting
that at this redshift H$\alpha$ is redshifted into the $K$-band window
and the extended $K$-band emission may be dominated by line emission
as has been found in other sources with extended optical/infrared
emission aligned with the radio component (e.g. Egami et
al. 1999). However, the $K$-band magnitude ($K = 19.2$ in an 8 arcsec
aperture, Jea01) is consistent with
this redshift using the $K-z$ diagram (e.g. Eales et al. 1997; Jea01).

{\bf 6C*0050+419} The radio structure of this source is compact at
4.9\,GHz with an angular size $\theta\, \ltsim\, 2$\,arcsec. We
pointed the telescope directly at the radio structure with a position
angle of 41$^{\circ}$ which was chosen to align with the slightly
extended radio structure in this direction. A total of 1 hour of
optical spectroscopy with the Keck telescope reveals continuum at the
position of the radio emission down to the limit of the spectral
window of 4000\AA, placing an upper limit on the redshift at $z \ltsim
3.4$ from the Lyman limit. We also identify four highly tentative
lines, which could correspond to CIV$\lambda 1549$\AA, HeII$\lambda
1640$\AA, CII]$\lambda 2326$\AA\, and [NeIV]$\lambda 2424$\AA\, at $z
= 1.748$. Using the $K-z$ diagram, the $K$-band magnitude of this
object ($K = 18.8$ in an 8 arcsec aperture) is also consistent with
this redshift.

Subsequent deep $I$-band imaging with the Keck telescope
reveals two faint components, both approximately 1.5\,arcsec away from
the centre of the radio emission, one to the south and one to the
north-west (see Jea01), which for our chosen PA, the slit would have passed between. Whether this emission is associated with the radio source
has yet to be confirmed. Further spectroscopy on the WHT in February
2000 with a PA aligned to encompass these two components failed to yield
any bright emission lines or detectable continuum (note that
Ly$\alpha$ at $z = 1.748$ would fall in a region of very low
signal-to-noise in our spectrum). In summary our
observations are consistent with a faint $K$-band ID at the position
of the southern $I-$band object (object `a' in Jea01) and with an additional $I-$band
component to the north-west.

There is also the possibility that the source of the radio emission
may be from an object $\approx 7$\,arcsec to the south-west of the
centre of the bright radio emission (object `b' in Jea01). This is roughly in the direction of the
slightly extended radio structure, and is co-spatial with a faint
radio component in our  8.4\,GHz map. Our existing spectroscopy
marginally passes through this object and there are hints of the same 
emission-lines as those detected towards the centre of the
bright radio emission, therefore this is a plausible candidate for the
ID of the radio source and suggests that there is an aligned component
in the optical/near-IR bands.

{\bf 6C*0052+471} Strong and broad CIV$\lambda 1549$\,\AA\, and
CIII]$\lambda 1909$\,\AA\, emission lines confirm this is a quasar at
$z=1.935$. There is also evidence for weaker Ly$\alpha$ emission and a
decreasing continuum blueward of the CIV emission. 
This may be due to light reddening of the quasar by dust
(e.g. Willott, Rawlings \& Jarvis 2000a).

{\bf 6C*0058+495} Strong [OII]$\lambda 3727$\,\AA\, emission along
with weaker lines corresponding to CIII]$\lambda 1909$\,\AA\,,
CII]$\lambda 2326$\,\AA\, and MgII$\lambda 2799$\,\AA\, identify this
as a radio galaxy at $z=1.173$. There is also a line which corresponds
to [SII]$\lambda 4072$\,\AA\, which is not usually as strong as the
line seen in our spectrum at the corresponding wavelength for this
redshift. This emission line does not have any contamination from
cosmic rays and is apparent in the 2-D spectra. Comparison of the
[OII] and [SII] line fluxes for radio galaxies using the ratios of
McCarthy (1993) gives a [OII] to [SII] ratio of 25:1, and the ratio in
our spectrum is $\approx\,$5:2. Thus, if real, this feature is
unusually strong for a radio galaxy spectrum.

{\bf 6C*0106+397} A blind spectrum was taken pointed at the mid-point
of the hotspots with LRIS on the Keck telescope and three definite
emission lines confirm
this as a radio galaxy at $z=2.284$. Emission-line fluxes are
estimated using a spectrophotometric standard taken on a different
night, so the emission-line luminosities may be unreliable. The
limited wavelength coverage of the LRIS spectrum means that we have no
data on the Ly$\alpha$ line. There is also a bright object ($K = 17.8$ in an
8 arcsec aperture) $\approx 5$ arcsec to the south-east of the radio galaxy ID which was 
aligned with our slit. A strong emission line at 6084 \AA\, which we
associate with [OII]$\lambda 3727$\,\AA, and a blue
continuum at the position of this object, lead us to conclude that it is a
foreground galaxy at $z = 0.632$ and is not associated with the radio
emission.

{\bf 6C*0112+372} Blind spectroscopy at the Lick telescope, pointed at the bright radio
component in our 4.9 GHz map, shows a bright emission line at the
same wavelength as a single line identification from spectra obtained
at the WHT in 1991 which we thought to be Ly$\alpha$ at $z =
2.535$. The detection of another line at 5476\,\AA\, from our Lick spectrum
corresponding to CIV$\lambda 1549$\,\AA\, confirms this as a radio galaxy at $z = 2.535$.

{\bf 6C*0115+394}
The redshift of this object is based on five emission lines.

{\bf 6C*0118+486}
This source has strong Ly$\alpha$ with corresponding CIV$\lambda 1549$\,\AA\, and
CII]$\lambda 2326$\,\AA\, emission at
$z=2.350$. We also identify three emission lines, marked (a), (b) and
(c) in Fig.~\ref{fig:spectra}, which correspond to [OII]$\lambda 3727$\,\AA\,,
H$\beta\lambda 4861$\,\AA\, and
possibly [OIII]$\lambda5007$\,\AA\, at $z=0.529$.
These lines suggest that there may be a foreground galaxy along or near the
line of sight of this source. We defer discussion of this to
Jea01. There also appears to be absorption in the Ly$\alpha$ profile
most likely due to associated \HI absorption (see Sec.~\ref{sec:sizevslinelum}).

{\bf 6C*0122+426}
A Ly$\alpha$ line at $z=2.635$ and a definite line corresponding to
CIII]$\lambda 2326$\,\AA\, place this source at $z = 2.635$. Observations on 
the Lick telescope confirm the other lines at this redshift.

{\bf 6C*0128+394} Spectra from both 1995 and 1998 show a clear
emission line at 7190 \AA\, which we associate with [OII]$\lambda
3727$\,\AA\, at $z = 0.929$. Further faint lines corresponding to
MgII$\lambda 2799$ and [NeIV]$\lambda 2424$ confirm this redshift.
 
{\bf 6C*0132+330}
The redshift of this source is based on three definite emission lines
and a possible fourth; Ly$\alpha$ is too far blueward of our spectrum
to detect.

{\bf 6C*0133+486} Our $K$-band image shows two objects aligned with
the radio emission. The source of the radio emission is therefore
ambiguous. Our tentative identification of [OII]$\lambda 3727$\,\AA\,
at $z=1.029$ at the position of the brighter optical ID is mildly
supported by very weak MgII$\lambda 2799$\,\AA\, and CII]$\lambda
2326$\,\AA\, at this redshift. This object has $K = 18.7$ (5 arcsec aperture), which,
using the $K-z$ diagram (Eales et al. 1997; Jea01) is consistent with this
redshift (although slightly fainter than mean relation). Further spectroscopic observations are needed to confirm
this redshift.

{\bf 6C*0135+313} $z=2.199$ radio galaxy with strong Ly$\alpha$ and
CIII]$\lambda 1909$\,\AA.  The Ly$\alpha$ emission is split into two
peaks at 3893\AA\, and 3909\AA\, with a trough between them. This is
possibly due to H{\small I} absorption associated with the host
galaxy. The projected linear size of the radio emission is small
($ D \approx 15$\,kpc), consistent with the results of
van Ojik et al. (1997) in which small ($< 50$\,kpc) radio sources are
more likely to exhibit associated H{\small I} absorption. The
CIV$\lambda 1549$\,\AA\, and HeII$\lambda 1640$\,\AA\, lines have FWHM
$< 10$\,\AA\, and could be spurious.
 
{\bf 6C*0136+388} The $K-$band identification for this source is
spatially coincident with the brightest feature in the 4.9\,GHz map,
which also has a spectral index $\sim 0$ and is most likely the
core. Optical spectra taken on the WHT in 1995 are inconclusive
regarding the redshift of this source, mainly due to the poor seeing,
although there is faint continuum visible down to at least $\sim
4000$\,\AA. Our more recent observation in Dec 1998 only provided
spectra from the red-arm of ISIS due to a technical fault in the
blue-arm observation. There is a strong emission line at 7858 \AA\, in
these spectra, and along with the evidence of continuum into the blue
from the previous observations, we take this line to be [OII]$\lambda
3727$\,\AA\, at $z=1.108$. The $K$-band magnitude of $\sim 17.3$
(8 arcsec aperture) also suggests a redshift around this value.

{\bf 6C*0139+344} The spectroscopic observations of this source
highlight the problems associated with assigning redshifts on the
basis of a single line.  Spectroscopy for this source was initially
attempted on the WHT in 1992 and again in 1995. The spectrum from the
1995 run showed a bright emission line at 7381\AA. This was initially
thought to be Ly$\alpha$ at $z=5.07$ as there was no obvious continuum
or emission lines blueward of this line. However, its proximity to a
$z=0.37$ foreground galaxy ($\sim 2.5$ arcsec to the north) and the
lack of a second emission line prevented conclusive proof.  We then
obtained an optical spectrum, dispersed at a longer wavelength to
search for CIV$\lambda 1549$\,\AA\, at $\lambda = 9402$\,\AA\, on the
Keck-II telescope.  However, the presence of a bright emission line at
9828\,\AA\, meant that the lines are now identified with MgII$\lambda
2799$\,\AA\, and [OII]$\lambda 3727$\,\AA\, at $z=1.637$.  The lack of
a spectrophotometric standard for the set-up used to take this
spectrum makes it impossible to flux calibrate the spectrum. Thus,
although we present a spectrum in Fig.~\ref{fig:spectra}, the
flux-density scale can only be used to estimate relative fluxes since
no absolute line fluxes are available at present.

{\bf 6C*0140+326}
The highest redshift source in the sample at $z=4.41$;
the radio emission is likely to be gravitationally amplified 
by a factor $\ltsim 2.0$ by a $z=0.927$ galaxy $\sim 1.6$ arcsec 
from the eastern radio lobe (see Rawlings et
al. 1996). Deeper Keck spectroscopy and imaging of this object can be
found in De Breuck (2000c) and van Breugel et al. (1998).

{\bf 6C*0142+427} Strong Ly$\alpha$ and very weak CIV$\lambda
1549$\,\AA\, co-spatial with the central core component in the radio
map confirm this as a radio galaxy at $z=2.225$.  There is also some
Ly$\alpha$ emission $\sim 5$\,arcsec to the south-east, which is
probably associated with the lobe.

{\bf 6C*0152+463} 
The presence of strong Ly$\alpha$ and HeII in our
spectrum confirm this as a radio galaxy at $z=2.279$.

{\bf 6C*0154+450} Broad CIV$\lambda 1549$\,\AA\, and MgII$\lambda
2799$\,\AA\, confirm this source as a quasar at $z=1.295$.

{\bf 6C*0155+424} Three bright objects lie close to the radio source
in the $R-$ and $K-$band images. The object closest to the centre of
the radio emission is a galaxy at $z=0.513$. This is separated from
another object to the north-west by $\ltsim 2$\,arcsec. The seeing for
 our spectroscopic observations was not sufficient to clearly resolve
these two objects. We do not have a spectrum for the third
spatially-resolved object $\sim 4$\,arcsec to the south-east of the
central object. These objects will be discussed in more detail in Jea01.

{\bf 6C*0158+315} It was unclear from our radio map whether this
source was a discrete object or a hot-spot of a larger source. In our
$R$- and $K$-band imaging we find an object co-spatial with the peak
of the radio
emission which we now take to be the radio galaxy. Definite lines
corresponding to MgII$\lambda 2799$\,\AA , [NeIV]$\lambda
2424$\,\AA\, and CII]$\lambda 2326$\,\AA\, confirm this source at
$z=1.505$. The NVSS radio map shows that the radio component
associated with the optical ID could conceivably be the core of a large ($\theta
\sim 14$\,arcmin) triple source, if this is the case then this source should be
excluded from the final sample, but if this were true the linear size
of the source would be huge ($\approx 7$\,Mpc for $\Omega_{\rm M}=1$, $\Omega_{\Lambda}=0$).

{\bf 6C*0201+499} The redshift of this source is based on five
emission lines. The single exposure including the Ly$\alpha$ emission
line has a cosmic ray in close proximity which may contribute to the
flux measurement in Table~\ref{tab:spectra}.

{\bf 6C*0202+478}
The near-infrared ID for this source is in very close proximity to two
bright
stars, consequently the 1-D extraction of the spectrum does not show
the emission lines very well. However, inspection of the 2-D spectrum shows
a definite emission line at 7314\AA, this would correspond to MgII at
$z=1.613$. There are also faint lines corresponding to [NeIV]$\lambda
2424$, CII]$\lambda 2326$\,\AA, CIII]$\lambda 1909$\,\AA\, and
CIV$\lambda 1549$\,\AA\, at this redshift, which, if on their own would be
inconclusive but collectively suggest that this redshift is
correct. 

{\bf 6C*0208+344}
Three definite emission lines place this source at $z=1.920$. The
Ly$\alpha$ emission also appears to have associated \HI absorption
(see Sec.~\ref{sec:sizevslinelum}).

{\bf 6C*0209+276} This source has an identification in $K$-band which
is coincident with the peak of the radio emission. Spectroscopy
reveals an emission line which we take to be [OII]$\lambda
3727$\,\AA\, at $z=1.141$, there is also possibly an emission line
corresponding to MgII$\lambda 2799$\,\AA\, at this redshift. A slight
worry was that the emission line appeared to be offset from the
expected target row on the ISIS detector (and there were no field
objects along the slit to bootstrap the astrometry). Further imaging
and spectroscopy may be needed to confirm whether this line emission
is associated with the radio source and also if it is definitely
[OII]. However, the $K$-band magnitude ($K=18.1$ in an 8 arcsec
aperture) is consistent with this redshift.

\begin{figure*}
{\hbox to \textwidth{ \null\null \epsfxsize=0.48\textwidth
\epsfbox{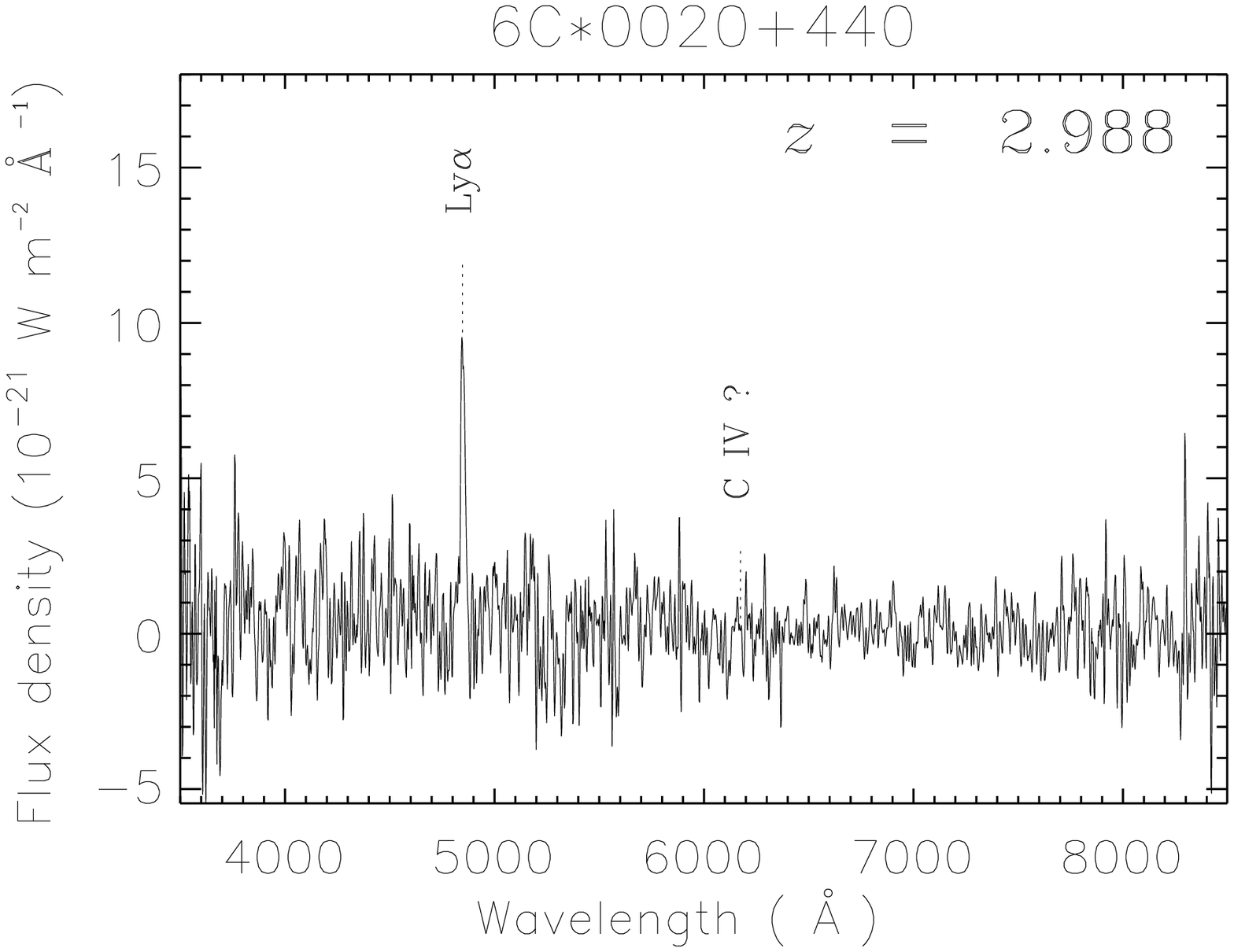}
\epsfxsize=0.48\textwidth
\epsfbox{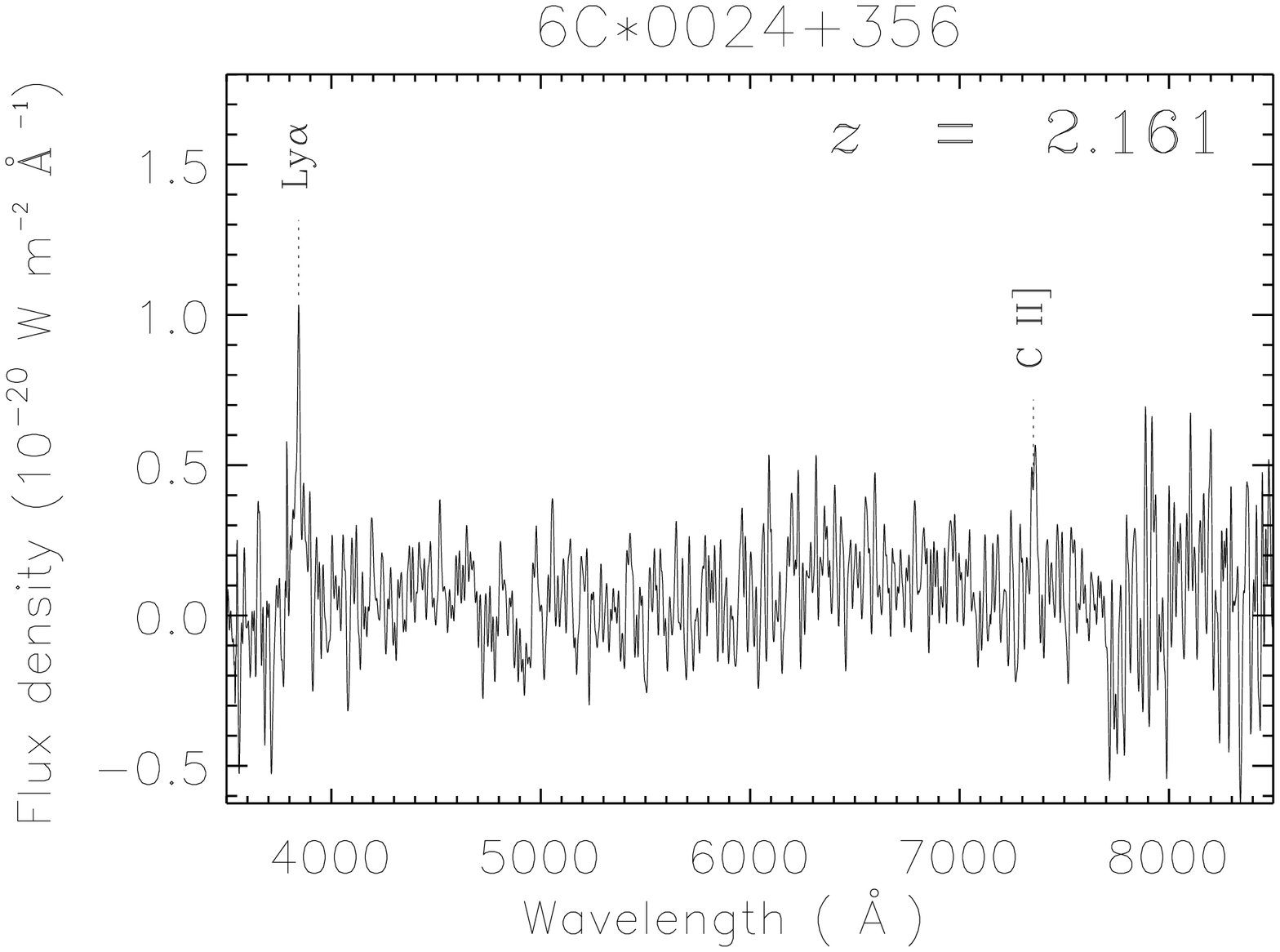} }}
{\hbox to \textwidth{ \null\null \epsfxsize=0.48\textwidth
\epsfbox{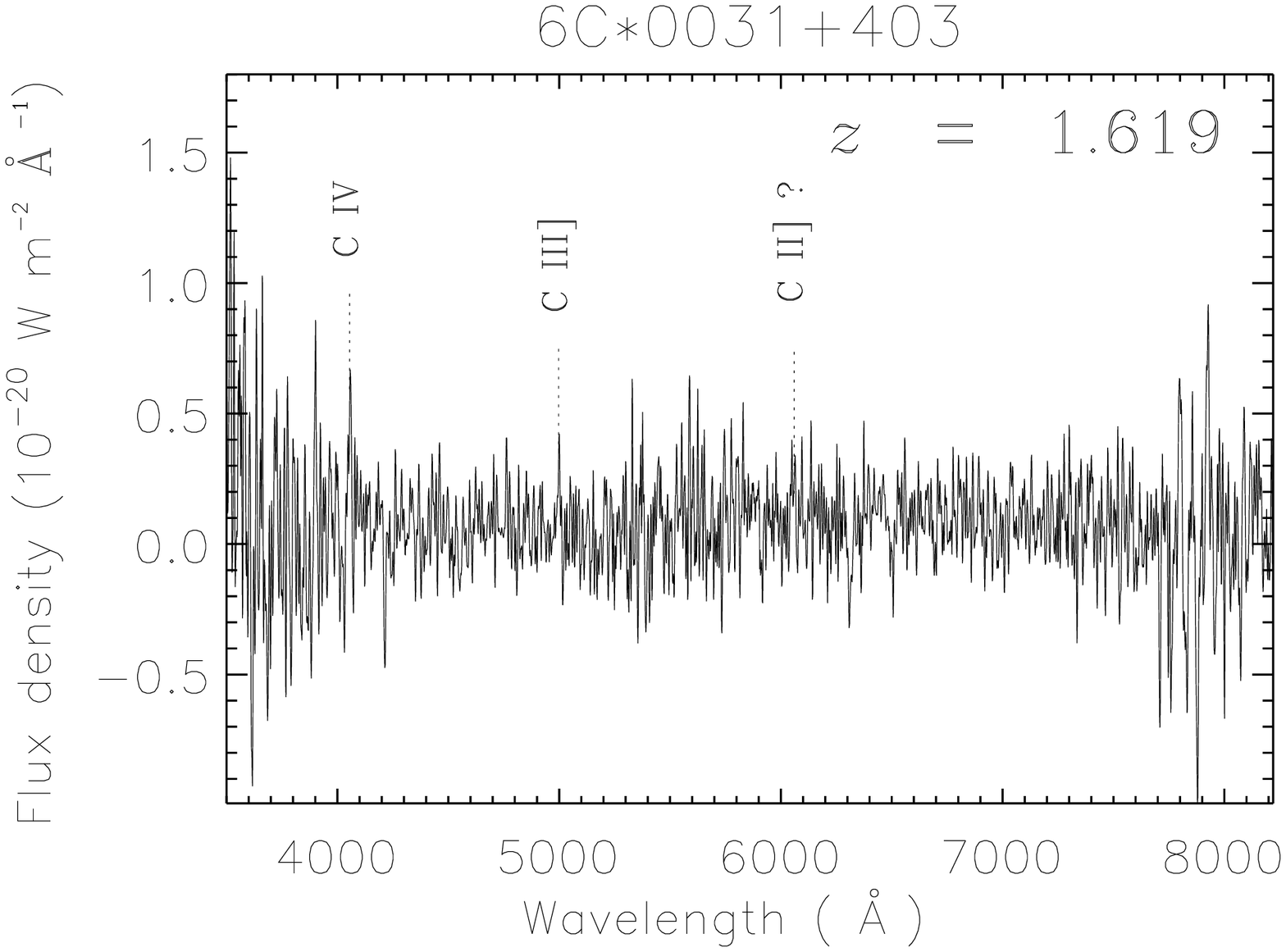}
\epsfxsize=0.48\textwidth
\epsfbox{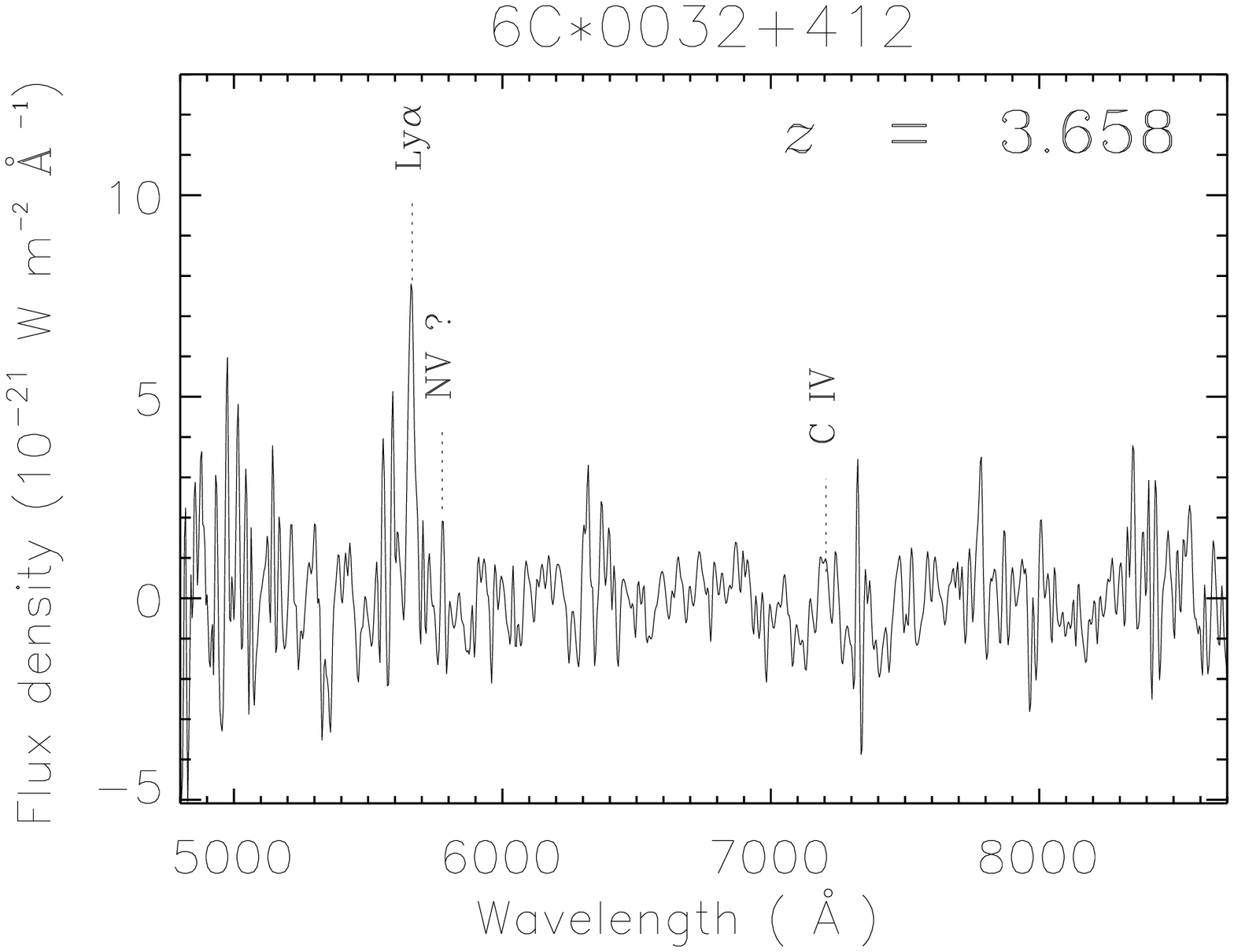} }}
{\hbox to \textwidth{ \null\null \epsfxsize=0.48\textwidth
\epsfbox{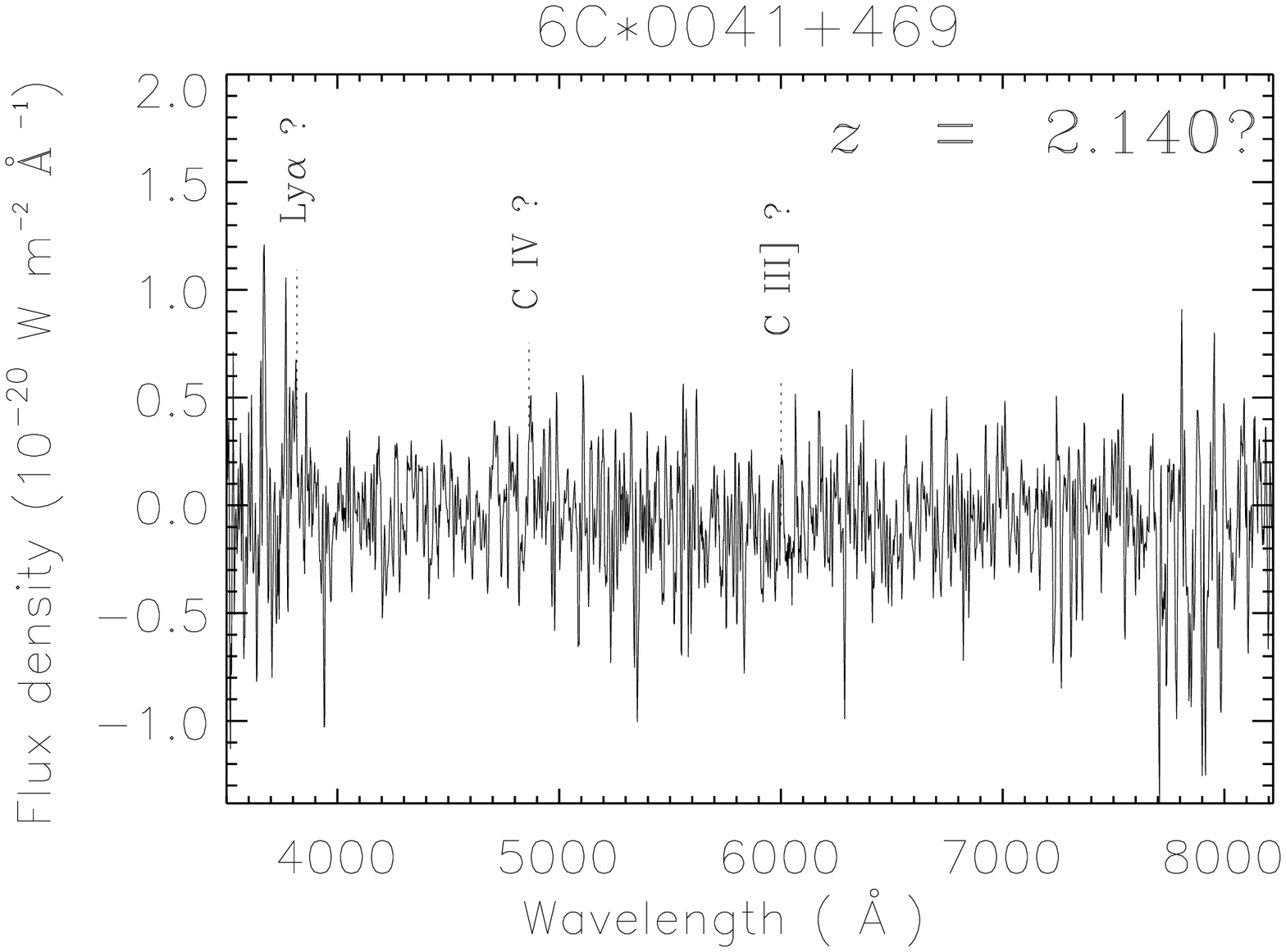}
\epsfxsize=0.48\textwidth
\epsfbox{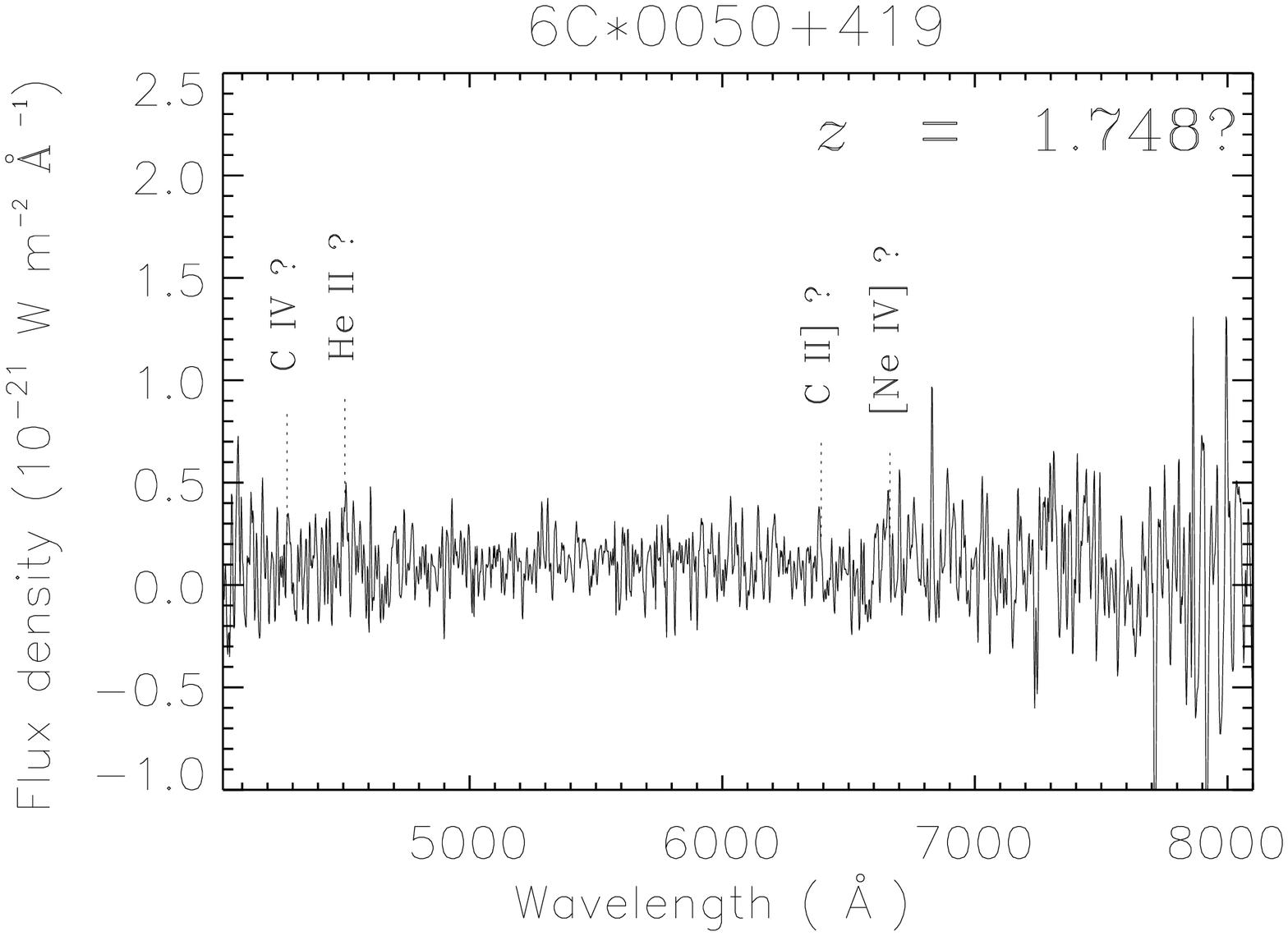} }}
{\hbox to \textwidth{ \null\null \epsfxsize=0.48\textwidth
\epsfbox{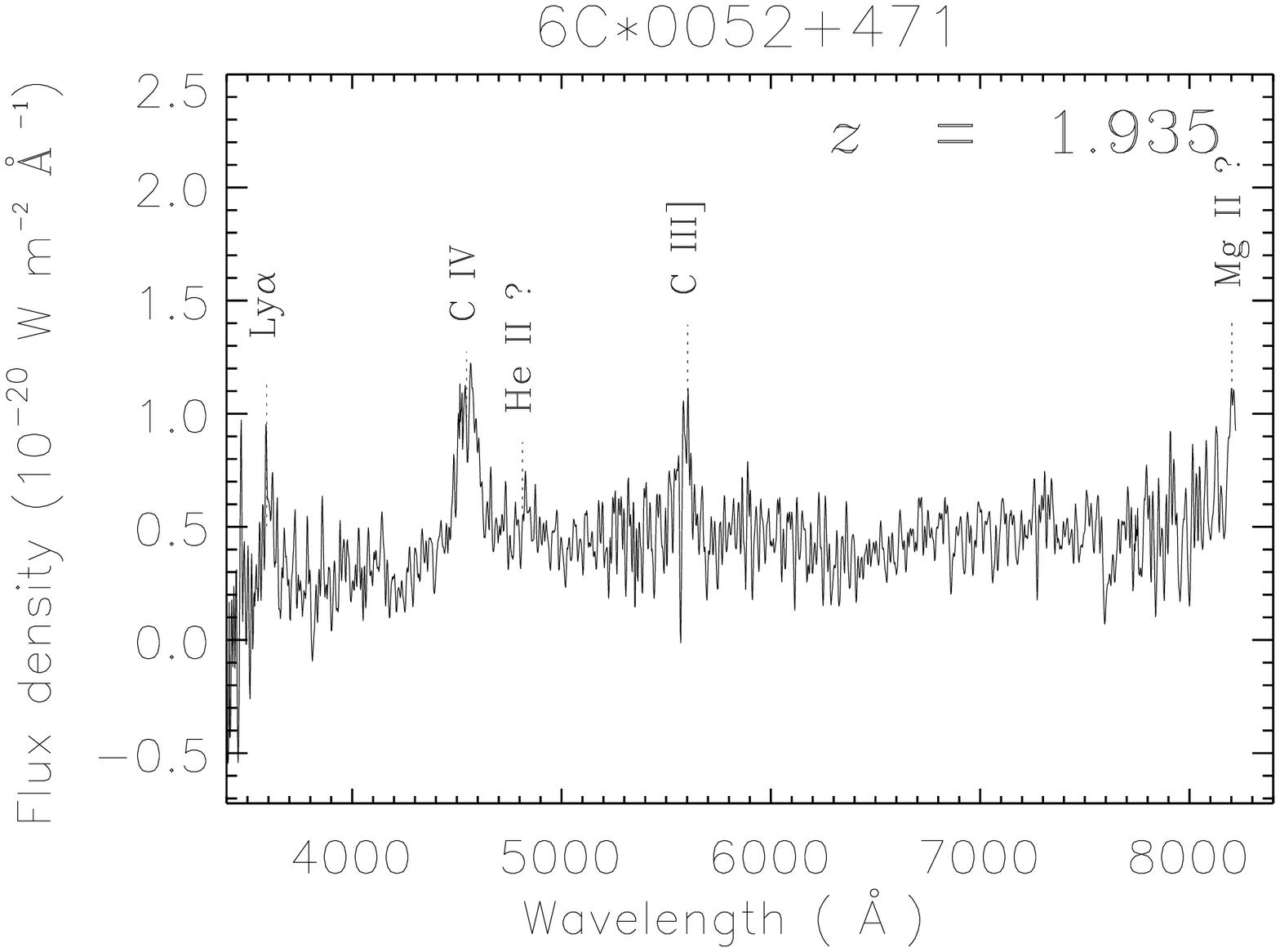}
\epsfxsize=0.48\textwidth
\epsfbox{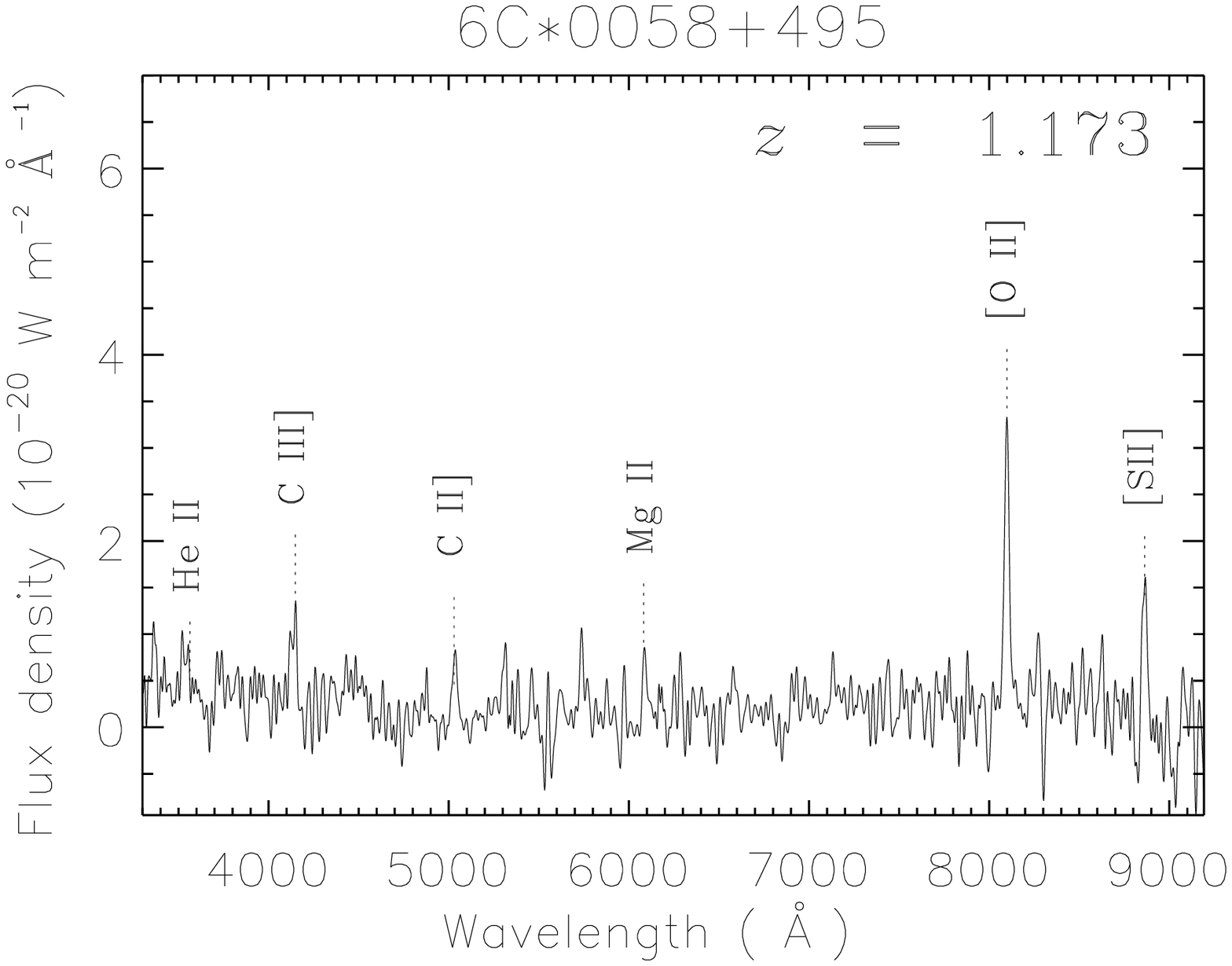} }}
\end{figure*}

\begin{figure*}
{\hbox to \textwidth{ \null\null \epsfxsize=0.48\textwidth
\epsfbox{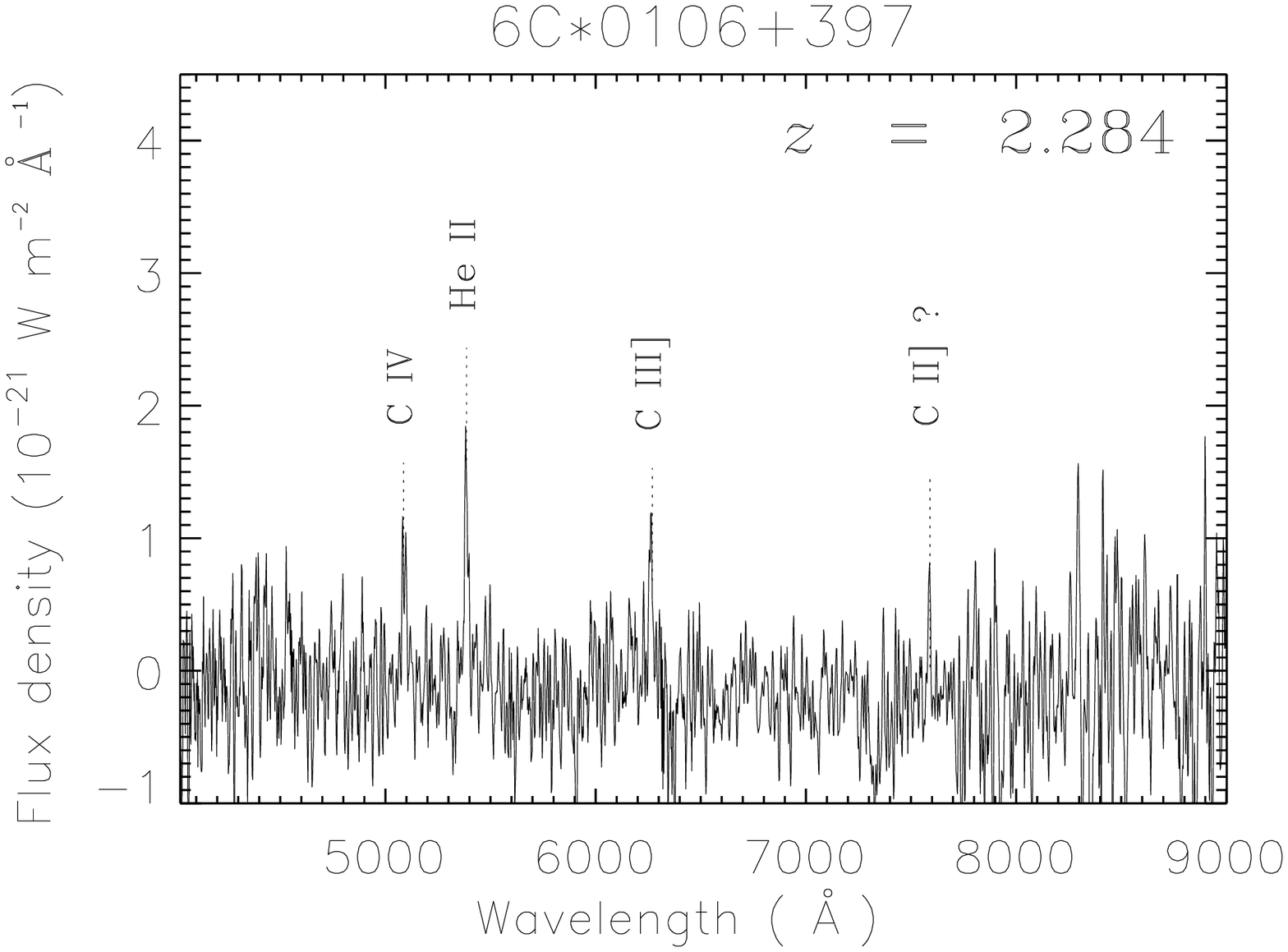}
\epsfxsize=0.48\textwidth
\epsfbox{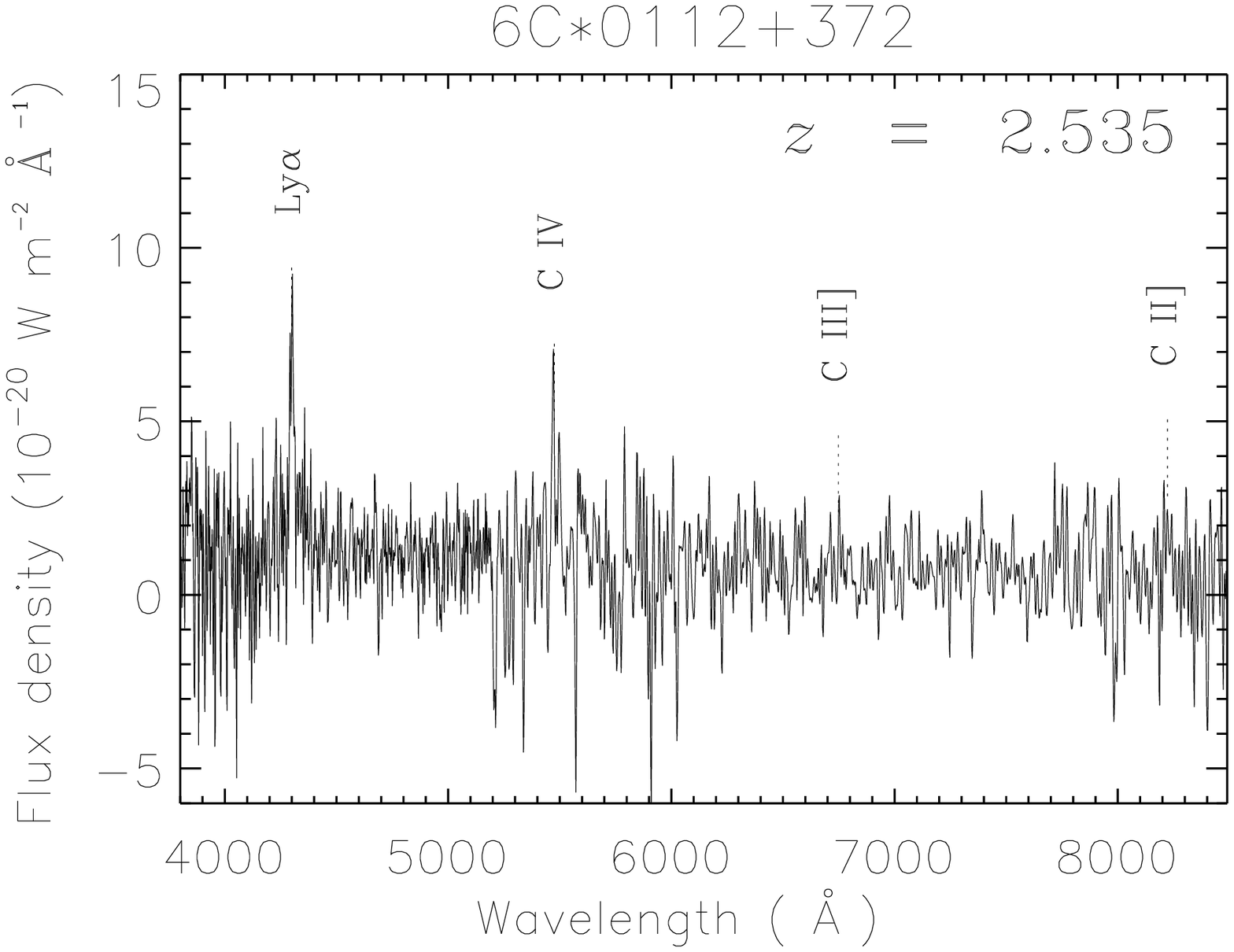} }}
{\hbox to \textwidth{ \null\null \epsfxsize=0.48\textwidth
\epsfbox{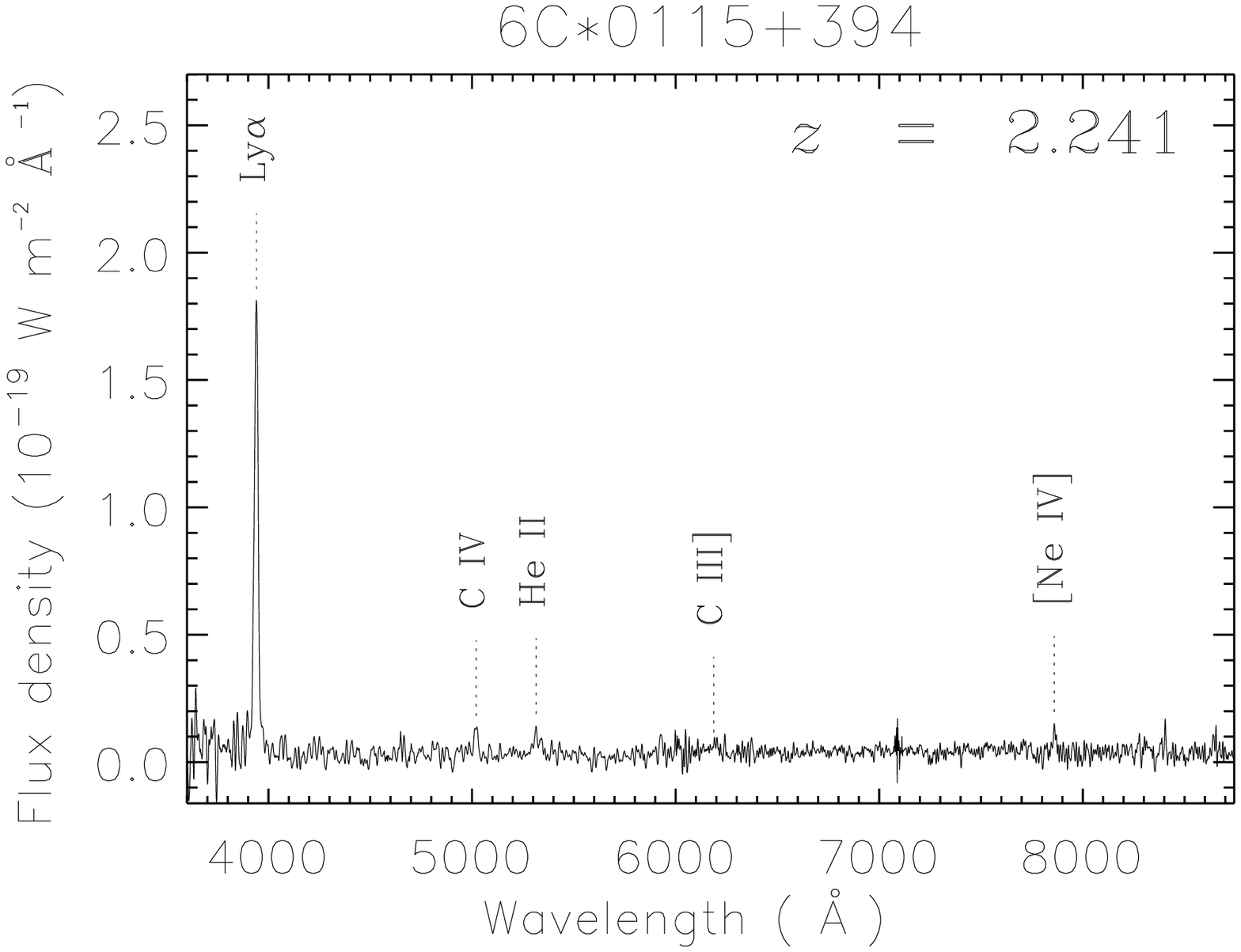}
\epsfxsize=0.48\textwidth
\epsfbox{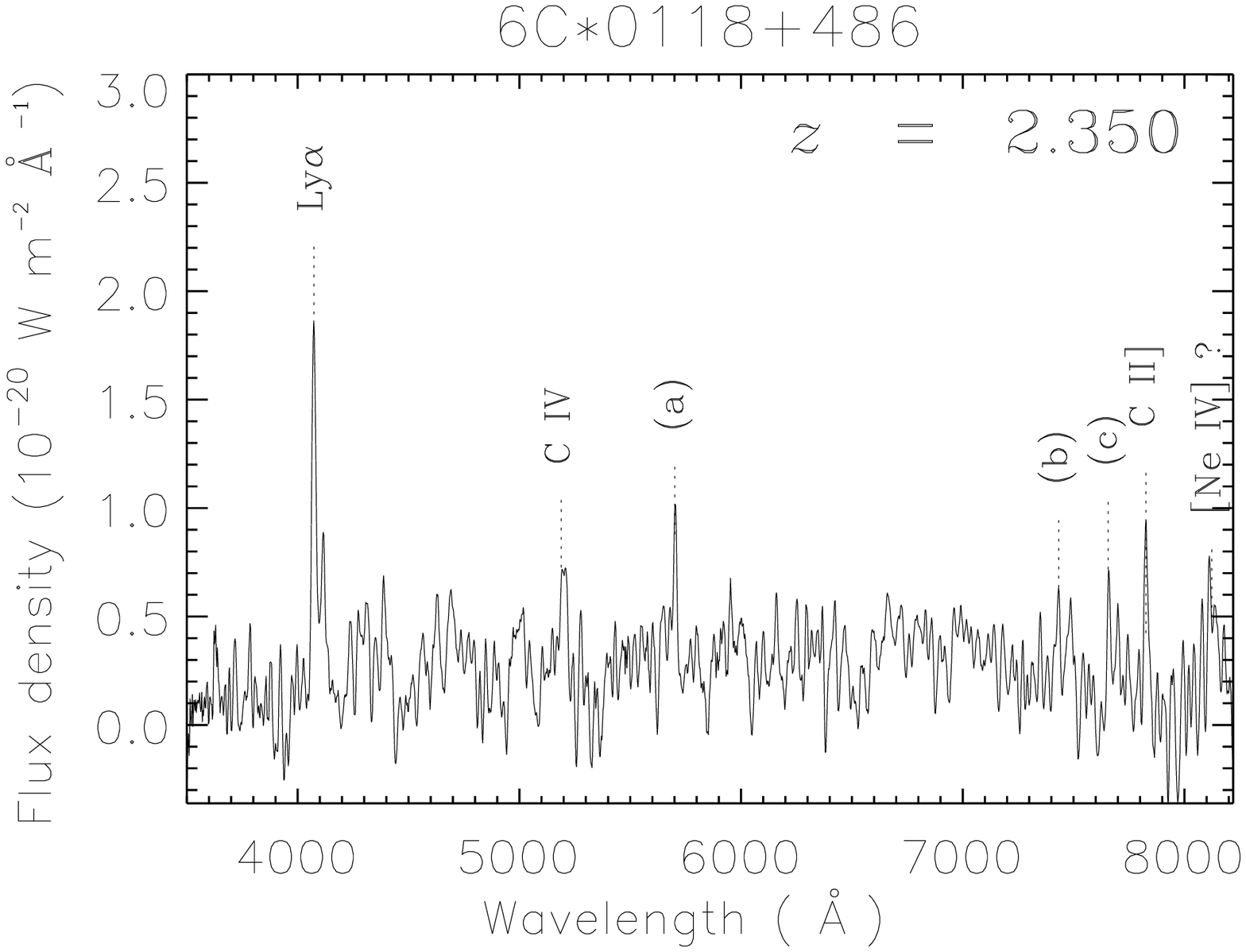} }}
{\hbox to \textwidth{ \null\null \epsfxsize=0.48\textwidth
\epsfbox{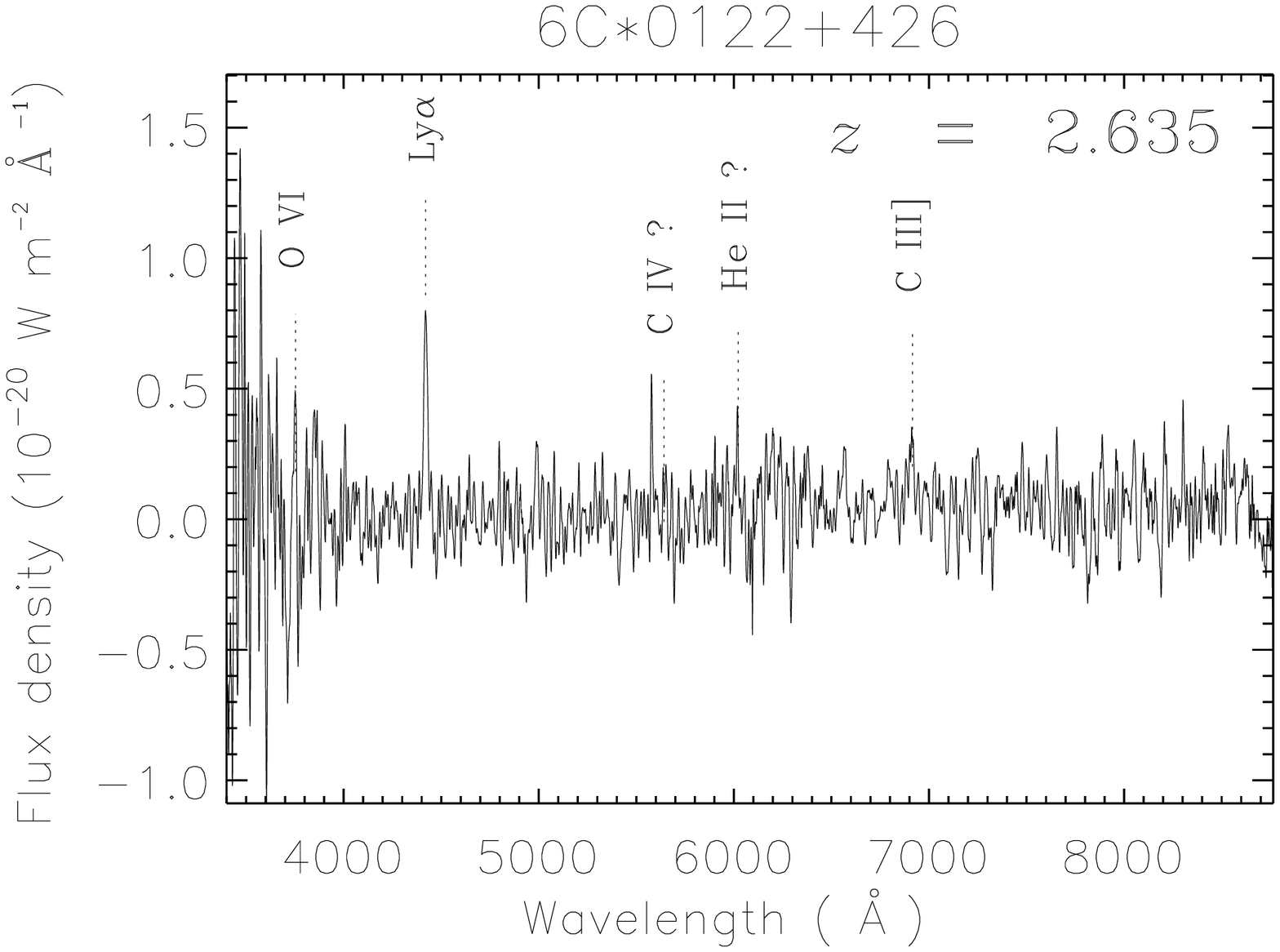}
\epsfxsize=0.48\textwidth
\epsfbox{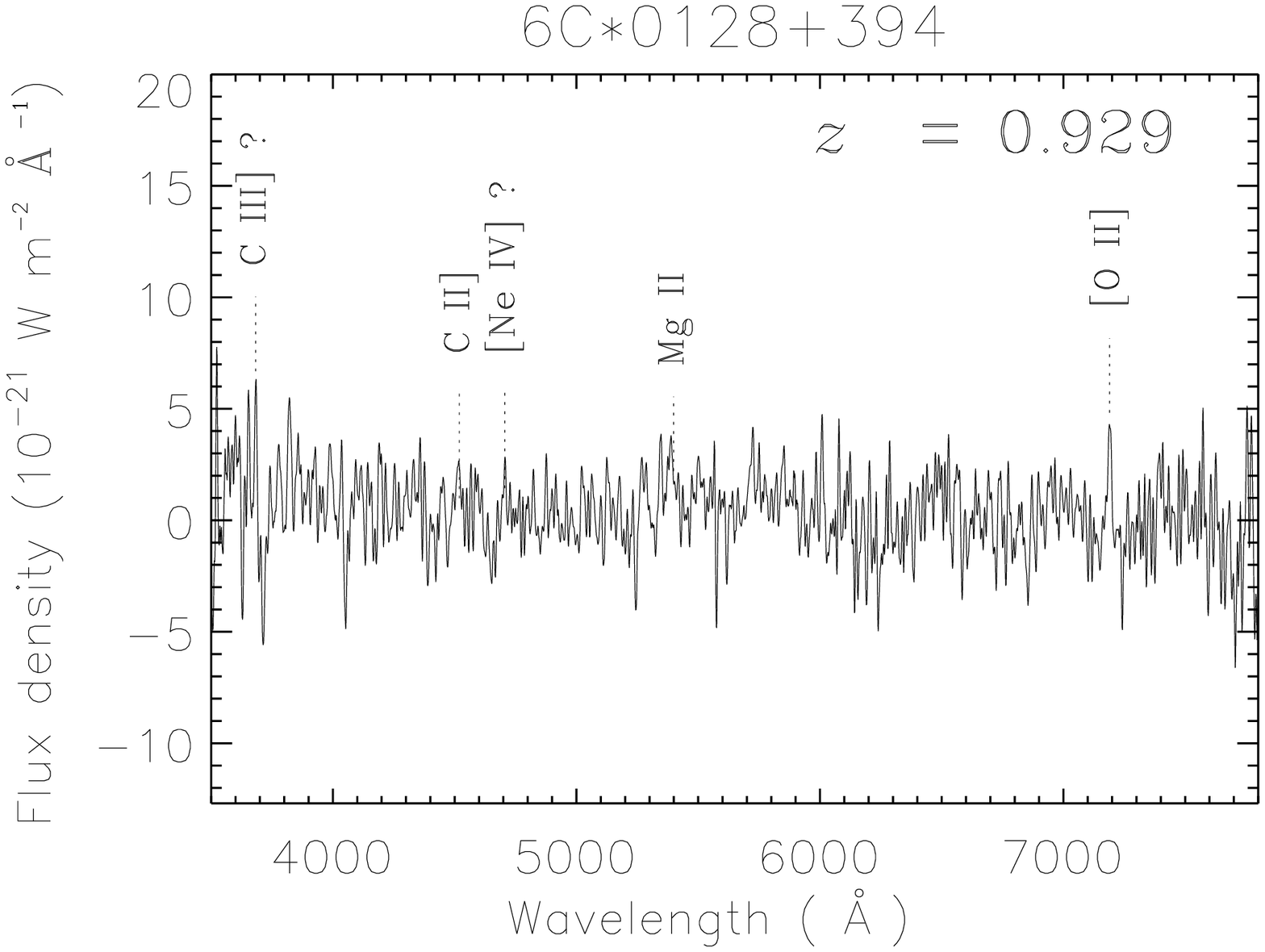} }}
{\hbox to \textwidth{ \null\null \epsfxsize=0.48\textwidth
\epsfbox{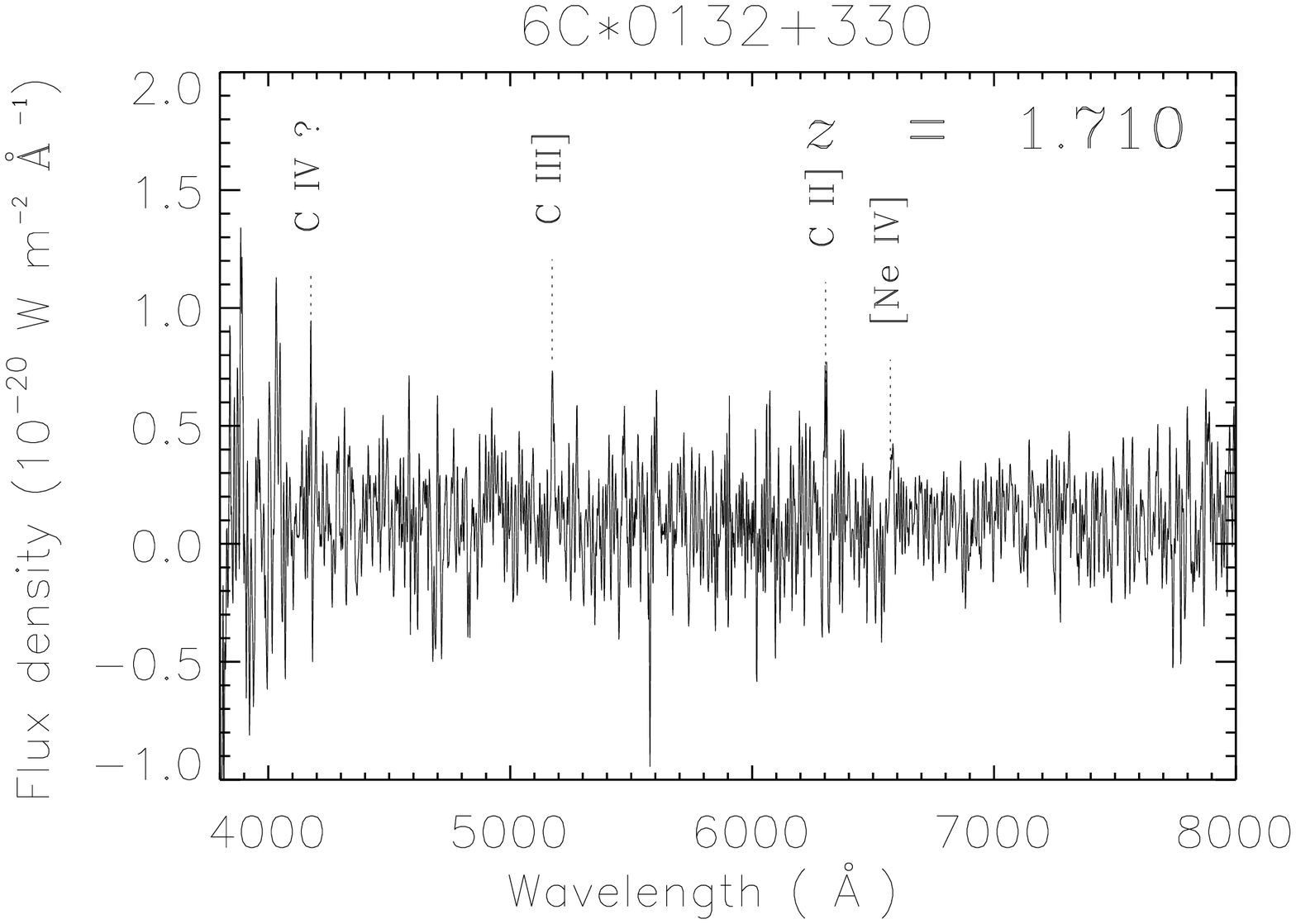}
\epsfxsize=0.48\textwidth
\epsfbox{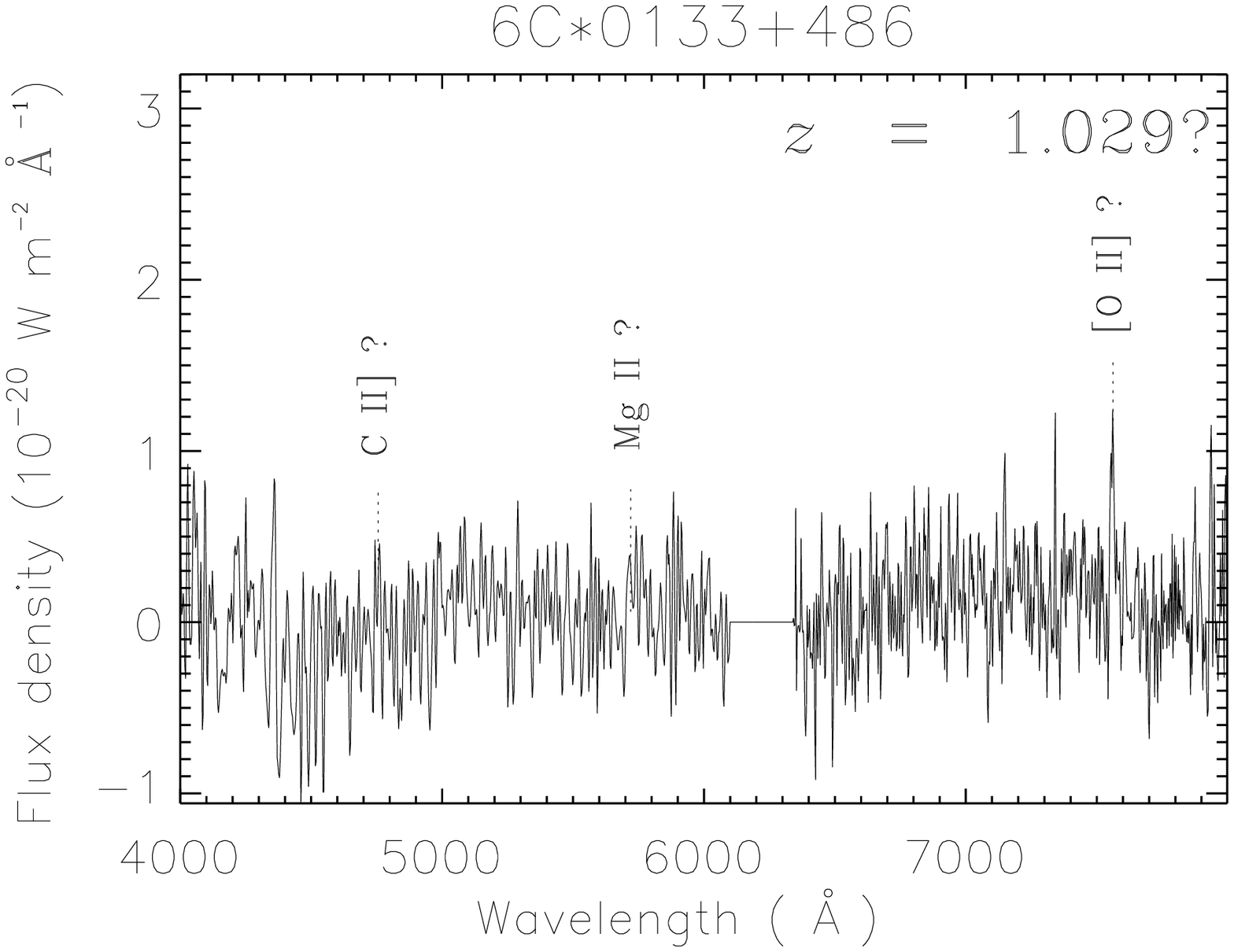} }}
\end{figure*}
\begin{figure*}
{\hbox to \textwidth{ \null\null \epsfxsize=0.48\textwidth
\epsfbox{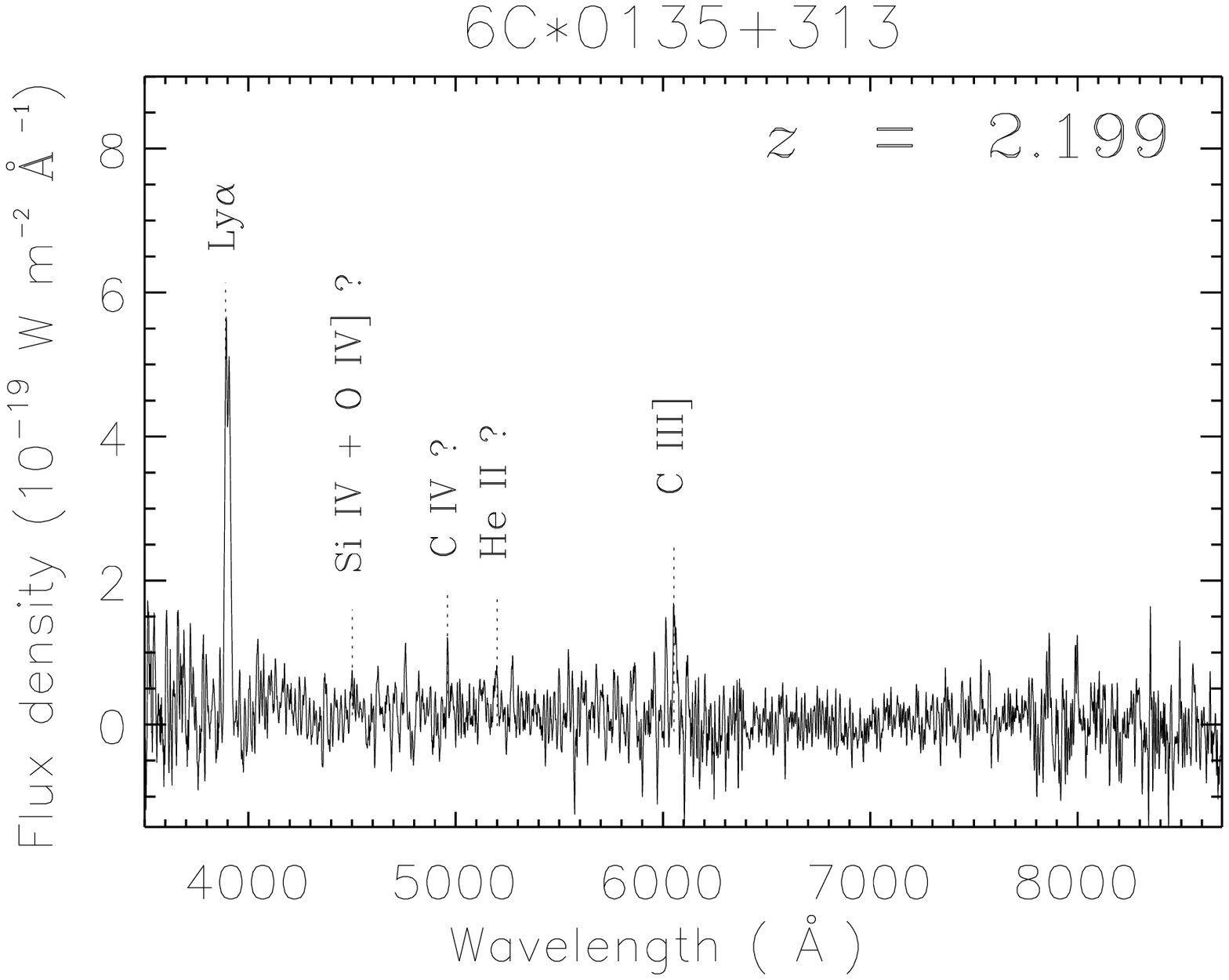}
\epsfxsize=0.48\textwidth
\epsfbox{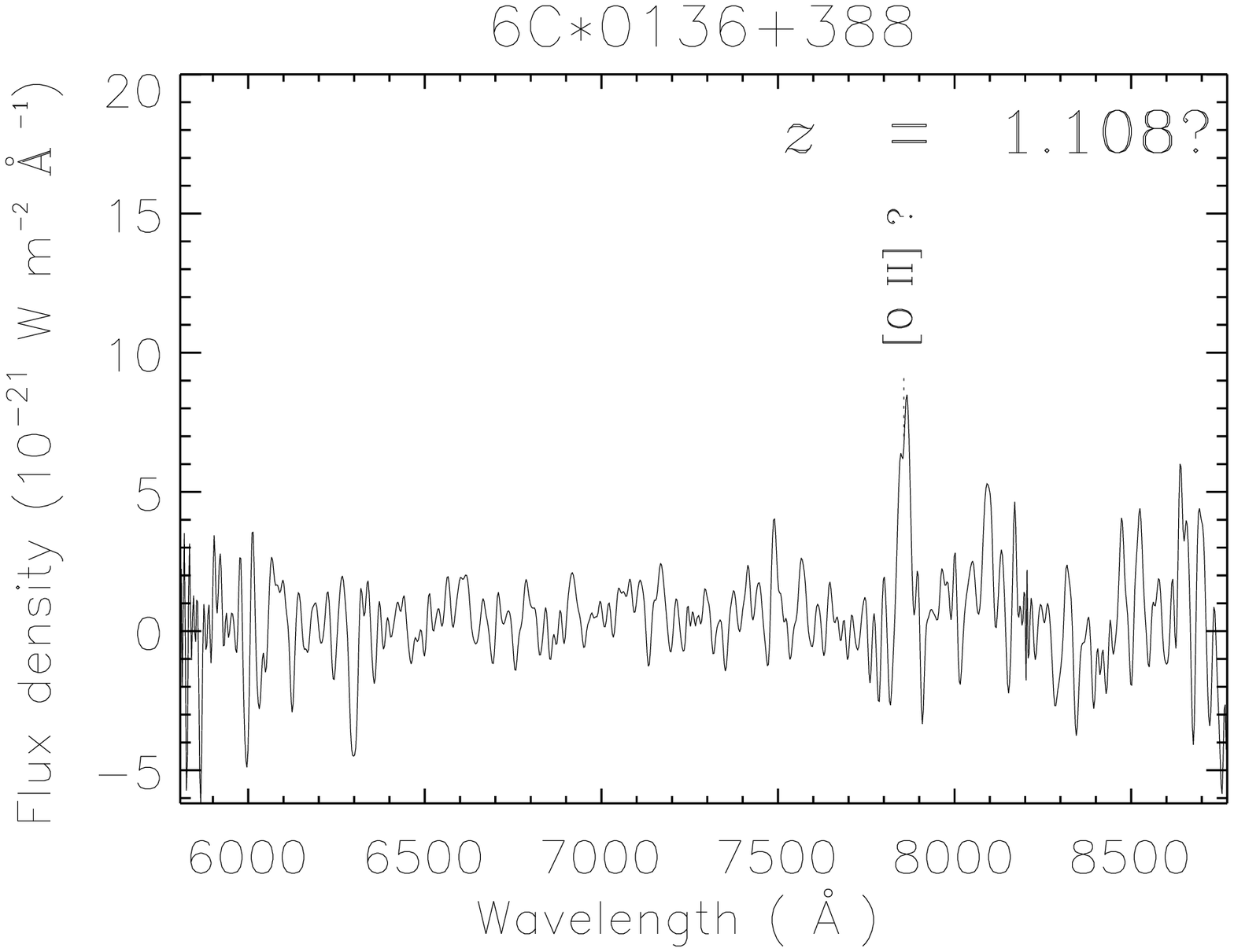} }}
{\hbox to \textwidth{ \null\null \epsfxsize=0.48\textwidth
\epsfbox{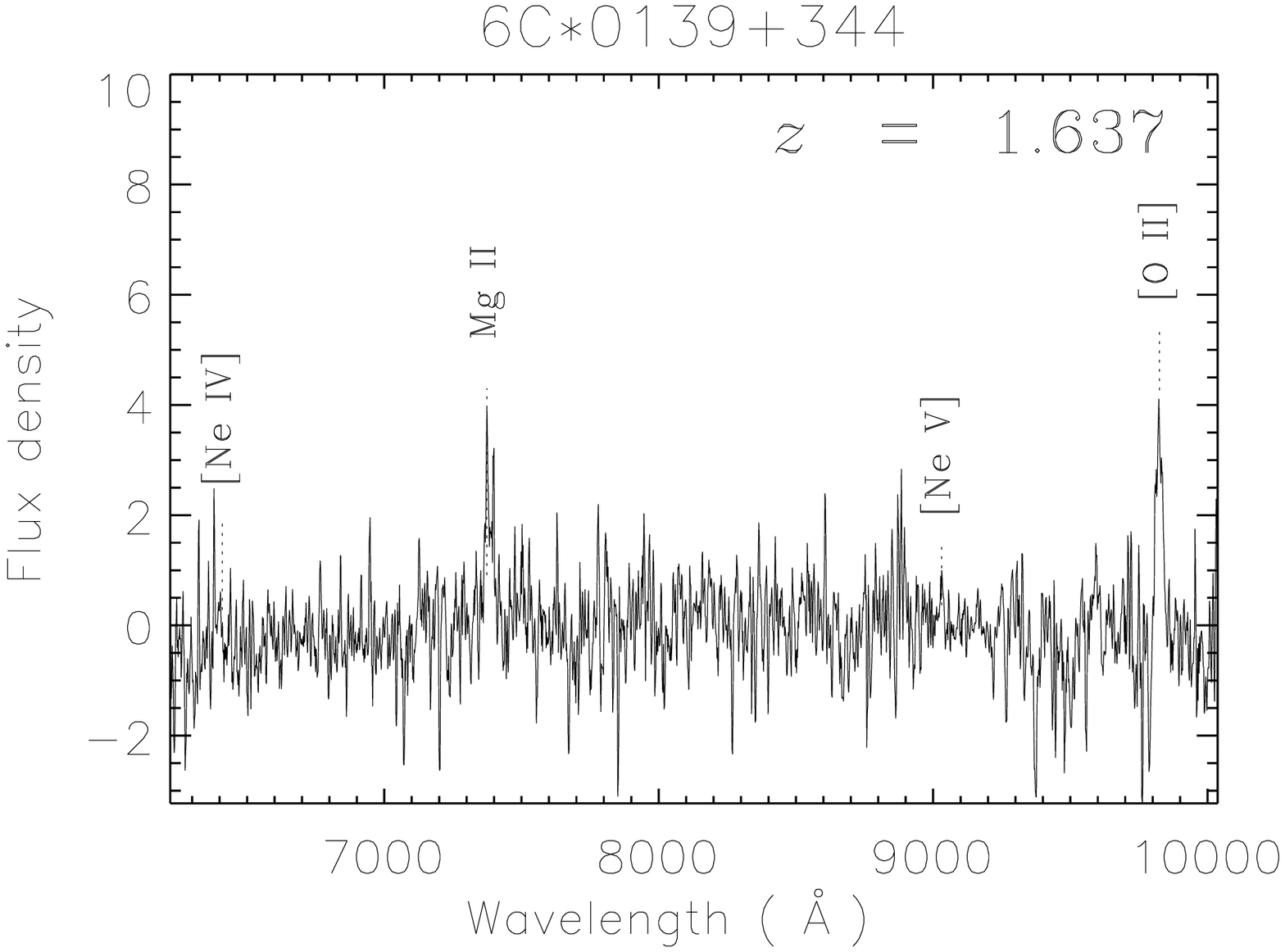}
\epsfxsize=0.48\textwidth
\epsfbox{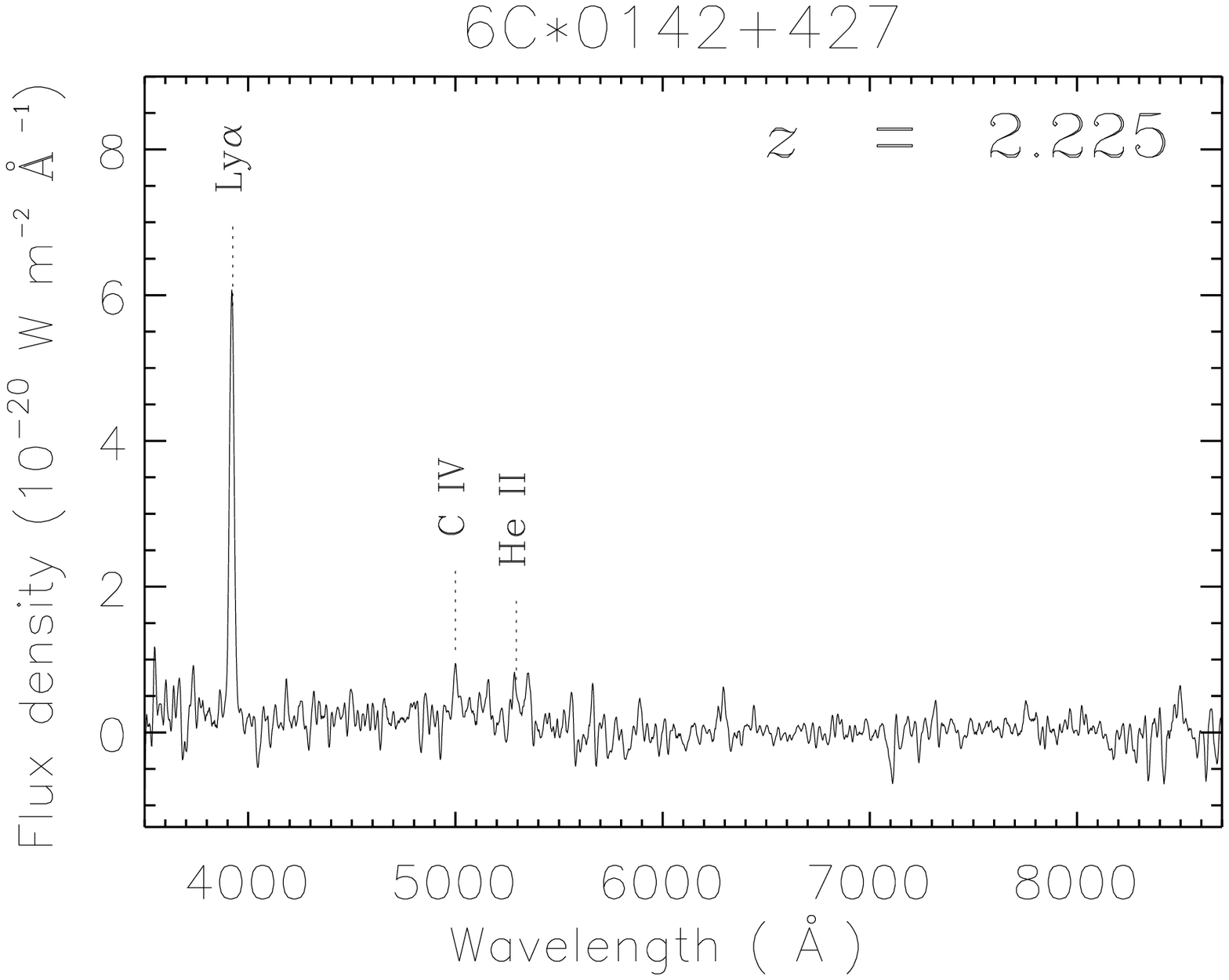} }}
{\hbox to \textwidth{ \null\null \epsfxsize=0.48\textwidth
\epsfbox{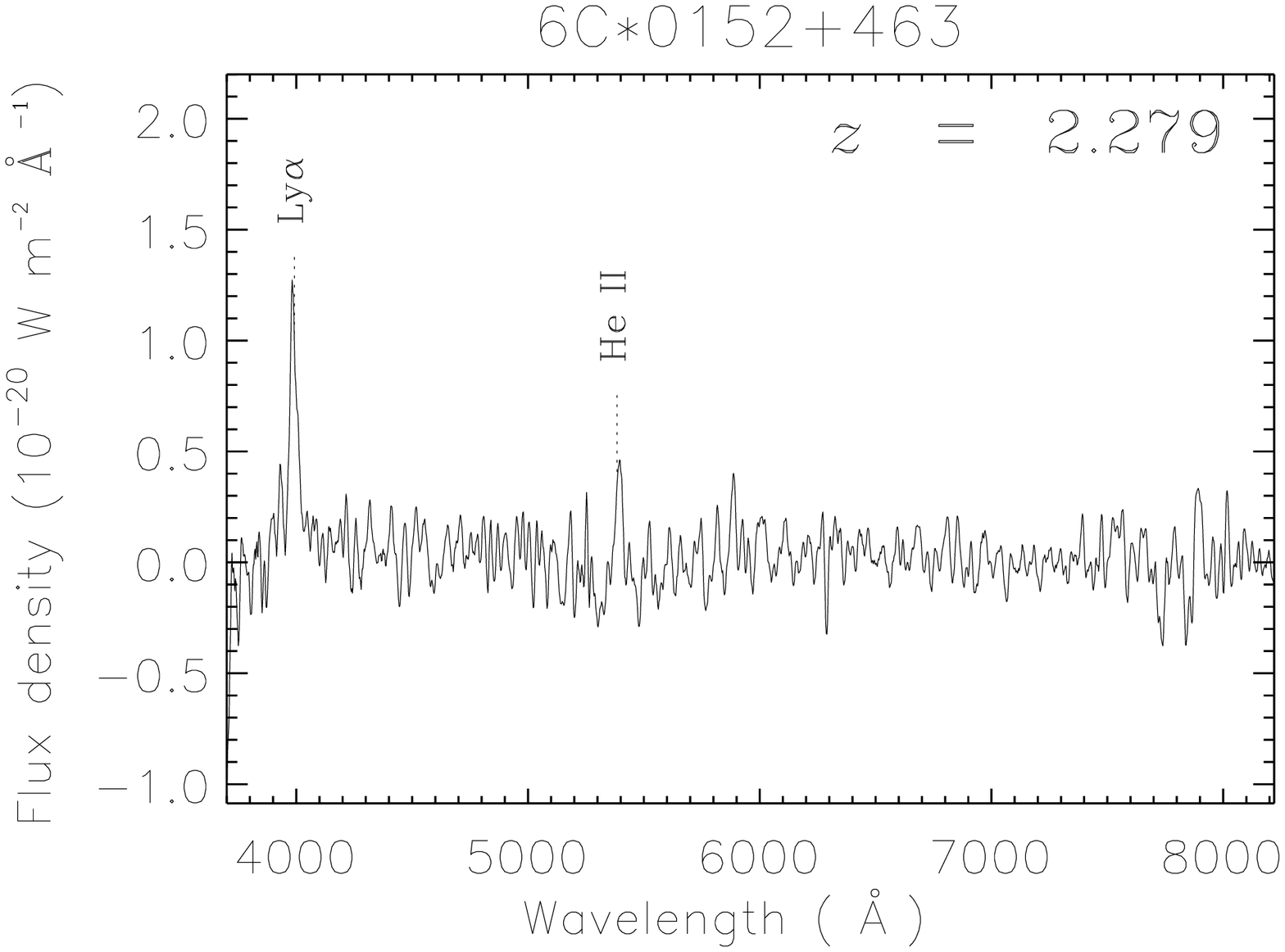}
\epsfxsize=0.48\textwidth
\epsfbox{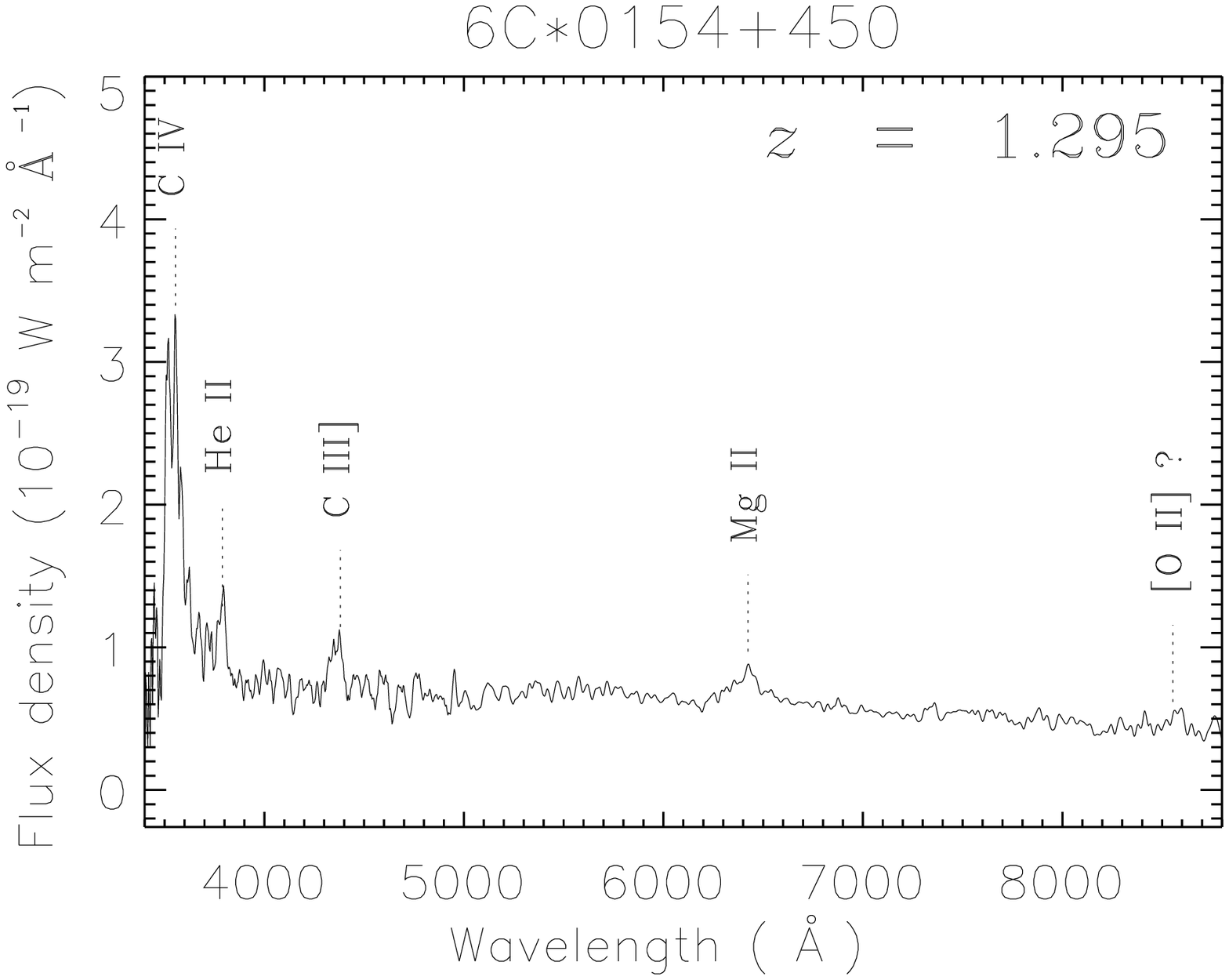} }}
{\hbox to \textwidth{ \null\null \epsfxsize=0.48\textwidth
\epsfbox{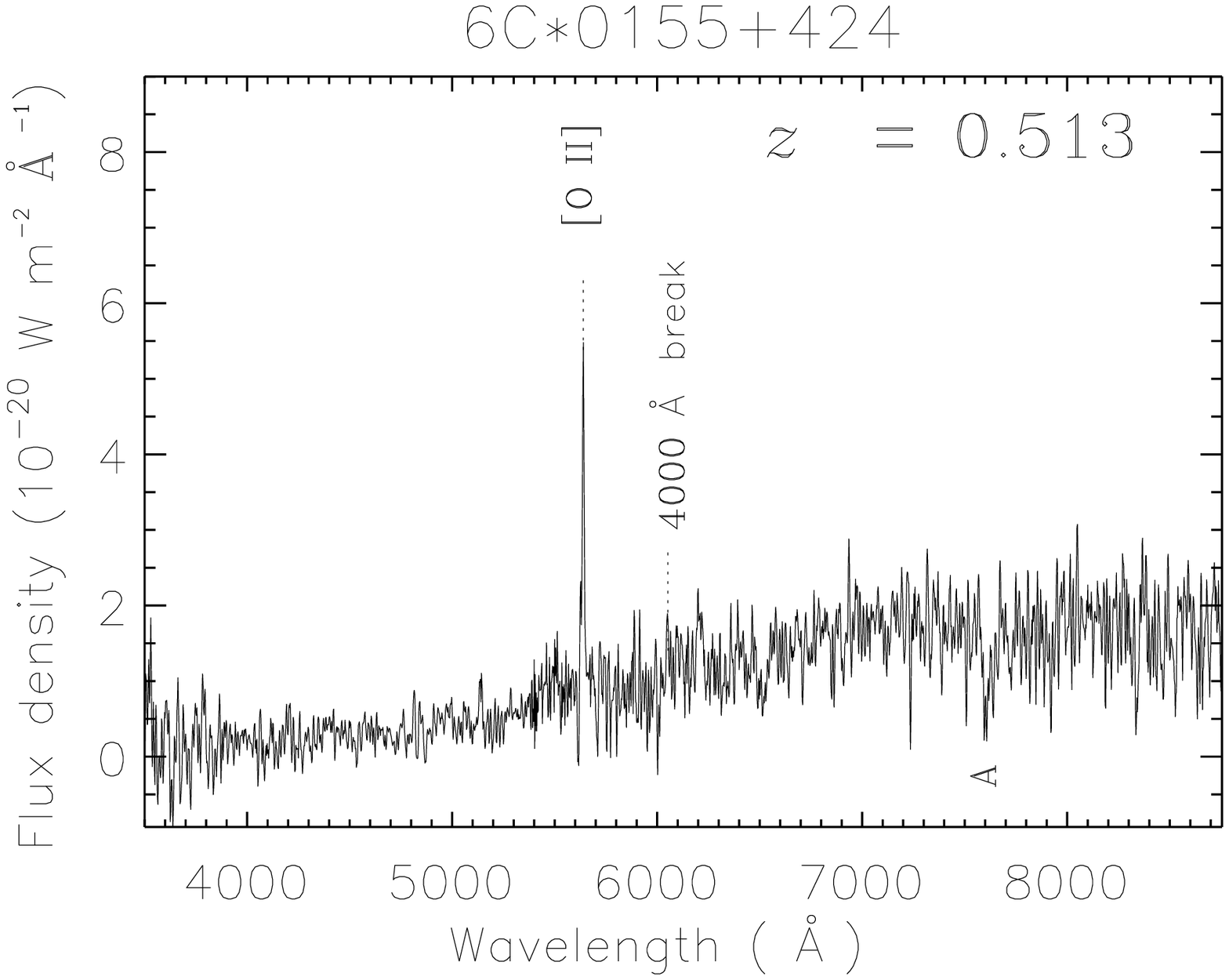}
\epsfxsize=0.48\textwidth
\epsfbox{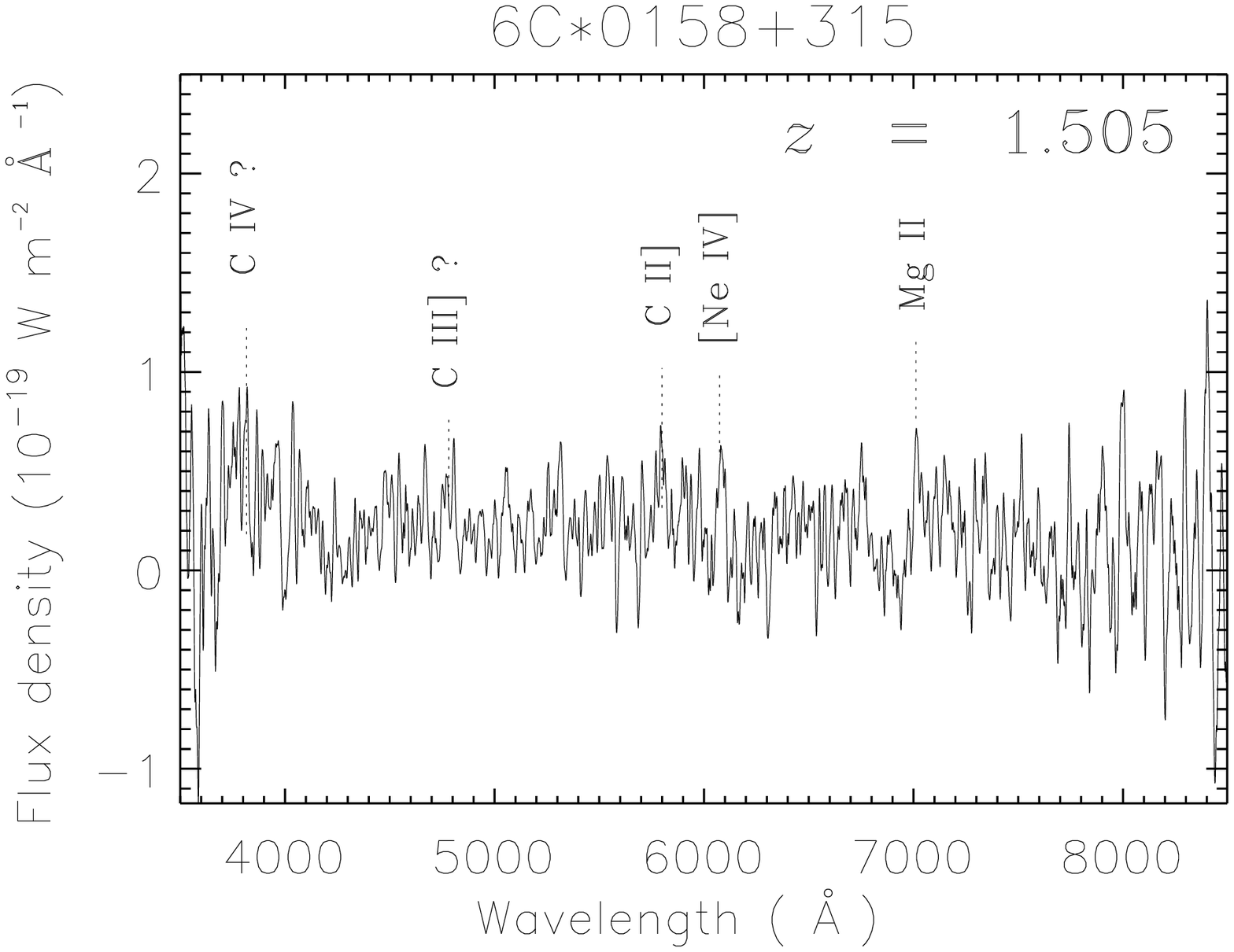} }}
\end{figure*}

\begin{figure*}
{\hbox to \textwidth{ \null\null \epsfxsize=0.48\textwidth
\epsfbox{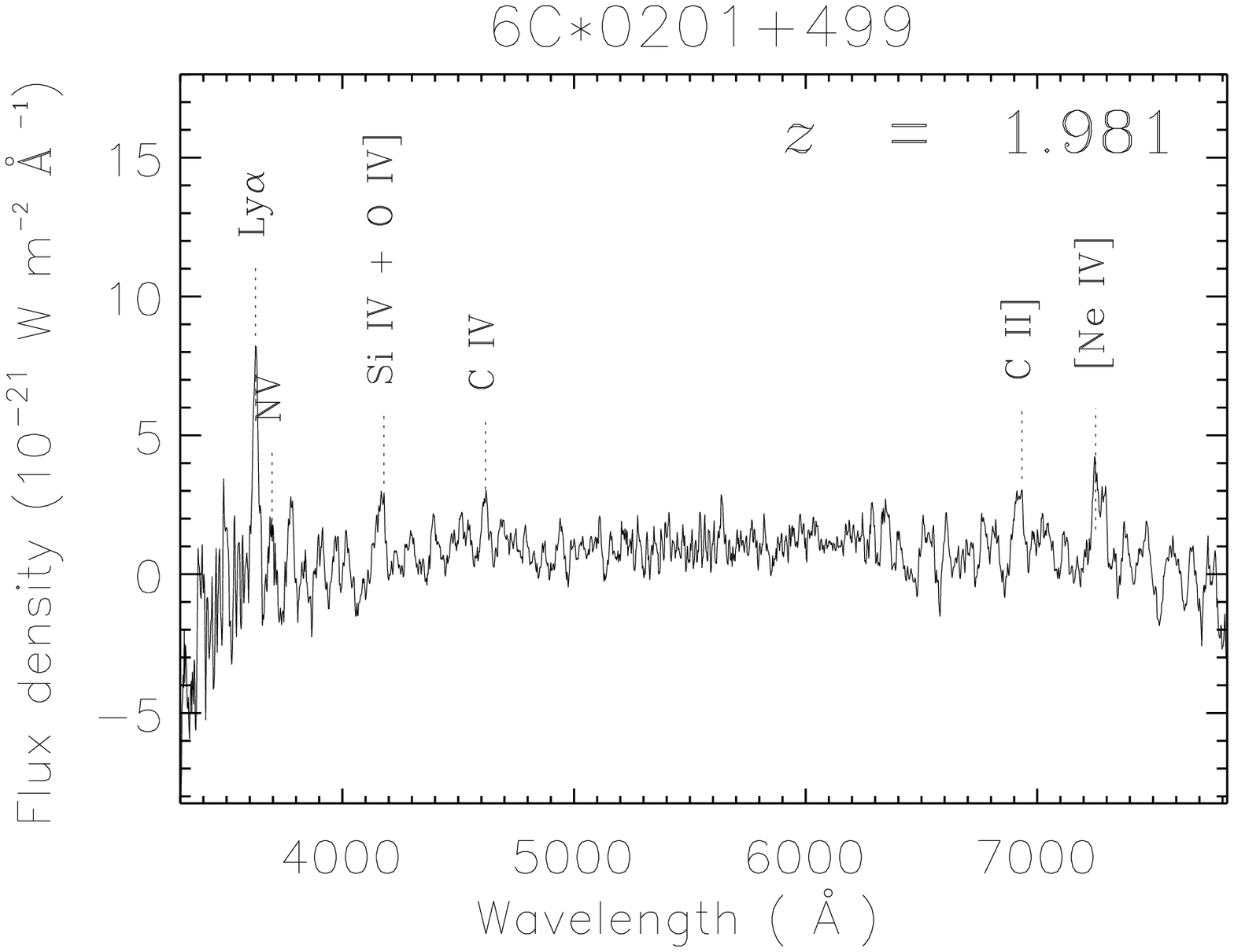}
\epsfxsize=0.48\textwidth
\epsfbox{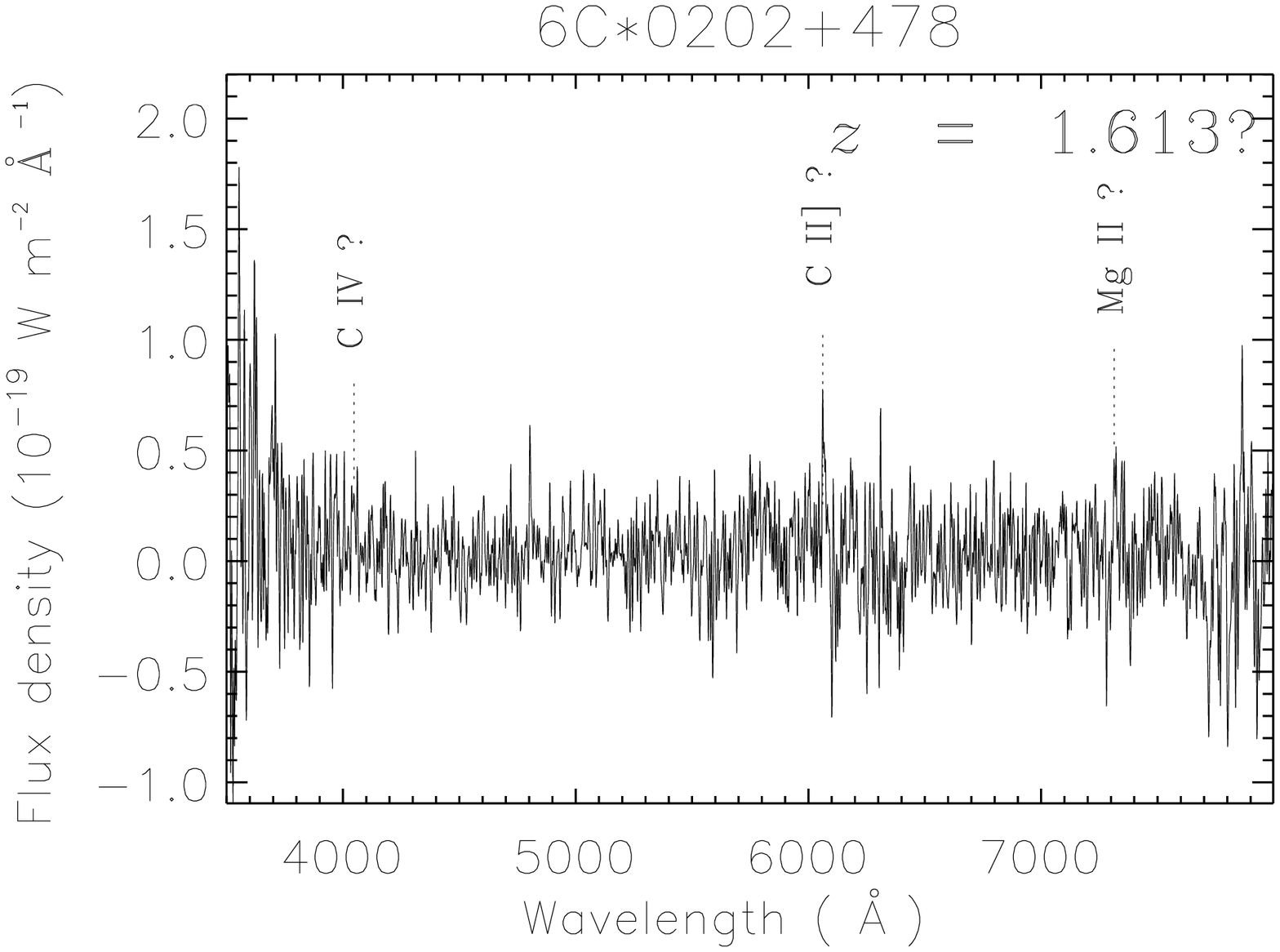}}} 
{\hbox to \textwidth{ \null\null \epsfxsize=0.48\textwidth
\epsfbox{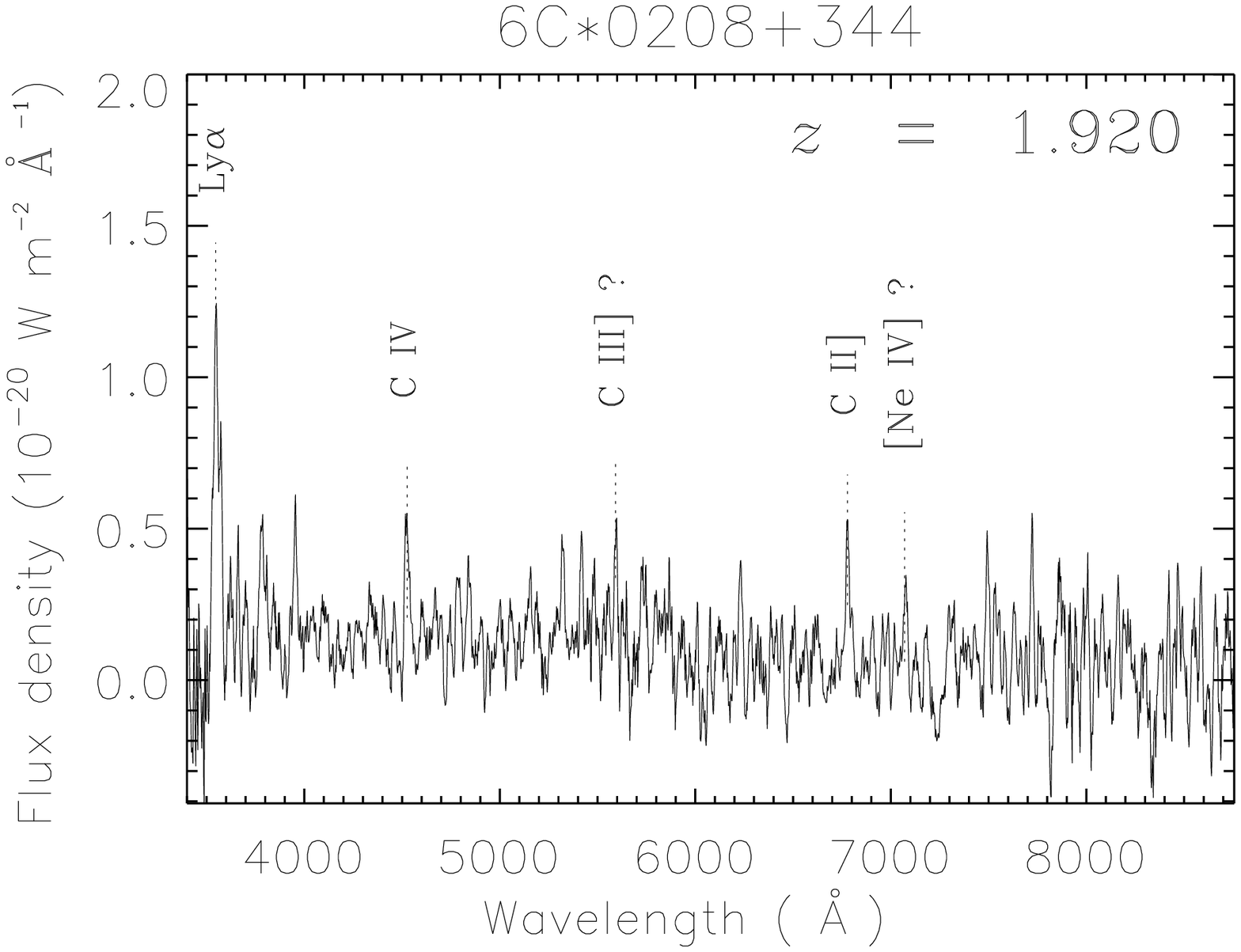}
\epsfxsize=0.48\textwidth
\epsfbox{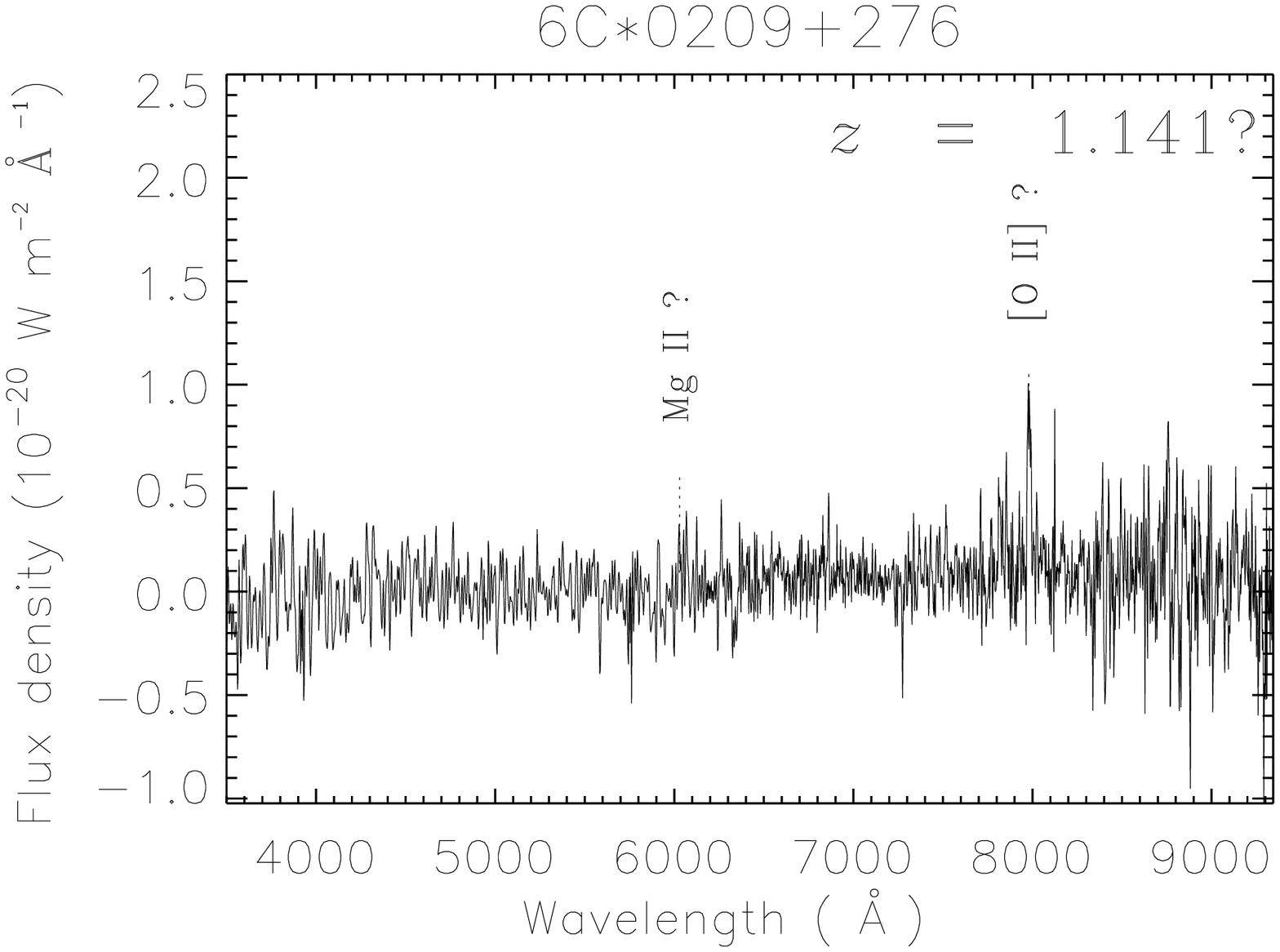}  }}
{\caption{\label{fig:spectra} Spectra of the 6C* sources obtained with
the various telescopes and instruments mentioned in
Sec.~\ref{sec:spectra}, with the emission line
identifications. Possible emission lines are marked with a `?'. The
emission lines which do not appear as significant detections in the
1-D spectra, and which are not marked with `?', are significant in the
2-D spectra.
The
gap in the spectrum of 6C*0133+486 is a result of combining blue- and
red-arm spectra where the short wavelength end of the red spectrum
suffered technical problems. The flux density scale for the
spectrum of 6C*0139+344 is measured in W m$^{-2}$ \AA$^{-1}$\, with
arbitrary normalisation, due to the lack of a spectrophotometric standard.}}
\end{figure*}

\begin{figure}[!ht] 
{\hbox to 0.48\textwidth{ \null\null \epsfxsize=0.45\textwidth
\epsfbox{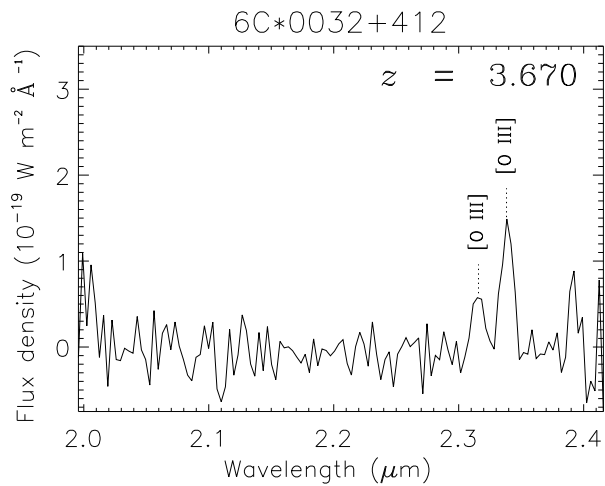} }}
{\caption{\label{fig:6cs0032Kspec} The UKIRT $K-$band spectrum of
6C\,0032+412. This spectrum was obtained with the CGS4 spectrometer on UKIRT
in Dec 1993 with the 75 lines mm$^{-1}$ grating, and employed the
standard nodding pattern (e.g. Eales \& Rawlings 1993) with a nod of
34 arcsec (11 detector rows); the 20s exposures were arranged in sets
of four shifted by 0, 0.5, 1, and 1.5 pixels in wavelength -- this
provided Nyquist wavelength sampling and ensured that a given
wavelength was sampled by two pixels; the total exposure time for the
final 126-pixel spectrum is $\approx 1800$\,seconds, taken with a slit
width of 3.1\,arcsec.
These data were extracted from a single detector row (3.1 arcsec)
and have not been smoothed.  The redshift of the [OIII]5007 line is $z
= 3.670 \pm 0.005$, other parameters of the [OIII]5007 line are:
measured FWHM $= 1350 ~ \rm km ~ s^{-1}$ (deconvolved value $\approx
1000 ~ \rm km ~ s^{-1}$); and line flux within $\pm 20$ per cent of
$1.6 \times 10^{-18} ~ \rm W ~ m^{-2}$.}}
\end{figure}

\section{6C* - what are we missing from a filtered sample?}\label{sec:6C*missing}

With the 6C* sample we are now able to compare the properties of this
filtered sample with the recently completed 6CE sample (Rawlings et
al. 2001) and 7CRS (Blundell et al. in prep; Willott et al. in
prep). The number of sources in the 6C catalogue within the 6C*
flux-density limits and over the same sky-area before filtering was
279, for which we would expect the median redshift to be the similar
to that of 6CE and 7CRS, i.e. $\sim 1.1$ (Rawlings et al. 2001;
Willott et al. in prep.).  The median redshift of the 29 6C* sources
is $z \approx 1.9$. As 6CE and 7CRS span flux-densities either side of
the flux-density range of 6C*, we can conclude that the filtering
criteria were effective in biasing the sample to objects at
high redshift. However, the main problem in dealing with and
interpreting a filtered sample such as 6C* is that it is difficult to
assess the properties of the population which are omitted from the
sample. In this section we use the most recently derived radio
luminosity function for low-frequency selected radio samples of
Willott et al. (2001a) to estimate the fraction of sources excluded
from 6C* over the redshift space probed. This is different to the
analysis of Jarvis et al. (2001a) which estimated the fraction of only
the most luminous population of radio sources which were
excluded from 6C*. Here we discuss the effects of the selection
criteria across the whole radio luminosity range.

\begin{figure*}[!ht] 
\begin{center}
{\hbox to \textwidth{ \epsfxsize=0.48\textwidth
\epsfbox{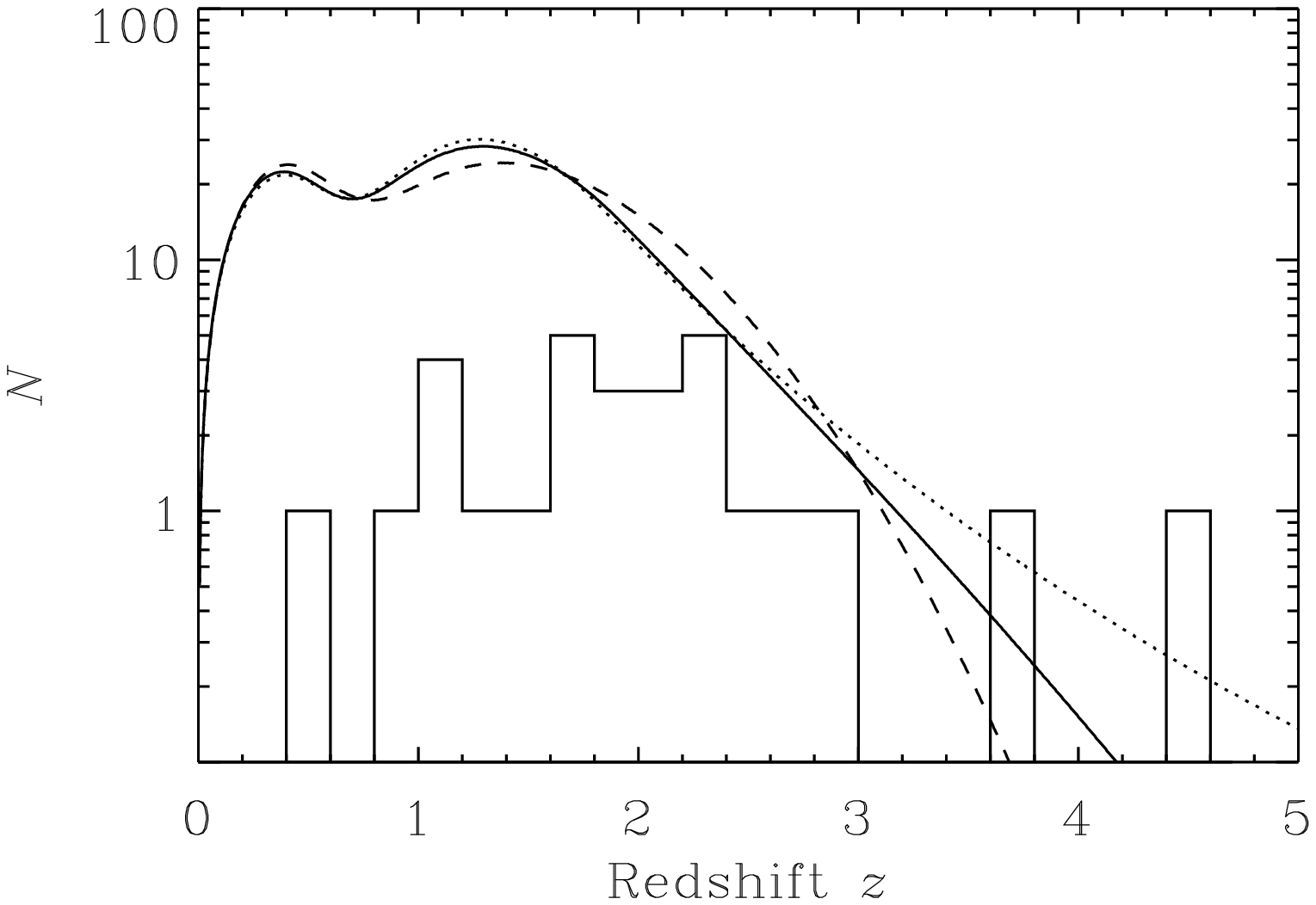}
\epsfxsize=0.48\textwidth
\epsfbox{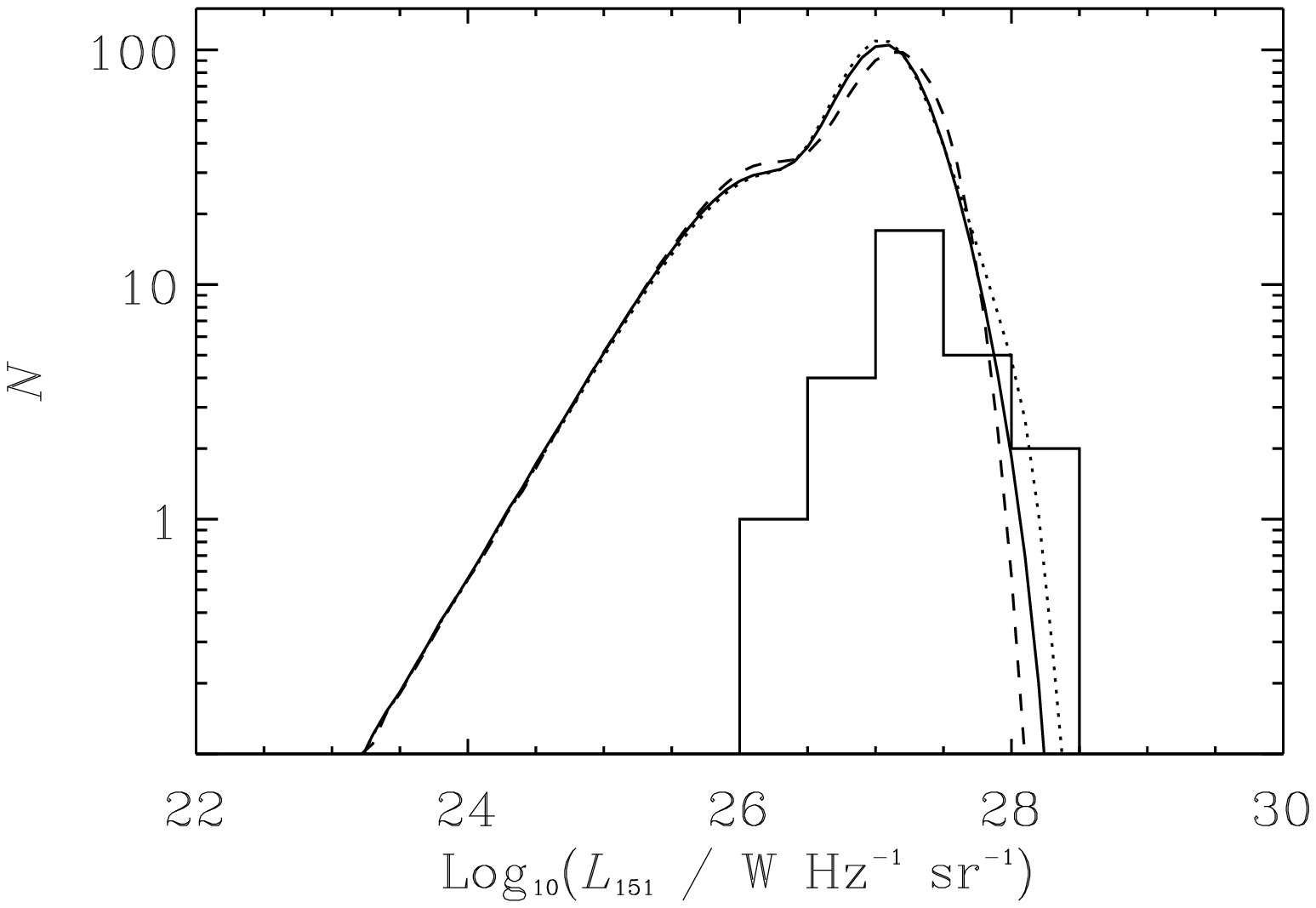}}}
{\caption{\label{fig:zdist6C*} (left) Histogram of the redshift
distribution $N$(z) of the 6C* sample; the bin width is $\Delta z=0.2$. (right) Histogram of the 151\,MHz rest-frame
luminosity distribution $N$($\log_{10}L_{151}$) of the 6C* sample; the
bin width is $\Delta (\log_{10}L_{151}) =0.5$.
Also shown are the source count predictions for the
flux-density range of 6C* ($0.96{\rm \,Jy} \leq S_{151} \leq 2.00$\,Jy) from
the three best-fit RLF models from Willott et al. (2001a) with no
filtering applied. The dashed line
corresponds to their model A (a cut-off at high redshift similar to
the decline at low-redshift); the dotted line to model B (constant
co-moving space density above a peak redshift); and the
solid line to model C (a Gaussian decline at high redshift decoupled
from the low-redshift evolution). }}
\end{center}
\end{figure*}

Fig.~\ref{fig:zdist6C*} shows the redshift distribution of the 6C*
sample on a logarithmic axis. One can immediately see that this
distribution is significantly different from the predicted
distribution from the three models for the RLF of Willott et
al. (2001a), with the distribution skewed towards objects at
$z\,\gtsim\,2.0$, as expected. Indeed, employing the selection
criteria has reduced the number of $z < 1.5$ sources by $\approx
90$\%, whereas in the redshift range $1.5 \ltsim z \ltsim 3.0$ the
fraction excluded varies from $\sim 70$\% to $\sim 30$\%. However,
this sample was initially designed to find objects at $z > 4$, and
with the discovery of the $z =4.41$ radio galaxy 6C*0140+326, (and
6C*0032+412 at $z = 3.658$) the sample has proven a success. The
expected number of sources at these high redshifts in the 6C* sky area
in the absence of the selection criteria is only $\sim 1 - 2$, even if
there is no redshift cut-off above $z \sim 2.5$ in the steep-spectrum
population (c.f. Dunlop \& Peacock 1990; Willott et al. 2001a). Thus,
the discovery of these two high-redshift objects from a sample of just
29 radio sources, relates both to a very high efficiency for the
detection of $z\, \gtsim\, 3.5$ objects in the selection techniques
used, and to the absence of an abrupt cut-off in the intrinsic
co-moving space density (Jarvis et al. 2001a). This has already led to
the reassessment of the evolution in the co-moving space density of
the most luminous steep-spectrum radio sources (Jarvis et al. 2001a)
resulting in a best-fit evolutionary scenario in which the space
density of these sources remains roughly constant out to at least the
limit of the 6C* data at $z = 4.41$.

The effects of the selection criteria on the luminosity distribution
are also illustrated in Fig.~\ref{fig:zdist6C*}. The skewness of the
distribution to high luminosity illustrates the effect of the biases
inherent in the sample. Our purpose, to find radio galaxies at
high redshift, necessarily skews the luminosity distribution to the bright
end if the filtering criteria are effective.
Thus the skewness is accentuated in the filtered sample compared to
complete samples.

We have shown that the filtering criteria of 6C* are efficient in
finding objects at high redshift, however it is extremely difficult
to decouple the two mechanisms which are causing this bias. The
steep-spectral index selection technique is widely used in targeting
high-redshift objects, but the effectiveness of this has yet to be
properly quantified (De Breuck et al. 2000a). The angular size constraint has been less widely used and
until the physics of radio sources and the cosmic evolution of radio
source environments are fully understood, the effect this may have on
radio samples will probably be similarly hard to quantify.  We do
know, however, that these two selection criteria work against finding
low-redshift, low-luminosity sources in flux-density-limited samples,
and thus, for searches for high-redshift sources they are proven to be
extremely effective. 

\section{Linking the radio and optical properties of luminous
high-redshift radio sources}\label{sec:linking}

There has been extensive work in recent years on linking the optical
properties of radio sources with their radio properties. A recent
review can be found in Willott et al. (1999), who used a combined
sample comprising 3CRR and 7CRS radio sources to decouple trends with
redshift from trends in luminosity. They found a strong correlation
between the narrow emission-line luminosity and the rest-frame 151-MHz
radio luminosity, but only weak correlations between the luminosity of
the narrow emission lines and projected linear size $D$ or rest-frame
spectral index. Best, R\"ottgering \& Longair (2000) from a sample of
3CR radio galaxies at $z \sim 1$ found evidence that the radio sources
with small linear sizes ($D \sim 100$\,kpc) have lower ionisation
states, higher [OII] emission-line fluxes and broader [OII] line
widths than the larger sources. Using a sample of 18 objects selected
to have good signal-to-noise spectra and UV/optical emission with
large spatial extents, van Ojik et al. (1997) found a correlation
between the radio source size and the extent of the Ly$\alpha$
emission-line gas in high-redshift ($z > 2.1$) radio galaxies with $10
\ltsim D \ltsim 100$\,kpc. They also found a high fraction ($\sim
90$\%) of \HI absorbed Ly$\alpha$ profiles in those with $D\,
\ltsim\,50$\,kpc.

With the 6C* sample in conjunction with the 3CRR and 6CE samples and
the 7CRS, we are now able to extend these investigations to higher
redshifts. We have chosen to confine our attention to radio galaxies
in the complete samples. According to simple unified schemes
(e.g. Willott et al. 2000b) we are therefore focusing on the sub-set
of sources with jet-axes within $\sim 53^{\circ}$ of the
plane-of-the-sky which has the obvious advantage that $D$ should be
within a factor $0.6 - 1.0$ of the true linear size\footnote{Note
however, from Fig.~\ref{fig:pzplane} that the quasar fraction at $z >
1.75$ in our combined dataset is somewhat lower than the 0.4 seen at
similar radio luminosities at lower redshift (Willott et
al. 2000b). Considering only flux-density-limited samples Rawlings et al. (2001)
have recently argued that this may be a statistical fluke, although an
epoch dependent increase in reddening (e.g. Willott, Rawlings \&
Jarvis 2000a) may play a r\^ole. We note for 6C* that the filtering
criteria both bias in favour of quasars (objects closer to the
line-of-sight will have a smaller angular size for a given true linear
size) and against them (objects closer to the line-of-sight may have
significant Doppler-boosted components which flatten their spectral
index), so it is difficult to interpret the quasar fraction of this
sample (see Jarvis 2000).}.

Correlations between the logarithms of various radio and optical
properties are presented: `best-fit' lines are calculated using an
algorithm which minimises the sum of the square of the perpendicular
distances from the data points to a line with adjustable slope and
intercept (see discussion in Jaynes 1991).

\subsection{$L_{{\rm Ly}\alpha}$ versus radio luminosity}\label{sec:lumlum}

\begin{figure}[!ht]
\begin{center} 
{\hbox to 0.48\textwidth{ \epsfxsize=0.45\textwidth
\epsfbox{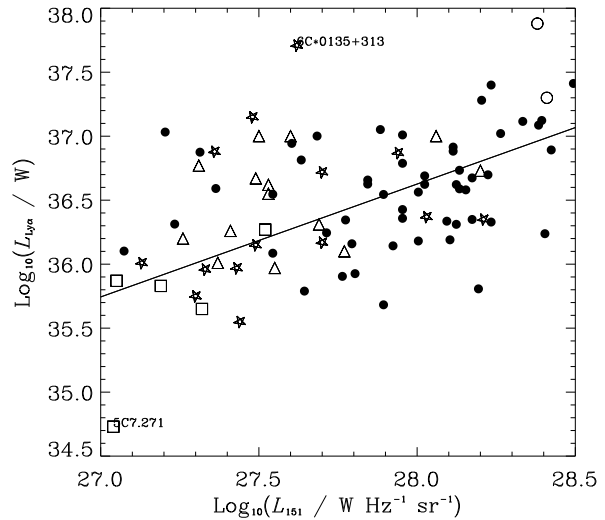} }}
{\caption{\label{fig:lumvslum} Ly$\alpha$ emission line luminosity,
$L_{{\rm Ly}\alpha}$, against rest-frame 151\,MHz radio luminosity,
$L_{151}$ for radio galaxies at $z > 1.75$ from the 3CRR (circles),
7CRS (squares), 6CE (triangles) and 6C* (stars) radio samples. The
filled circles are the high-redshift steep-spectrum sources from de
Breuck et al. (2000b). The solid line shows the best-fit to all of the
points shown.}}
\end{center}
\end{figure}

In Fig.~\ref{fig:lumvslum} we plot the Ly$\alpha$ emission-line
luminosity $L_{\rm Ly\alpha}$ against the 151-MHz radio luminosity
$L_{151}$ for a sample of 35 radio galaxies with $z > 1.75$, derived
from the 3CRR, 6CE and 7CRS samples along with the 6C*
sample.\footnote{As illustrated in Fig.~\ref{fig:pzplane} there are
38 radio galaxies in our dataset at $z > 1.75$.  In this analysis we
include only those with a measured Ly$\alpha$ flux.  This excludes the
6C* radio galaxy 6C*0106+397 (for which we do not have spectra around
the Ly$\alpha$ line) and the 7CRS radio galaxies 5C6.242 (7C0218+3103)
and 5C7.208 (7C0820+2506). The two 7CRS exclusions are because these
galaxies have no emission lines in their optical spectra and hence only photometric redshifts (Willott, Rawlings \&
Blundell 2001b): their
exclusion does not affect any of the results as even if they truly lie
at $z > 1.75$ their line strengths will be weak, placing them at the
bottom left of the plots in Fig.~\ref{fig:lumvslum} in line with the correlation discussed in
Sec.~\ref{sec:lumlum}.}  

It is apparent that a strong correlation
exists between the $L_{\rm Ly\alpha}$ and $L_{151}$ for the radio
galaxies present in this subsample. The strength of this correlation
is quantified using the Spearman partial rank correlation coefficient
(e.g. Macklin 1982) and the results are presented in
Table~\ref{tab:ranklumlum}; note that the use of a combined 3CRR, 6CE,
7CRS and 6C* dataset severely weakens the strong correlations between
$L_{151}$ and $z$ in a single flux-density-limited sample (as
illustrated in Fig.~\ref{fig:pzplane}).

The best-fit line has a slope of $1.49 \pm 0.25$ which is steeper than the
power-law index found by Willott et al. (1999) for the [OII] luminosity
versus $L_{151}$ for sources from the 3CRR and 7CRS [which using the
same algorithm is $1.00 \pm 0.04$, for the analysis which includes 6CE (see Willott 2000)].

Any `least-squares' algorithm is sensitive to anomalous outliers, so
we have investigated this by excluding 5C7.271 (7C0826+2504), which
has $\log_{10} L_{{\rm Ly}\alpha} = 34.73$ and is thus more than 0.5
dex lower than any of the other sources, and is below the radio luminosity at which the quasar fraction of radio sources is no longer
independent of luminosity and/or redshift (Willott et al. 2000b). We
also exclude 6C*0135+313 which appears to be anomalously bright in
Ly$\alpha$. Excluding these sources we find that the correlation
becomes $1.33 \pm 0.21$. Thus, although we are spanning only 1.5 dex
in radio luminosity we find a tight correlation between $L_{{\rm
Ly}\alpha}$ and $L_{151}$ (with a Gaussian $1\sigma$ spread of 0.12
perpendicular to the best-fit line) which is marginally steeper than that found
by Willott et al. (1999) for the correlation between $L_{\rm [OII]}$
and $L_{151}$.

\begin{table}
\begin{tabular}{l|r|r|r|r|}
\hline\hline
\mc{1}{c|}{Correlated} & \mc{1}{c|}{$r_{\rm AB}$} &
\mc{1}{c|}{$P_{r}$} & \mc{1}{c|}{$r_{\rm AB, C}$} & \mc{1}{c|}{$\sigma$} \\
\mc{1}{c|}{variables: A, B} & \mc{1}{c|}{} &
\mc{1}{c|}{} & \mc{1}{c|}{} & \mc{1}{c|}{} \\
\hline\hline
$L_{151}$, $L_{{\rm Ly}\alpha}$ & 0.619 & $<0.1$\% & 0.679 & 4.603 \\
$L_{151}$, $z$ & 0.242 & 16\% & 0.421 & 2.497 \\ 
$L_{{\rm Ly}\alpha}$, $z$ & $-0.138$ & 43\% & $-0.378$ & 2.212 \\ 
\hline\hline
\end{tabular}
{\caption{\label{tab:ranklumlum} Spearman rank correlation analysis of
the correlations present between $L_{151}$, $L_{{\rm Ly}\alpha}$ and
$z$ for the 35 objects in our subsample at $z > 1.75$. $r_{\rm AB}$ is
the Spearman rank coefficient between the parameters A and B, $P_{\rm
r}$ is the two-tailed probability that a given value of $r$ is
predicted by chance, given the null hypothesis that A and B are
unrelated and $\sigma$ is equivalent to the deviation from a unit
variance normal distribution if there is no correlation
present. $r_{\rm AB, C}$ is the partial rank coefficient (e.g. Macklin
1982), this assesses the statistical significance of correlations
between the variables A and B in the presence of the third variable,
C. The significance of the partial rank correlation is equivalent to
the deviation from a unit variance normal distribution if there is no
correlation present. }}
\end{table}

If we also include the high-redshift sample of De Breuck et
al. (2000b), which includes high-redshift sources selected on account
of their steep radio spectral indices (which is similar, but not
identical to the spectral index criterion of 6C*), then we find that
the slope of this correlation flattens to $0.87 \pm 0.10$ \footnote{The radio
luminosities calculated by De Breuck et al. were for a rest-frame
frequency of 325 MHz so we have estimated $L_{151}$ assuming a
spectral index of $\alpha = 1.0$. This flatter spectral index should
thus compensate for any intrinsic spectral curvature in these objects
between the relatively high frequency $\alpha$ measured and that at
rest-frame 151\,MHz.}. At $z > 1.75$ Ly$\alpha$ lies
within the observable optical window, and is typically detected at many
times the detection limit, so there should not be a large number of $z >
1.75$ objects missing in a systematic way from the De Breuck et
al. compilation. Because of the restricted size of our 3CRR/6CE/7CRS/6C*
dataset, and the consequences of differing filtering criteria in larger compilations, it is
difficult at this stage to conclude much beyond a rough
proportionality between Ly$\alpha$ and $L_{151}$ for luminous radio
sources.


The results presented here further strengthen the evidence for a link
 close to a proportionality between $L_{151}$ and narrow-line
 luminosity (e.g. Willott et al. 1999), and suggest its roughly
 redshift-independent form is maintained to $z \sim 4.5$. The physical
 origin of this link is somewhat controversial: following Rawlings \&
 Saunders (1991), Willott et al. (1999) favour a close tie between the
 luminosity of the quasar nucleus and the bulk power carried by the
 radio jets; however, other authors (e.g. Bicknell, Dopita \& O'Dea
 1997; Koekemoer \& Bicknell 1998) favour models in which the
 line-emitting gas is photoionised by shocks driven by the radio
 emitting components.

\subsection{Radio source size and the emission line luminosities}\label{sec:sizevslinelum}

Best et al. (2000) find that the theoretical predictions of shock
ionisation models agree with the emission line ratios of
CIII]1909/CII]2326 and [NeIII]3869/[NeV]3426 for smaller sources,
whereas the emission-line ratios in larger sources are consistent with
the ionising photons being emitted directly from active nuclei.

\begin{figure*}[!ht] 
\begin{center}
{\hbox to 0.86\textwidth{ \epsfxsize=0.85\textwidth
\epsfbox{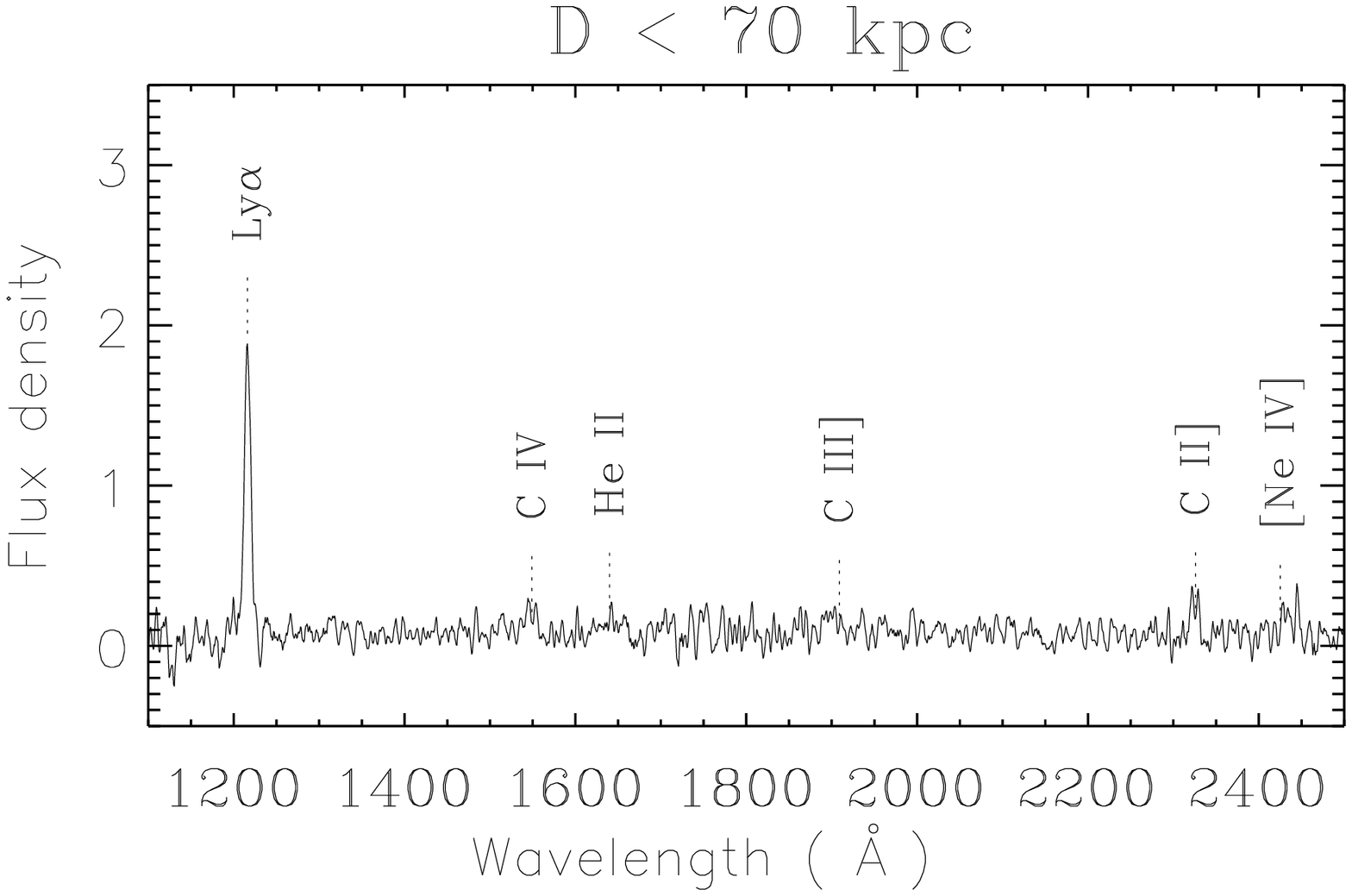}}}
{\hbox to 0.86\textwidth{ \epsfxsize=0.85\textwidth
\epsfbox{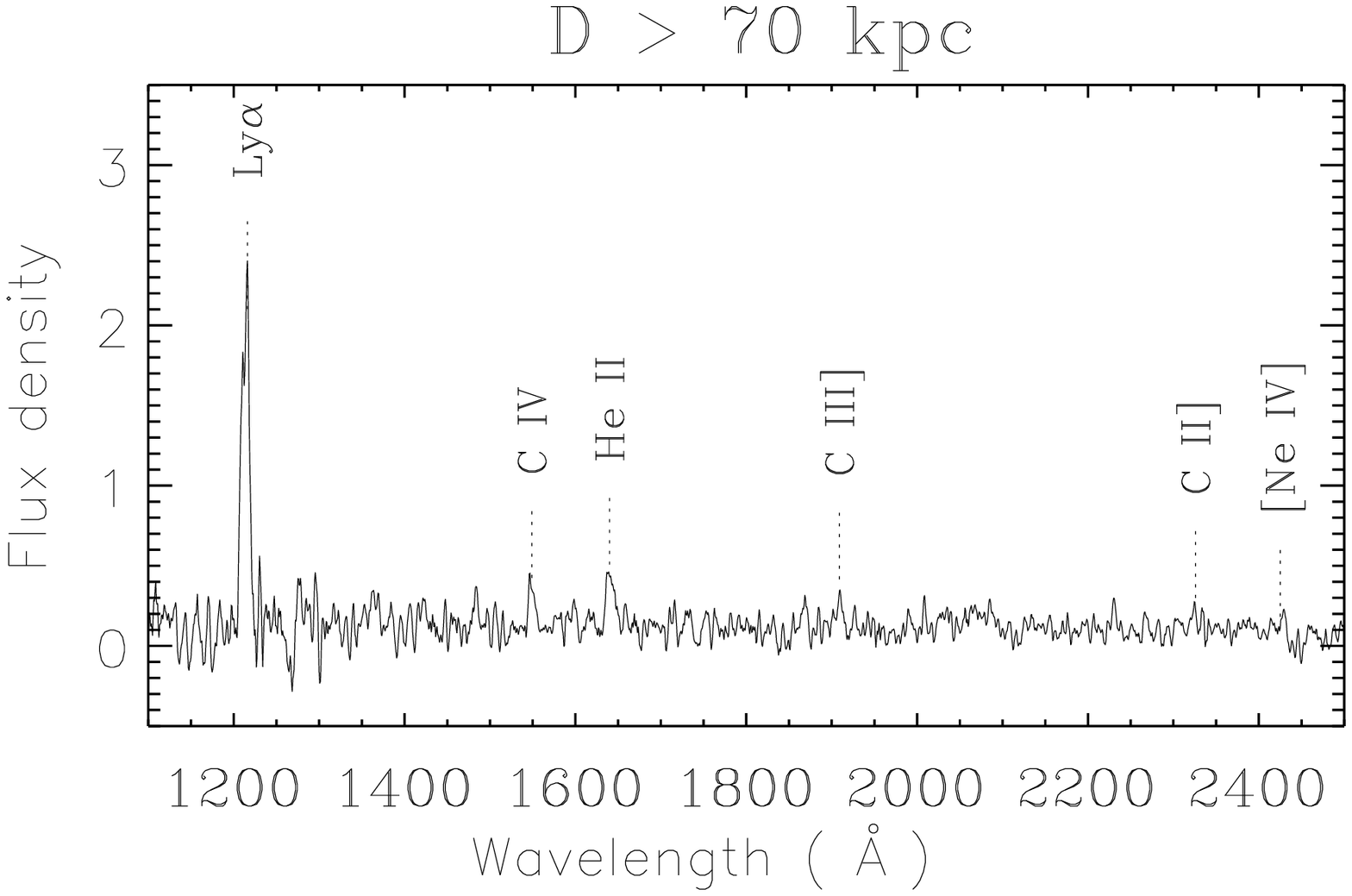}}}
{\caption{\label{fig:composites} Composite spectra resulting from 14
sources (8 from 6C*, 2 from 6CE and 4 from 7CRS) with projected linear
sizes $D < 70$\,kpc (top) and 13 (4 from 6C*, 8 from 6CE and 1 from
7CRS) sources with $D > 70$\,kpc (bottom). Only the sources with
Ly$\alpha$ in the spectrum were used as the integrated 
Ly$\alpha$ flux was used to normalise the different spectra, the method of
combination is that described in Rawlings et al. (2001) and the
resultant spectrum has been smoothed with a 3-bin boxcar filter. The
flux-density is measured in units of W m$^{-2}$~\AA$^{-1}$ with an
arbitrary normalisation.}}
\end{center}
\end{figure*}

Using the 7CRS, 6CE and 6C* samples we investigate this result for
sources at $z > 1.75$. The angular sizes and projected linear sizes of
these sources are summarised in Blundell et al. (1998) and Blundell et al. (in
prep). At these redshifts the [NeIII] 3869 and [NeV] 3426 emission
lines are redshifted out of the optical window, thus we are forced to
use only the CIII] 1909 and CII] 2326 emission lines. Although the
majority of spectra for these sources span the waveband which would
capture both the CIII] and CII] lines, the depth and signal-to-noise
ratio of the observations are insufficient to calculate ratios for
individual sources. Therefore, we chose to split the data into two
broad bins in projected linear size. The median projected linear size
of the 28 sources\footnote{Of the 38 sources in our dataset this
sub-sample excludes objects lacking spectra available in an electronic
format; namely both 3CRR sources (3C239 and 3C294) and 6CE0902+3419,
6CE1141+3525, 6CE1204+3708 and 6CE1232+3942 from the 6CE
sample. 6C*0041+469 and 6C*0106+397 from 6C* are also excluded;
6C*0106+397 does not cover the wavelength of the redshifted Ly$\alpha$
and 6C*0041+469 lacks sufficient signal-to-noise. Qualitatively, the
spectra of the excluded 3CRR, 6CE sources and 6C*0106+397 look similar to the composites, so no bias is
expected. The few weak-line objects (6C*0041+469, 5C6.242, 5C7.208)
are not concentrated in one bin in $D$ and so should not seriously skew
the result.} used in this analysis is $\approx 70$\,kpc, thus we use
this as the division between two bins. We then combine the spectra in
each bin to form a composite radio galaxy spectrum for sources in the
two bins, following the same method of combination as Rawlings et
al. (2001). These composites are shown in
Fig.~\ref{fig:composites}. It is apparent that there is a difference
between the two, with the smaller sources having a CIII] 1909/CII]
2326 ratio of $\sim 0.5$ and the larger sources having CIII] 1909/CII]
2326 $\sim 1.3$. It is also apparent that in the composite of the
larger sources, the CIV and particularly the HeII emission lines are
much stronger than in the small source composite.  This is in
quantitative agreement with the findings of the Best et al. study of
3C sources at $z \sim 1$ (which as shown in Fig.~\ref{fig:pzplane},
are of comparable $L_{151}$). We also split the data into two radio
luminosity bins to investigate any dependence of the line ratios on
the radio luminosity of the source. We found a factor of $\sim 1.7$ higher
CII]$\lambda 2326$\,\AA\, flux in sources with $\log_{10}L_{151} >
27.5$ than for sources at $\log_{10}L_{151} < 27.5$. However, there
was no discernible difference between the CIV$\lambda 1549$\,\AA\, and
HeII$\lambda 1640$\,\AA\, line fluxes in the two bins. Thus, we
conclude that there is no {\it strong} correlation between the low-frequency
radio luminosity and the ionisation state of the emission-line
region. This is in agreement with the results of Tadhunter et
al. (1998) for a low-redshift sample ($z < 0.7$) of 2 Jy radio
sources, where they also found no evidence for a correlation between
ionisation state measured by the ratio of [OII]$\lambda 3727$ /
[OIII]$\lambda 5007$ and radio luminosity. If any such correlation
exists it may well be best seen amongst the sources with the lowest
photoionising power $Q_{\rm phot}$ in complete samples (Saunders et
al. 1989); in Fig.~\ref{fig:ionise}, the luminosity of all lines even
those of low luminosity like [OII] scale with incident photoionising luminosity. However, Stern et al. (1999) have found a very tentative
correlation between CIV / CIII] ratio and radio luminosity within a
sample of 30 radio galaxies, mainly from the MIT-Green Bank survey
(Bennett et al. 1986; Lawrence et al. 1986) and the ultra-steep sample
of R\"ottgering et al. (1994). However, this correlation is extremely
weak, being significant at the $\approx 2\sigma$ level. This
correlation was also not seen in the work of De Breuck et al. (2000b)
and should be viewed with caution.

Returning to the sub-sample of 35 objects from our dataset considered
in Sec.~\ref{sec:lumlum} we also find a highly tentative correlation
between the residual Ly$\alpha$ emission-line luminosity (after
subtraction of the $L_{{\rm Ly}\alpha}$ calculated from the radio
luminosity of each source using the best-fit derived in
Sec.~\ref{sec:lumlum}) and the projected linear size of the radio
sources in our dataset.  The Spearman rank coefficient for this
correlation is $r_{\rm s} \approx 0.23$, with the two-tailed
probability that this arises by chance, given the null hypothesis of
no correlation of $P_{\rm r} \approx 17$\% for the best-fit with all
sources included. If 5C7.271 and 6C*0135+313 are excluded then the
correlation becomes slightly more significant with $r_{\rm s} \approx
0.27$ with $P_{\rm r} \approx 13$\%.

\begin{figure}[!ht]
{\hbox to 0.5\textwidth{\epsfxsize=0.48\textwidth
\epsfbox{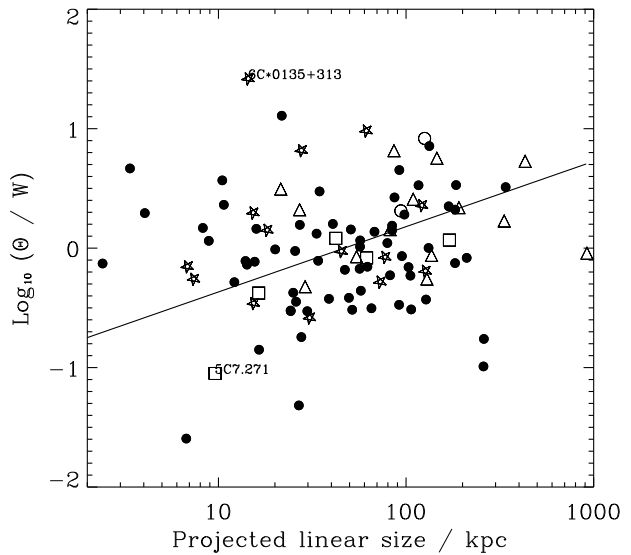}}}
{\caption{\label{fig:deBresid} The residual emission-line luminosity (after
subtraction of the best-fit $\log_{10}L_{{\rm Ly}\alpha} - \log_{10}L_{151}$
slope of Fig.~\ref{fig:lumvslum}) against projected linear size of the radio galaxies from
our 3CRR (open circles), 6CE (triangles), 7CRS (squares) and 6C* (stars) datasets combined with that of De Breuck et
al. (2000b) (filled circles). The solid
line shows the best-fit to the data.}}
\end{figure}

\begin{figure}[!ht]
{\hbox to 0.5\textwidth{\epsfxsize=0.48\textwidth
\epsfbox{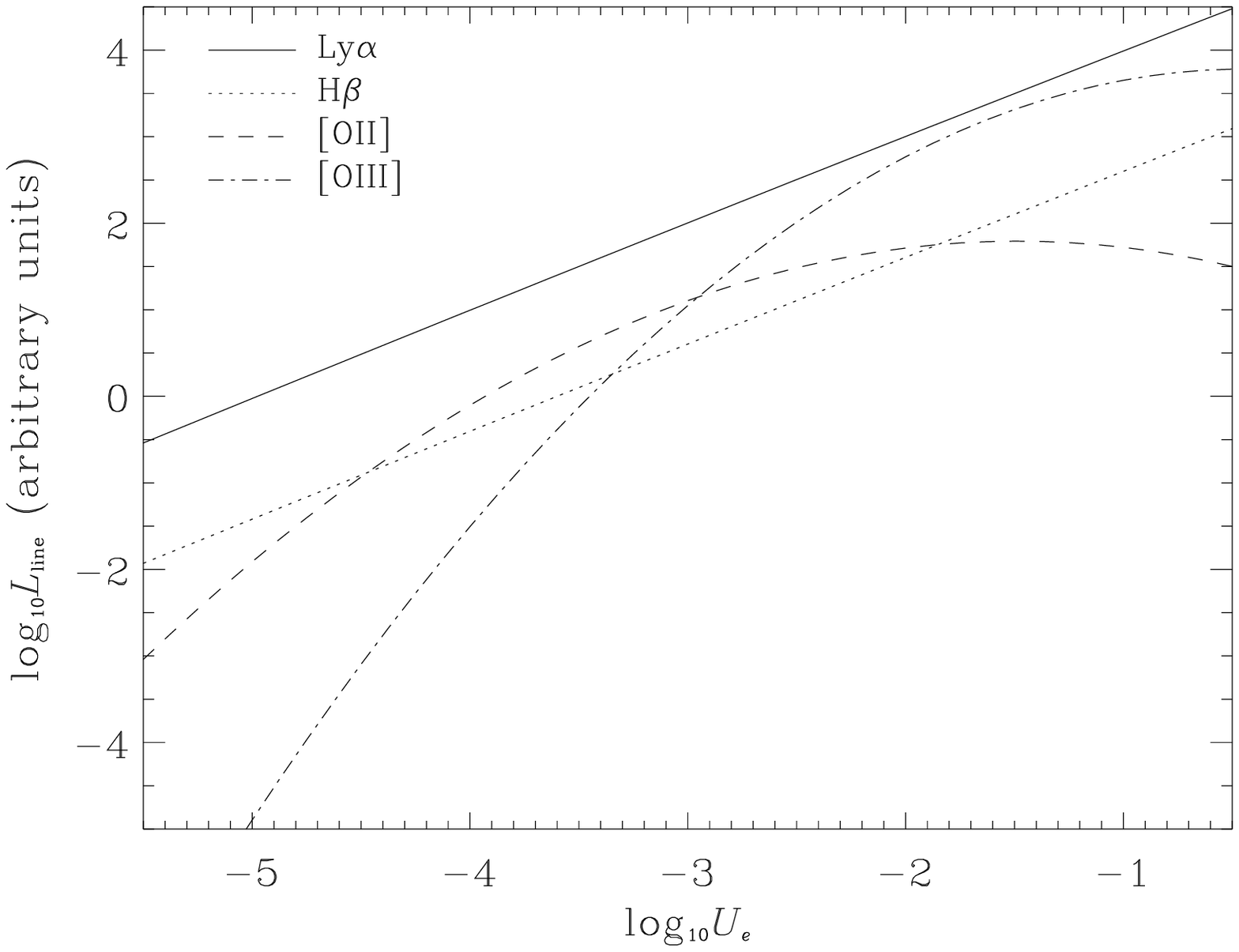}}}
{\caption{\label{fig:ionise} Luminosity of the Ly$\alpha$ (solid
line), H$\beta$ (dotted line), [OII] (dashed line) and [OIII]
(dotted-dashed line) emission lines as a function of effective
ionisation parameter, $U_{e}$ for an ionisation-bound plane-parallel
slab with a hydrogen number density of $10^{9}$\,m$^{-3}$; $U_{e}$ is
given as the ratio of the photon number density to particle number density at
the incident face of the cloud, i.e. $U_{e} = Q_{\rm phot}/(4\pi
r^{2} c n)$, where $r$ is the distance to the cloud from the ionising
source and $n$ is the particle density. This was modelled with {\sc
CLOUDY} Version 94.00 (van Hoof, Martin \& Ferland 1999; {\bf
http://www.pa.uky.edu/$\sim$gary/cloudy/}) using the AGN continuum
spectrum described by Mathews \& Ferland (1987), for Ly$\alpha$,
H$\beta$, [OII] and [OIII].
Both Ly$\alpha$ and H$\beta$ respond linearly to increases in the
effective ionisation parameter $U_{e}$, so that if quasars of all luminosities
had narrow-line regions with similar radial density profiles then one
might expect a proportionality between $L_{\rm Ly\alpha}$ (or $L_{\rm
H\beta}$) and quasar luminosity over the full range in the
photoionising luminosity $Q_{\rm phot}$, although (as will be discussed in
Sec.~\ref{sec:sizesize}) it should be noted that the {\sc
CLOUDY}-based modelling takes no account of the resonant \HI
absorption of Ly$\alpha$.}}
\end{figure}

Although normally we would be highly suspicious of what is clearly a
marginal correlation, we will explore a physical model in
Sec.~\ref{sec:age} involving the quantity $\Theta \equiv L_{\rm
Ly\alpha}/L_{151}^{6/7}$ (the form of the
residual after taking out the correlation of Fig.~\ref{fig:lumvslum}
as detailed in Sec.~\ref{sec:lumlum}). We plot $\log_{10}\Theta$ against $\log_{10}D$ in
Fig.~\ref{fig:deBresid}. With our dataset combined with the sources
from De Breuck et al. (2000b) the correlation between $\Theta$ and $D$
is significant at the 94\% level ($r_{\rm s} \approx 0.19$), with a
best fit slope of $0.55 \pm 0.10$. 
Although evidence for a positive correlation in
Fig.~\ref{fig:deBresid} is slight, it is interesting in the light of
the weak but significant (at the 99\% level) {\it negative}
correlation seen in the plot of residual [OII] luminosity versus $D$
in Willott et al. (1999).

One explanation of this difference is that the smaller sources have
stronger associated \HI absorption (van Ojik et al. 1997) and hence,
systematically suppressed Ly$\alpha$ than in the larger
sources. Indeed, van Ojik et al. (1997) found that 90\% of small
($<50$\,kpc) radio galaxies have strong associated absorption, whereas
only 25\% of the larger ($>50$\,kpc) radio galaxies exhibit such
strong absorption. Our spectroscopic dataset has insufficient
resolution and sensitivity to check this result with the combined
3CRR, 6CE, 7CRS and 6C* samples. However, if we assume that
approximately 50\% of the Ly$\alpha$ flux is absorbed for sources with
$D \leq 50$\,kpc (van Ojik et al. 1997), then the correlation of
Fig.~\ref{fig:deBresid} could
be weakened or extinguished. To test this, we doubled the Ly$\alpha$
fluxes of all sources with $D\,\ltsim\,50$\,kpc and re-calculated the
correlations finding the significance of the combined sample drops to
$\sim 60$\% with a best-fit slope of $\approx -0.2$. We conclude that \HI
absorption is a plausible source of the weak positive correlation in
Fig.~\ref{fig:deBresid}, but it does not explain why residual [OII]
shows an anti-correlation with $D$. 

Why then do Willott et al. (1999) find an anti-correlation 
between the residual [OII] luminosity (after taking out the
strong $L_{\rm [OII]} - L_{151}$ correlation) and $D$? One possible
reason is that we do not yet have enough data to strengthen the
correlation, both from a lack of high resolution spectroscopy around the
Ly$\alpha$ emission-line to constrain the \HI absorption, and from the
intrinsically small dataset of sources at $z > 1.75$. However, assuming
that this is not the case, and that our small sample is representative
of the population as a whole, we conclude this section by developing a physical argument to explain this.

Note first that our study finds tentative evidence that, at a given
$L_{151}$ at $z > 1.75$, the smallest ($D \ltsim 70$\,kpc) radio
sources have relatively low line luminosities of lines like Ly$\alpha$
(and HeII and CIV): the evidence for systematically low Ly$\alpha$
luminosity comes from the correlation of Fig.~\ref{fig:deBresid} and
for low CIV/HeII luminosities from Fig.~\ref{fig:composites} and
Table~\ref{tab:composites}. Since in the absence of \HI absorption
Ly$\alpha$ tracks $Q_{\rm phot}$ (see Fig.~\ref{fig:ionise}), this is
in qualitative agreement with simple radio
source models (e.g. Willott et al. 1999) in which the smallest sources
(of a given $L_{151}$) are young objects of systematically lower $Q$
than their older (larger $D$) counterparts, and in which the
photoionising power faithfully tracks $Q$ which is assumed to be
independent of radio source age (see Sec.~\ref{sec:age}).

Note second that the relative strength of low-ionisation lines like
CII] and [OII]\footnote{Simpson (1998) has emphasised that
the luminosity of this line is almost independent of $Q_{\rm phot}$ at
high ionisation levels (i.e. effective ionisation parameter, $U_{e}\,\gtsim\,10^{-3}$), making it an
unreliable estimate of $Q_{\rm phot}$, if (as is most likely for the
most luminous quasars) the bulk of the narrow-line emitting material
is at high $U_{e}$.}
in the small (low $D$) sources can then be viewed as a boosted
low-ionisation component. From our Fig.~\ref{fig:composites} and Table~\ref{tab:composites}, and Fig. 4 of Willott et al. (1999), and indeed Fig. 1
of Best et al. (2000), we can infer that the flux
from low-ionisation lines needs to be boosted by a factor $\sim 5$. It
seems plausible that this can be achieved by compressing and
potentially shredding emission-line clouds, as the source passes
through radii $\ltsim 50$\,kpc, decreasing the $U_{e}$ at each cloud
and potentially increasing the covering factor for photoionising
radiation from the nucleus (see e.g. Bremer, Fabian \& Crawford 1997).
As illustrated by Fig.~\ref{fig:ionise}, the ratio of [OII]/Ly$\alpha$
increases by about the required factor if $U_{e}$ drops from about
$10^{-2}$ to $10^{-3}$ due to compression of the gas clouds by a
factor $\sim 10$ within the photoionising cone of the quasar
(e.g. Lacy et al. 1998). Although the radio source induced
shocks could also yield in situ photoionisation (e.g. Dopita \&
Sutherland 1996; Best et al. 2000) these would in general boost all
narrow-line emission and do not seem to be required by the data. The negative correlation between $D$ and the FWHM of [OII]
(Best et al. 2000) show that the shocks do appear to be affecting the
kinematics of the line-emitting clouds. Kaiser, Schoenmakers \& R\"ottgering
(2000) argue that clouds like those we deem to be responsible for the boosted
low-ionisation component will disperse as the radio source propagates
to larger radii; the data of Best et al. on large-$D$ 3C
sources (namely line widths appropriate to gravitationally-induced motions and simple quasar-excited line ratios) are in accord with this picture.

The complicating effects on line ratios introduced by \HI absorption
and/or radio source induced changes in $U_{e}$ mean that it is
probably dangerous to use a single emission line to measure $Q_{\rm
phot}$. It is not surprising that the details of emission line/radio
correlations
appear to depend on the line studied.
As advocated by Rawlings \& Saunders (1991), an estimate of
total line-luminosity (ideally by summing the flux from all of the
narrow emission lines and accounting for absorption and reddening) is required to probe, for example, the true slope of the
emission line/radio correlation - but the requisite comprehensive
UV/optical/near-infrared dataset is not yet available for samples of
radio sources.

\begin{table}
\begin{center}
\begin{tabular}{c|c|c|c}
\hline\hline
\mc{1}{c|}{Emission} & \mc{1}{c|}{$\lambda$} & \mc{1}{c|}{$D <
70$\,kpc} & \mc{1}{c|}{$D > 70$\,kpc} \\
\mc{1}{c|}{line ID} & \mc{1}{c|}{(\AA)} & \mc{1}{c|}{(\% of
Ly$\alpha$)} & \mc{1}{c|}{(\% of Ly$\alpha$)} \\
\hline\hline
Ly$\alpha$ & 1216 & 100 & 100 \\
CIV & 1549 & 7 $\pm$ 1 & 14 $\pm$ 2 \\ 
HeII & 1640 & 4 $\pm$ 2 & 20 $\pm$ 2 \\
CIII] & 1909 & 5 $\pm$ 2 & 4 $\pm$ 1 \\
CII] & 2326 & 10 $\pm$ 2 & 3 $\pm$ 1 \\
NeIV & 2424 & 4 $\pm$ 2 & 2 $\pm$ 1 \\
\hline\hline
\end{tabular}
{\caption{\label{tab:composites} Emission line ratios to Ly$\alpha$ for
our composite radio galaxy spectra (Fig.~\ref{fig:composites}) for
radio galaxies with $D < 70$\,kpc and $D > 70$\,kpc. The uncertainties
on the ratios are $\approx 1\sigma$ errors and arise from several
factors, e.g. the uncertainties of the continuum
level.}}
\end{center}
\end{table}

\subsection{Radio source size and the extent of the Ly$\alpha$
emission-line region}\label{sec:sizesize}

\begin{figure}[!ht]
{\hbox to 0.5\textwidth{\epsfxsize=0.48\textwidth
\epsfbox{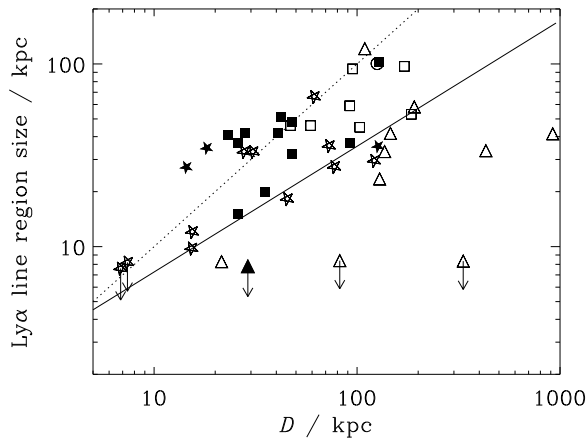}}}
{\caption{\label{fig:lyaextentvssize} The extent of the Ly$\alpha$
emission line region versus projected radio size. The 6C* objects are
denoted by the stars; the 6CE objects by triangles; and 3CRR by
circles. The squares are the points of van Ojik et al. (1997). The
objects with strong associated absorption are denoted by the filled
symbols.  The solid line is the best-fit to our data. The dotted line
is the line of equality between the Ly$\alpha$ extent and $D$. Note
that two 6C* sources (6C*0135+313 and 6C*0208+344) are significantly above this line implying that the
extent of the Ly$\alpha$ emission exceeds the extent of the radio
emission. It is interesting to note that these two sources both
appear to exhibit \HI absorption. The 6CE
source with associated absorption is 6CE0930+3855 (Rawlings et
al. 2001). Note that the upper limit symbols mean that the emission
line detections are unresolved in our data.}}
\end{figure}

We now investigate whether there is a link between the size of the
radio source and that of the Ly$\alpha$ emission-line region which
might be expected if the radio source itself is capable of
producing additional UV opaque material at radii $\ltsim 50$\,kpc
(e.g. via the shredding mechanism of Bremer et al. 1997). The extent
of the Ly$\alpha$ emission line was measured at the full-width
zero-intensity of a cross-cut through the emission line for all
sources for which we have 2-D spectra and deconvolving them from the
seeing [Sec.~\ref{sec:spectra} for the 6C* sources; Rawlings et
al. (2001) for the 6CE sources]\footnote{26 of the 38 objects in our
dataset are included: sources excluded because the 2D spectra are
unavailable are all those from the 3CRR and 7CRS samples, and for the same 6CE and 6C* sources
excluded from the composite spectra of Fig.~\ref{fig:composites};
namely, 6CE0902+3419, 6CE1141+3525 and 6CE1232+3942 from the 6CE
sample and 6C*0106+397 from 6C*.}. Our
results are illustrated in Fig.~\ref{fig:lyaextentvssize} where we
plot the extent of the Ly$\alpha$ region along the slit versus the
projected linear size of the radio source.  It is apparent that there
exists a strong correlation between the extent of the Ly$\alpha$
emission and the projected size of the radio source, with a $r_{\rm e}
= 0.54$ with a 0.5\% probability of this correlation occurring by
chance.  This is in agreement with previous work addressing this by
van Ojik et al. (1997) for objects at $z > 2.1$, although the majority
of their objects (which are also plotted in
Fig.~\ref{fig:lyaextentvssize}) lie above our best-fit to the
correlation (Fig.~\ref{fig:lyaextentvssize}). There are two possible
reasons for this. First, the sample of van Ojik et al. is not complete
in the sense of the 3CRR, 6CE and 7CRS samples, and does not have the well defined
filtering criteria of the 6C* sample: their sample was selected on
the basis of bright Ly$\alpha$ emission lines and there is an obvious
danger of them missing compact low-luminosity Ly$\alpha$
emitters. Second, the length of our exposures, which were kept
relatively short to permit more pointings in a single observing run,
might have led to some of the Ly$\alpha$ flux falling below our
detection limit.  

The results of Fig.~\ref{fig:lyaextentvssize} are in agreement with
the idea that additional UV-opaque emission line material is produced
at radii $\leq 50$\,kpc by interactions between the expanding radio
source and its environment. For linear sizes in the range $10 \leq D
\leq 100$\,kpc the sizes of the corresponding Ly$\alpha$ emitting
regions are scattered up to a maximum value which closely tracks the
size of the radio source, and thus provides the envelope driving the
correlation. 

There are two 6C* sources (6C*0135+313 and 6C*0208+344)
which appear to lie clearly above the envelope given by the (dotted)
line of equality in Fig.~\ref{fig:lyaextentvssize} [One of these
(6C*0135+313) was the prominent outlier noted in
Fig.~\ref{fig:lumvslum}.] These both lie at sufficiently low values of
$D$ that it is certainly plausible that there is sufficient dense
narrow-line material at the radii in question to produce detectable
Ly$\alpha$ emission before any boosting effects caused by the
expanding radio source (see also van Ojik et al. 1997). In the case of
6C*0135+313, detection of Ly$\alpha$ at large radii (for a small $D$
source) might be linked to the anomalous high Ly$\alpha$ luminosity
of this source (Fig.~\ref{fig:lumvslum}).

The existence of at least
one $D \sim 100$\,kpc source with a compact region is also not too
surprising: the size of the Ly$\alpha$ emitting region will clearly
involve a competition between narrow-line covering factor enhancement processes
(e.g. Bremer et al. 1997) and reduction processes (e.g. Kaiser et
al. 2000) which might be expected to vary considerably from
source-to-source given perhaps only subtle environmental
differences. We expect from the results of
Best et al. (2000) that the correlation between Ly$\alpha$ size and $D$ will
break down at $D \geq 100$\,kpc, presumably because of the absence of
the seed material for luminous narrow-line clouds at very large radii.

\section{The importance of source age over environment}\label{sec:age}

In this section we develop an argument based on the correlations
discussed in Secs. ~\ref{sec:lumlum}, ~\ref{sec:sizevslinelum} \&
~\ref{sec:sizesize} which suggests that source age has a much
more profound influence on radio source properties than environment.

We adopt the simple model\footnote{More complicated models exist which
account for synchrotron and inverse-Compton cooling (e.g Kaiser,
Dennett-Thorpe \& Alexander 1997), and the r\^ole of the hotspot
in governing the electron energy spectrum injected into the lobe
(Blundell et al. 1999), but for sources in the flux-density limited
samples this simple model suffices.} for radio source evolution used by Willott
et al. (1999), including an assumed power-law environmental density
profile $n(r)$ given by
\begin{equation}\label{eqn:n}
n(r) = n_{100} \left (\frac{r}{100 {\rm kpc}} \right )^{-\beta},
\end{equation}
where $\beta \simeq 1.5$ and is a dimensionless parameter.
In addition to the normalising factor $n_{100}$, the other key
physical variables influencing the evolution of a radio source are bulk
jet power $Q$ and source age $t$. These are linked to the observables
$L_{151}$ and $D$ via the equations
\begin{equation}\label{eqn:D}
D \propto \left ( \frac{t^{3} Q}{n_{100}} \right )^{\frac {1}{5-\beta}}
\propto Q^{2/7} n_{100}^{-2/7} t^{6/7}\,\,\,\,\, ({\rm for}\, \beta=1.5)
\end{equation}
\begin{eqnarray}\label{eqn:L151}
L_{151} \propto
Q^{(26-7\beta)/[4(5-\beta)]}n_{100}^{9/[4(5-\beta)]}t^{(8-7\beta)/[4(5-\beta)]}
\\ \nonumber \propto Q^{31/28} n_{100}^{9/14}t^{-5/28}\,\,\,\,\, ({\rm for}\,
\beta = 1.5).
\end{eqnarray}
The physical basis for these equations are the dimensional arguments
of Falle (1991), and energy conservation (assuming $Q$ to be
independent of $t$) following the ideas of Scheuer (1974) [see Willott
et al. (1999) and Blundell \& Rawlings (1999) for further details].

If we add to this model the assumption that $Q \propto Q_{\rm phot}
\propto L_{\rm Ly\alpha}$ [e.g. Rawlings \& Saunders (1991), Willott
et al. (1999) and Fig.~\ref{fig:ionise}] we can combine
eqns. ~\ref{eqn:D} \&~\ref{eqn:L151} to show that
\begin{equation}\label{eqn:reslya}
\Theta \equiv \frac{L_{\rm Ly\alpha}}{L_{151}^{6/7}} \propto D^{5/28}n_{100}^{-1/2},
\end{equation}
where our discussion of Sec.~\ref{sec:sizevslinelum} shows that the
log of the left-hand side of this equation is essentially the residual Ly$\alpha$
quantity plotted against $D$ in Fig.~\ref{fig:deBresid}.

The tentative correlation seen in Fig.~\ref{fig:deBresid} is in quantitative
agreement with the slow dependence of $\Theta$ on $D$ [within the
uncertainties, e.g. including some correction for \HI absorption
(Sec.~\ref{sec:sizesize}), the slope is $\sim 5/28$]. We can then conclude that the
range in $n_{100}$ cannot be so large that it would disrupt this
correlation, i.e. $\Delta n_{100} \leq 50$. 
Although the existence of any $\Theta - D$ correlation due to these
effects is highly tentative (see discussion of \HI absorption effects
in Sec.~\ref{sec:sizevslinelum}) the data in Fig.~\ref{fig:deBresid} are
consistent with a very gradual dependence of $\Theta$ on $D$ and require a
restricted range in density $n_{\rm e}$.

We are therefore led to a picture in which all $z > 1.75$ luminous
radio galaxies have similar environments (and values of
$n_{100}$). Thus, as a source ages, the linear size $D$ is driven
largely by the time dependence in eqn.~\ref{eqn:D} so that the spread
in $D$ is dominated by the spread in ages, which for the low-frequency
selected objects in question mean $\sim 2$ dex spread in
Fig.~\ref{fig:deBresid}. $L_{151}$ drops slowly with age
(eqn.~\ref{eqn:L151}) so, at a given $L_{151}$, the smallest sources
will be drawn from objects with a lower $Q$ and have lower Ly$\alpha$:
this is a possible origin for the tentative correlation between
$\Theta$ and $D$ in Fig.~\ref{fig:deBresid}. As the shocks pass
through the $r\,\ltsim\, 50$\,kpc region, compression of cool clouds
(e.g. Lacy et al. 1998) will temporarily increase the ratios of lines
like [OII] and CII] to Ly$\alpha$, thus explaining the
anti-correlation of residual [OII] and $D$ (Willott et al. 1999), the
differences between the composite spectra in Fig.~\ref{fig:composites}
(Table~\ref{tab:composites}) and the results of Best et al. (2000).
Shredding of emission-line clouds can increase the
effective covering factor, producing a moderate amount of additional
extended Ly$\alpha$ emission. This explains the correlation between
Ly$\alpha$ size and $D$ in Fig.~\ref{fig:lyaextentvssize}. As the
source passes to $D \,\gtsim\,100$\,kpc the seed material for the
clouds is now completely contained within the high-pressure radio
lobes, and as argued by Binette et al. (2000) the clouds which
condense are not dense enough, or too far from the ionising nucleus,
to produce detectable Ly$\alpha$ emission; they will also shadow too
small a fraction of the inner Ly$\alpha$ emission to produce
detectable \HI absorption.

Despite our data agreeing in all important respects with those of van
Ojik et al. (1997) we reach an opposite conclusion. They favour an
environmental/frustrated scenario in which the smallest high-redshift
radio sources reside in the densest (e.g. cluster) environments. While
we cannot rule out subtle effects of this type, we have argued that
the variation in the age of radio sources is the dominant effect
on its size, and variations in age, not environment are (after
$Q$) the most important drivers of the emission line properties.

The main argument used by van Ojik et al. (1997) against the scenario
we now prefer was a
perceived difficulty explaining the decrease in \HI absorption as the
radio source propagates to large $D$. As Binette et al. (2000) have recently
argued, this was probably because van Ojik et al. (1997) assumed that the emission and
absorption gas was co-spatial with similar physical
conditions. Binette et al. (2000) present a compelling argument that
absorption is caused by ionised material in an outer halo which may well provide the seed
material for Ly$\alpha$-emitting gas, but which has a much lower
density and much higher filling factor prior to the passage of the
radio source shocks.

Our contention that variations in source age are more important than
variations in source environment has some interesting
implications. First, it fits in naturally with the explanation for the
well-known anti-correlation between $D$ and $z$ preferred by Blundell et al. (1999) and
Blundell \& Rawlings (1999) - the so-called `youth-redshift
degeneracy' - and argues against the `evolving environments' explanation preferred
by Neeser et al. (1995). Second, it suggests that there is not a
significant spread in the environmental properties amongst luminous high-redshift radio
galaxies. It should be noted that Bremer et al. (1997) reach similar
conclusions from their starting hypothesis that luminous radio sources
inhabit the central regions of massive cooling-flow
clusters. 

Irrespective of the details, our broad conclusion is perhaps
not too surprising. The hypothesis that radio galaxies pinpoint
recently collapsed structures is not new (e.g. Haehnelt \& Rees 1993),
and since the density inside such systems depends only on the epoch of
collapse, the small range in $z$ probed by this and similar studies
should yield only a small variation in environmental properties.

\section{Conclusions}\label{sec:conclusions}
In this section we reiterate our main conclusions derived from new
studies made possible by the
6C* sample.

\begin{itemize}

\item The filtering criteria and flux-density limits employed in the
6C* sample to find objects at $z > 4$ are such that the fraction of
high-redshift objects is significantly larger than if no filtering were
employed, and this allowed the discovery of 6C*0140+326 at $z = 4.41$ and
6C*0032+412 at $z = 3.66$. The median redshift of the 6C* sample is
$\sim 1.9$ whereas the median redshift of a complete sample at the
flux-density of 6C* is $\sim 1.1$.

\item We find a rough proportionality between the luminosity of the
narrow Ly$\alpha$ emission line $L_{\rm Ly\alpha}$ and the rest-frame
151-MHz radio luminosity $L_{151}$, for sources at $z > 1.75$. This
correlation is similar to that found by Willott et al. (1999) for the
luminosity of [OII] versus $L_{151}$ for radio sources and may be
interpreted in the same way, namely as evidence for a primary
relationship between the accretion rate and the jet power of radio
sources (Rawlings \& Saunders 1991; Willott et al. 1999).

\item We present tentative evidence that high-redshift radio sources with
large projected linear sizes ($D > 70$\,kpc) have higher ionisation
states than their smaller counterparts. This is in agreement with the
work of Best et al. (2000) for similarly radio luminous 3CRR radio
sources at $z \sim 1$.

\item We find a weak correlation between residual $L_{\rm Ly\alpha}$
(having corrected for the strong correlation between $L_{\rm
Ly\alpha}$ and $L_{151}$) and $D$, whereas previous work on samples
with a lower median-redshift (Willott et al. 1999) found a weak
anti-correlation between the residual luminosity of [OII] and the size
of the radio source. This may be explained in two ways: (i) the
smaller sources have stronger associated absorption of Ly$\alpha$ than the larger
sources, and (ii) the slope of the residual $L_{\rm Ly\alpha}-D$ is
similar to the power-law dependency of radio jet-power on radio
source size from simple evolutionary models of radio sources (Willott
et al. 1999). If effect (ii) is real then the Ly$\alpha$ luminosity
seems to trace the underlying ionising continuum much better than
[OII], which may be boosted in small sources by compressions
associated with shocks caused by the radio source as
it propagates through regions close ($\ltsim 50$\,kpc) to the
nucleus.

\item There exists a strong correlation between the size of the
Ly$\alpha$ emission region and the size of the radio source in the
sense that the total Ly$\alpha$ size is $\ltsim\, D$ for $D \ltsim 100$\,kpc. This
suggests that extra Ly$\alpha$ in emission is being promoted by the
expansion of the radio source, perhaps via a shredding mechanism
(Bremer et al. 1997) which increases the covering factor to quasar
radiation from the nucleus.

\item We combine the observational results to develop an argument that
suggests variations in source age have a stronger influence
on the radio/emission-line properties of high-redshift radio luminous
galaxies than variations in environment. A small range in
environmental properties is deduced which is in line with the
hypothesis that luminous high-redshift radio sources are triggered in
recently collapsed massive structures.

\end{itemize}

\section*{Acknowledgements}
We thank Alan Stockton and Susan Ridgway for making initial
spectroscopic observations of some of the objects in the 6C* sample.
We are grateful to Adam Stanford and Arjun Dey for assistance in
obtaining the Keck spectroscopy, and to Steve Dawson for his help
during the observing at Lick. Thanks also to the referee, Philip Best for
some very useful comments.
The WHT is operated on the island of La
Palma by the Isaac Newton Group in the Spanish Observatorio del Roque
de los Muchachos of the Instituto de Astrofisica de Canarias.  The
W.M.Keck Observatory is operated as a scientific partnership among the
University of California, the California Institute of Technology, and
the National Aeronautics and Space Administration. The Observatory was
made possible by the generous financial support of the W.M.Keck
Foundation.

\section*{Appendix : Sources excluded from 6C*}
We show in Figs.~\ref{fig:rad0111} \& ~\ref{fig:rad0141} NVSS radio
maps of the five radio sources from the original 6C* sample
excluded from the final version because their angular sizes are now
known to be $> 15$\,arcsec. We also present 1-D optical spectra for
three of these sources (Fig.~\ref{fig:excspectra} \&~\ref{fig:excspectra2} ) taken before it was
clear that their true angular sizes exceeded the $\theta < 15$\,arcsec
filtering criterion of the 6C* sample.

\makebox{}

{\bf 6C*0100+312} This source is a quasar at $z=1.189$ (see
Fig.~\ref{fig:excspectra}), but is excluded from the 6C* sample as the
angular size of the radio source is $\approx 90$\,arcsec
[Fig.~\ref{fig:rad0111} and the  radio map in Blundell et al. (1998)].

{\bf 6C*0107+448} Our 8.4 GHz map of this sources (Blundell et
al. 1998) revealed radio structure with four components on a scale
$\ltsim 10$\,arcsec. The NVSS map suggested that these components may
be part of a radio lobe of a large (1.5 arcmin) double radio
source. However, a deep $R-$band image revealed emission co-spatial
with the bright radio component in our 8.4 GHz and led us to believe
that it was indeed an isolated source. Spectroscopy failed to resolve
the redshift of the source, however continuum and a strong emission
line were observed, which we associated with the faint $R-$band
object. A subsequent radio map at 1.4 GHz (Jea01) has now revealed
that this source is indeed a large classical double and the emission
we see in our optical imaging data is either associated with the
hot-spot or a foreground object.

{\bf 6C*0111+367} We have now determined that the brightest component
on the original 4.9\,GHz radio map (Blundell et al. 1998) is a hot-spot associated
with a larger source. The faint radio emission in the 8.4\,GHz map is
now identified with an object on our $R$-band image (Jea01) at 01
11 29.8 +36 43 27.5 (B1950), and we take this as the core, the other
weak feature on the 8.4\,GHz map at 01 11 27.8 +36 43 19.0 (B1950) is now
assumed to be a second lobe as these three components are roughly
co-linear. Thus, the radio structure is $\approx 41$\,arcsec; its
redshift is unknown.

{\bf 6C*0120+329} This source has an angular size $\theta \approx
95$\,arcsec (Parma et al. 1986), so is excluded from the
sample. The redshift of this source is $z=0.0164$ (de Ruiter et al. 1986).

{\bf 6C*0141+425} Blundell et al. (1998) mentioned that this is possibly a
double radio source with $\theta \approx 113$\,arcsec and the NVSS
map confirms this. The redshift of this source is $z = 0.0508$ and the
1-D spectrum is presented in Fig.~\ref{fig:excspectra2}.

\begin{figure*}[!ht]
{\hbox to 0.95\textwidth{\epsfxsize=0.4\textwidth
\epsfbox{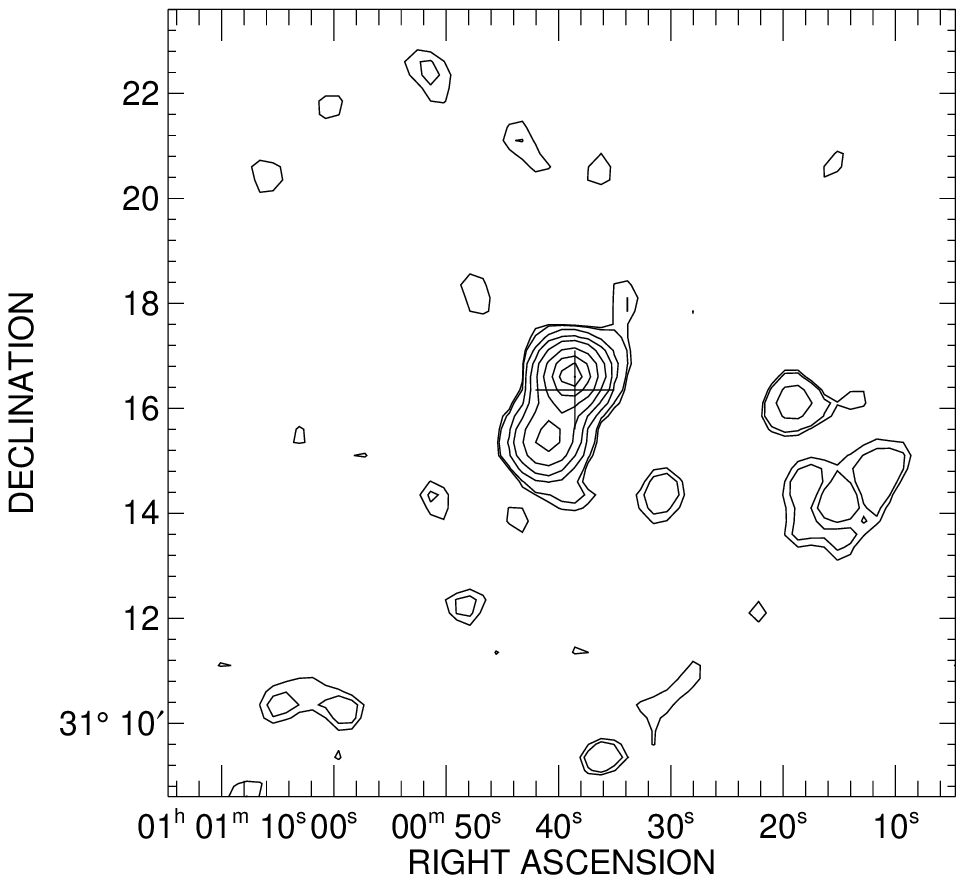}
\epsfxsize=0.4\textwidth
\epsfbox{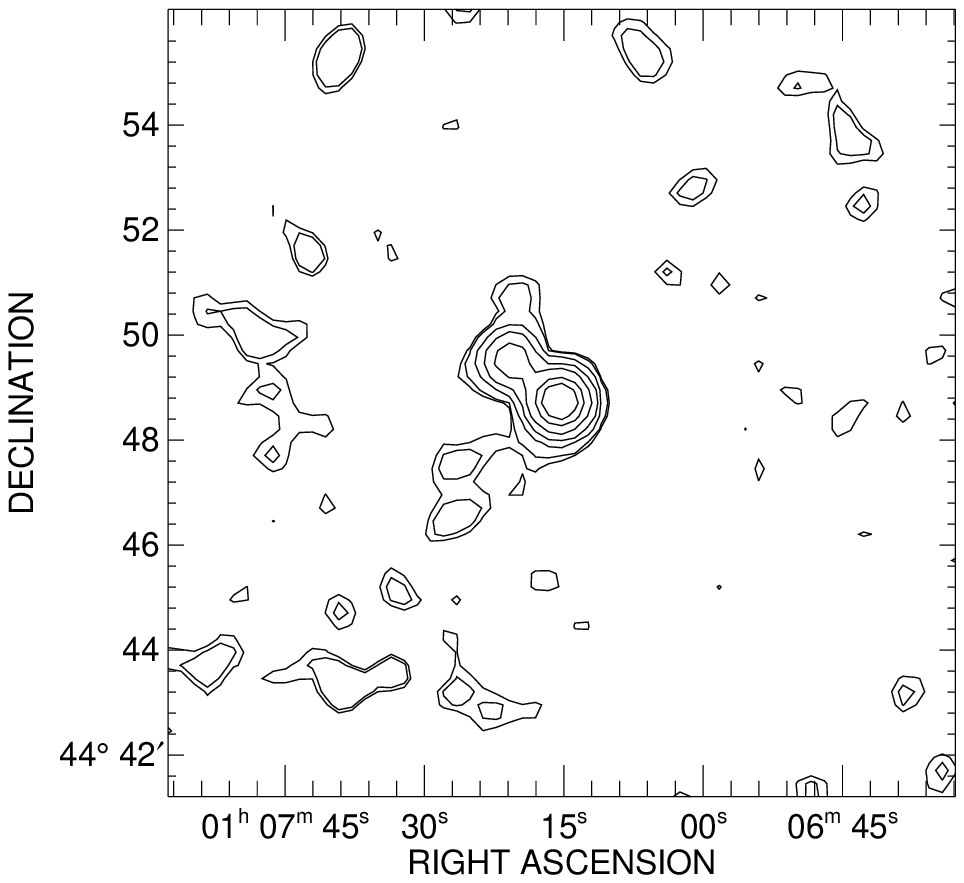} }}
{\caption{\label{fig:rad0100} NVSS radio maps of two 6C* radio sources
excluded from the final sample; the cross marks the optical
positions.  6C*0100+312 (left) is now identified with a quasar
(Fig.~\ref{fig:excspectra}) with an angular size of $\approx
1.5$\,arcmin from our 1.5\,GHz map (Blundell et al. 1998). The NVSS 1.4\,GHz map
shown resolves the source with the 45\,arcsec beam and has a peak flux
of 63.7 mJy/beam (contour levels in mJy per synthesised beam are 62, 50, 35, 20, 10, 5,
2.5, 1, 0.75). 6C*0107+448 (right), the optical identification for
this source is still ambiguous from our $R-$ and $K-$band images and the digitised sky
survey, with no obvious bright identification. This $\approx
1.5$\,arcmin\, source has a peak flux of 50.8\,mJy/beam (contour levels in
mJy per synthesised beam are 45, 38, 32, 26, 20, 14, 7, 2).}}

\end{figure*}

\begin{figure*}[!ht]
{\hbox to 0.95\textwidth{\epsfxsize=0.4\textwidth
\epsfbox{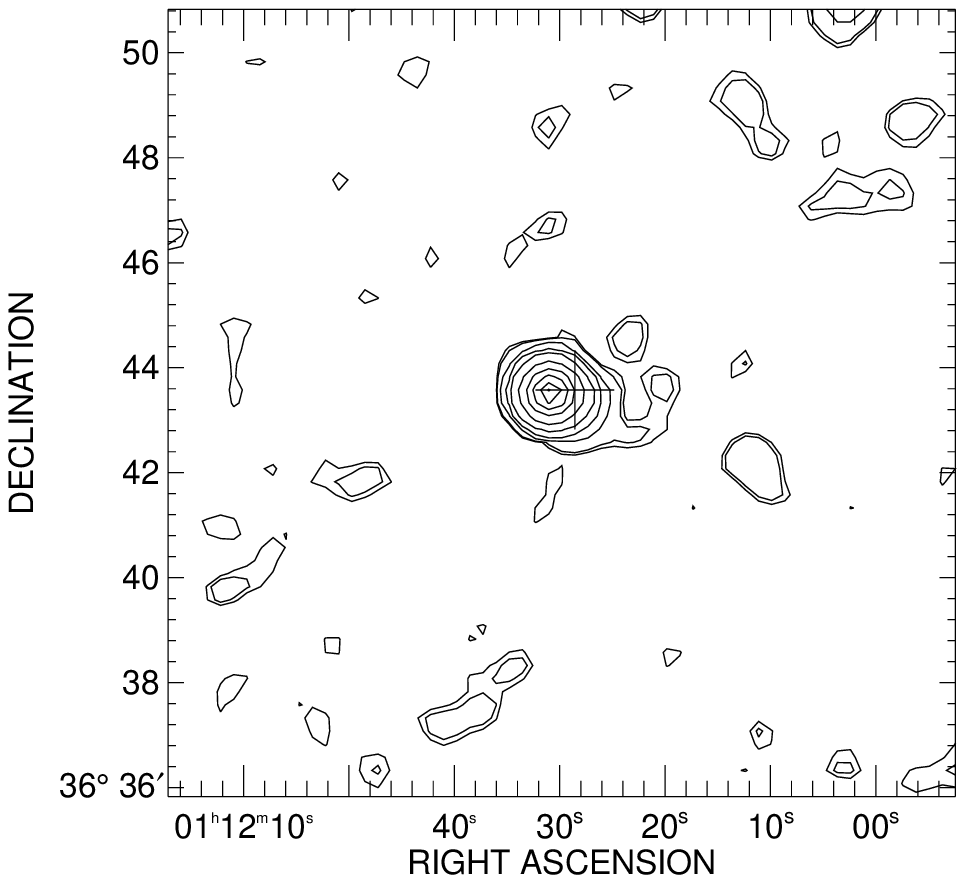}
\epsfxsize=0.4\textwidth
\epsfbox{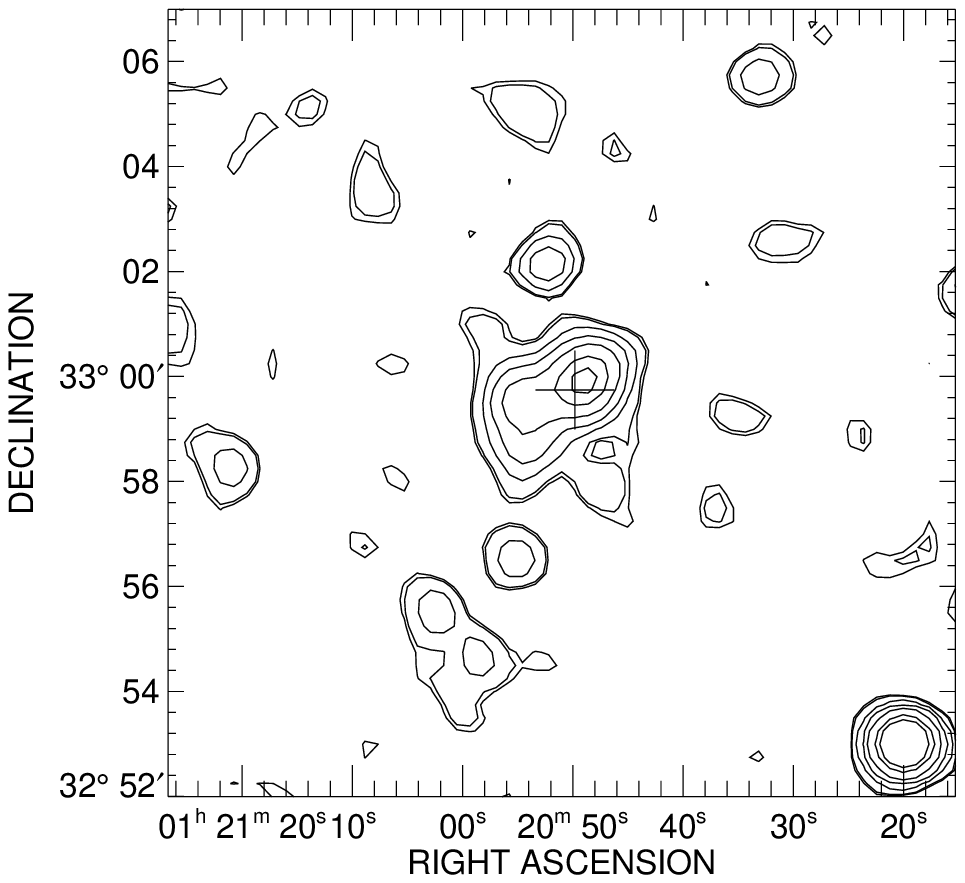} }}
{\caption{\label{fig:rad0111} 
NVSS radio maps of two 6C* radio sources
excluded from the final sample; the crosses mark the optical positions.
6C*0111+367 (left) is resolved in this NVSS map and
our 1.49\,GHz map suggests an angular size of $\approx 41$\,arcsec, it
has a peak flux of 123.9\,mJy/beam (contour levels in mJy per synthesised beam are 120, 100, 75,
50, 25, 10, 5, 2.5, 1, 0.75). 6C*0120+329 (right) is resolved with the 45\,arcsec beam of
NVSS and has a peak flux of 36.4\,mJy/beam (contour levels in mJy per
synthesised beam are 30, 20,
10, 5, 2.5, 1, 0.75). }}
\end{figure*}

\begin{figure*}[!ht]
{\hbox to 0.95\textwidth{\epsfxsize=0.4\textwidth
\epsfbox{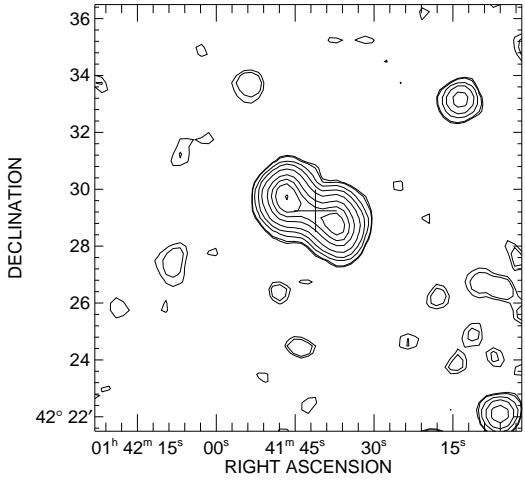} }}
{\caption{\label{fig:rad0141} 6C*0141+425 (right) is also resolved
and has a peak flux of 54.9\,mJy/beam (contour levels in mJy per
synthesised beam are 55, 40, 30,
20, 10, 5, 2.5, 1, 0.75).}}
\end{figure*}

\begin{figure*}
{\hbox to \textwidth{\epsfxsize=0.48\textwidth
\epsfbox{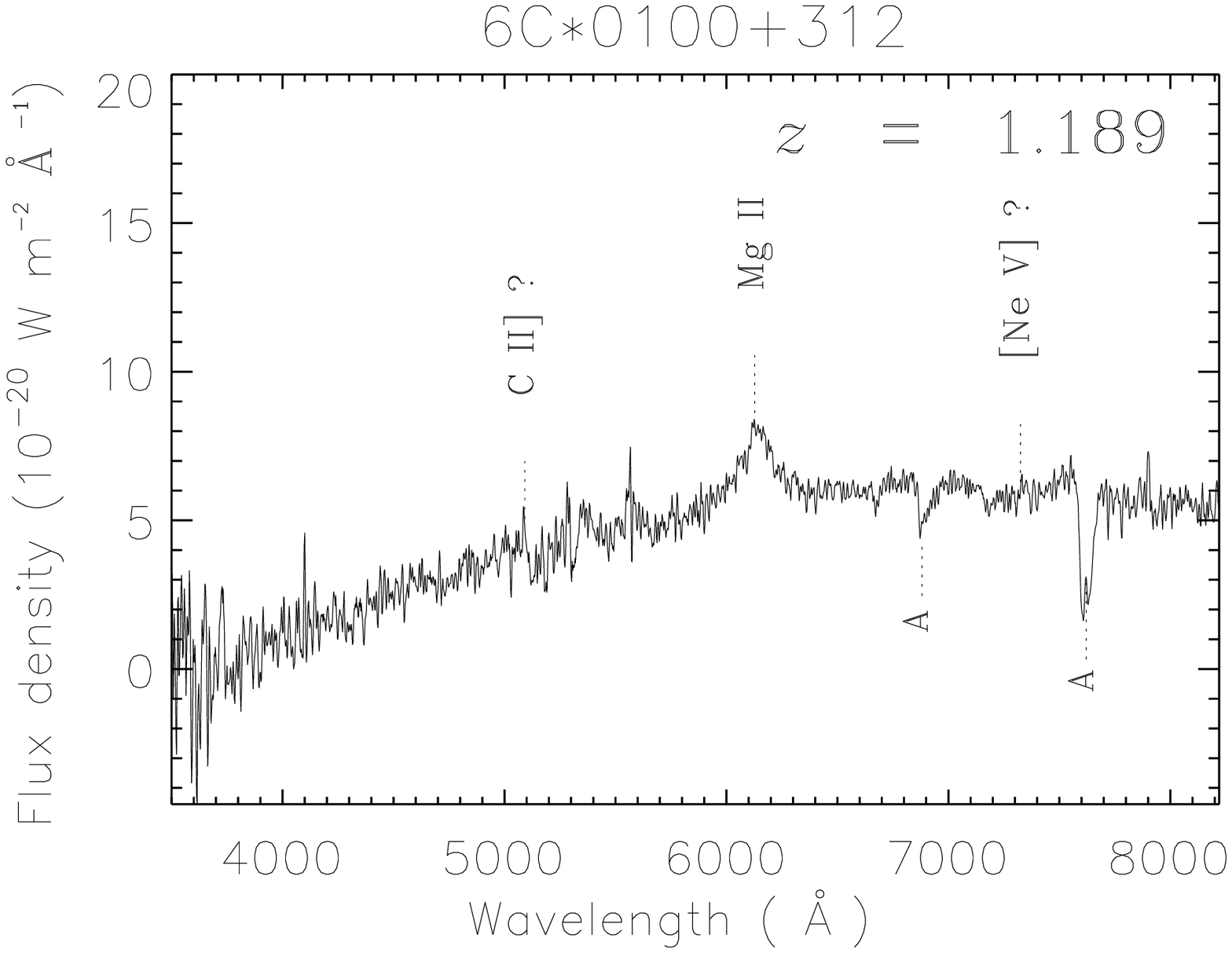}
\epsfxsize=0.48\textwidth
\epsfbox{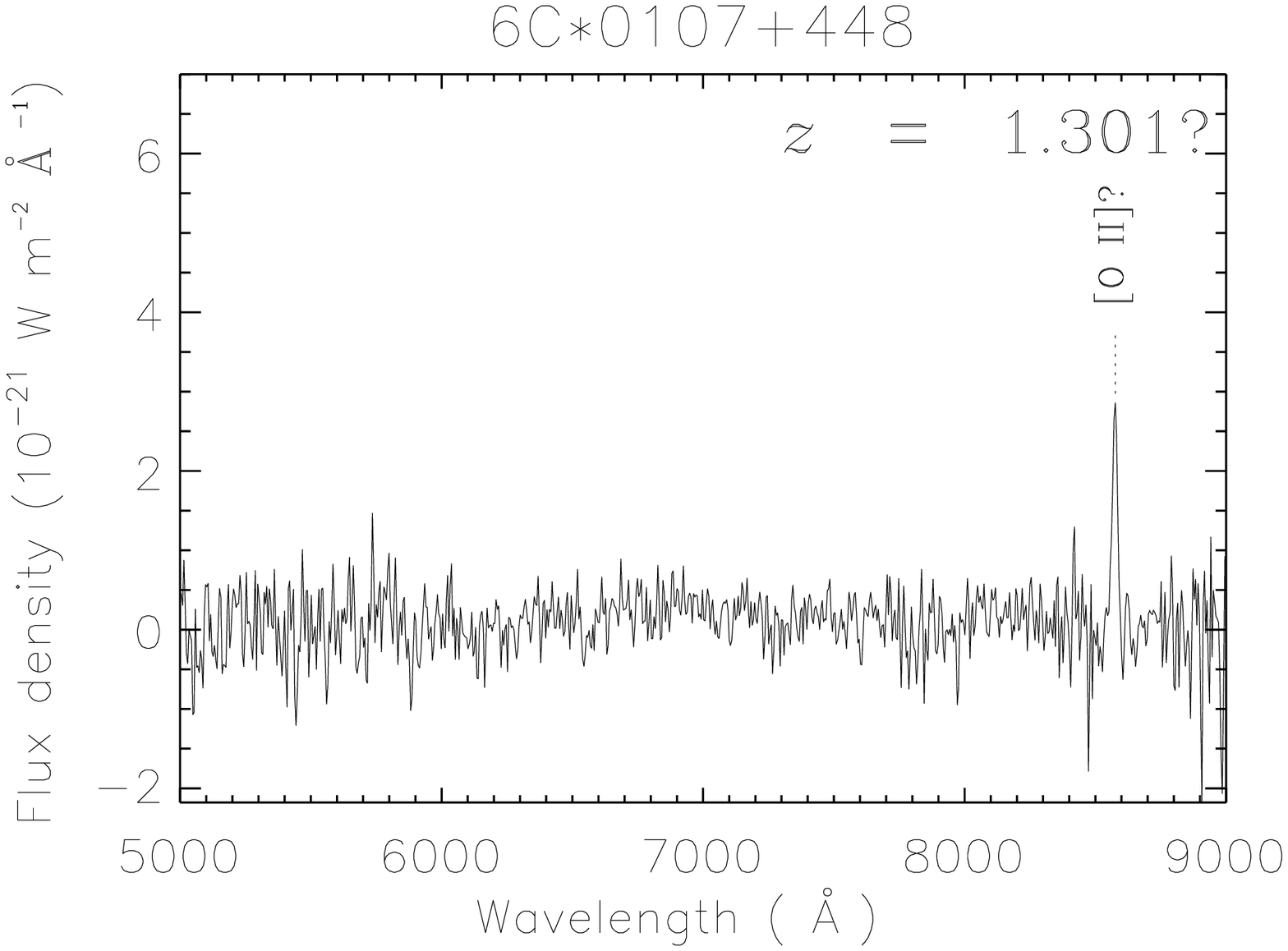} }}
{\caption{\label{fig:excspectra} Spectra of two of the sources
excluded from the 6C* sample on account of their angular size; the
absorption bands marked `A' are atmospheric bands which have not been
fully removed from the data. 6C*0100+312 (left) is a quasar with a
broad line most plausibly associated with MgII$\lambda 2799$\,\AA\,
(FWHM = 133 $\pm$ 21 \AA\, and flux = 3.9$\times 10^{-18}$ W m$^{-2}$ $\pm$
20\%) at $z = 1.189$. This spectrum was taken at the WHT on the 1995
observing run with an integration time of 500 seconds with a slit
width of 3.2\,arcsec; the PA of the slit was 160$^{\circ}$ to align
with the radio structure. 6C*0107+448 (right) is a spectra of what is
possibly emission from the hotspot of this large classical double. The
identification of the emission line at 8575\,\AA\, is undetermined,
but could plausibly be [OII] at $z = 1.301$.}}
\end{figure*}

\begin{figure*}
{\hbox to \textwidth{\epsfxsize=0.48\textwidth
\epsfbox{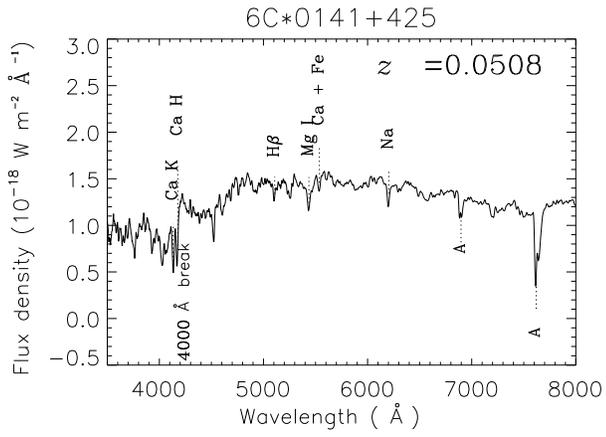} }} 
{\caption{\label{fig:excspectra2} 6C*0141+425
(right) is a galaxy at $z = 0.0508$ with the spectrum dominated by
stellar absorption features. This spectrum was taken at the Lick-3m
telescope in 1998 with an integration time of 1200 seconds and a slit
width of 2.0\,arcsec.}}
\end{figure*}

\end{document}

%% file: macro.tex
\newcommand{\gtsim}{\mbox{{\raisebox{-0.4ex}{$\stackrel{>}{{\scriptstyle\sim}}
$}}}}
\newcommand{\ltsim}{\mbox{{\raisebox{-0.4ex}{$\stackrel{<}{{\scriptstyle\sim}}
$}}}}

\newcommand{\figstart}[1]
    { \begin{figure}[htb]
      \begin{picture}(0,#1) }
\newcommand{\figend}[4]
    { \end{picture}
      \special{#1}
      \caption[#2]{#3}
      \label{#4}
      \end{figure} }